\documentclass[useAMS,usenatbib]{mn2e}
\usepackage{amsmath}
\usepackage{amssymb}
\newcommand{\gdet}{\sqrt{-g}}
\newcommand{\p}{\partial}
\usepackage{graphics}
\usepackage{graphicx}
\usepackage{epstopdf}
\usepackage{footnote}
\usepackage{tablefootnote}
\usepackage{txfonts}
\usepackage{lipsum}
\usepackage{float}

\makesavenoteenv{tabular}

\DeclareSymbolFont{cmletters}{OML}{cmm}{m}{it}
\DeclareMathSymbol{v}{\mathalpha}{cmletters}{"76}

\usepackage[usenames,dvipsnames,svgnames,table]{xcolor}
\usepackage{hyperref}
\definecolor{darkblue}{rgb}{0.0,0.0,0.3}
\hypersetup{colorlinks,breaklinks,
            linkcolor=darkblue,urlcolor=darkblue,
            anchorcolor=darkblue,citecolor=darkblue}

\newcommand{\dd}{{\rm d}}
\title[Electron Thermodynamics in Black Hole Accretion]{Electron Thermodynamics in GRMHD Simulations of Low-Luminosity Black Hole Accretion }
\author[S. M. Ressler, A. Tchekhovskoy, E. Quataert, M. Chandra, C. F. Gammie]{S. M. Ressler$^{1},$ A. Tchekhovskoy$^{1}$\thanks{Einstein Fellow}, E. Quataert$^{1}$, M. Chandra$^{2}$, C. F. Gammie$^{2,3}$\\
$^{1}$Departments of Astronomy \& Physics, Theoretical Astrophysics Center, University of California, Berkeley, CA 94720 \\
$^{2}$Department of Astronomy, University of Illinois, 1002 West Green Street, Urbana, IL 61801\\
$^{3}$Department of Physics, University of Illinois, 1002 West Green Street, Urbana, IL 61801}

\begin{document}
\pagerange{\pageref{firstpage}--\pageref{lastpage}} \pubyear{2015}
\maketitle
\label{firstpage}
\begin{abstract}
Simple assumptions made regarding electron thermodynamics often limit the extent to which general relativistic magnetohydrodynamic (GRMHD) simulations can be applied to observations of low-luminosity accreting black holes. We present, implement, and test a model that
  self-consistently evolves an entropy equation for the electrons and
  takes into account the effects of spatially varying electron heating
  and relativistic anisotropic thermal conduction along magnetic
  field lines.  We neglect the back-reaction of electron pressure on
  the dynamics of the accretion flow.  Our model is appropriate for
  systems accreting at $\ll 10^{-5}$ of the Eddington accretion rate,
  so radiative cooling by electrons can be neglected.  It can be
  extended to higher accretion rates in the future by including
  electron cooling and proton-electron Coulomb collisions.  We present
  a suite of tests showing that our method recovers the correct
  solution for electron heating under a range of circumstances,
  including strong shocks and driven turbulence.  Our initial
  applications to axisymmetric simulations of accreting black holes
  show that (1)~physically-motivated electron heating rates that
  depend on the local magnetic field strength yield electron
  temperature distributions significantly different from the constant
  electron to proton temperature ratios assumed in previous work, with higher electron temperatures concentrated in the coronal region between the disc and the jet;
  (2)~electron thermal conduction significantly modifies the electron temperature in
  the inner regions of black hole accretion flows if the
  effective electron mean free path is larger than the local
  scale-height of the disc (at least for the initial conditions and
  magnetic field configurations we study).  The methods developed in
  this work are important for producing more realistic predictions for
  the emission from accreting black holes such as Sagittarius A* and
  M87; these applications will be explored in future work.

 \end{abstract}

\section{Introduction}

A wide variety of low luminosity accreting black holes are currently
interpreted in the context of a Radiatively Inefficient Accretion
Flow (RIAF) model that describes a geometrically thick, optically thin disc with a low accretion rate and luminosity.  In particular, this is true of the black hole at the center of our galaxy, Sagittarius A* \citep{Narayan1998}, the black hole at the center of Messier 87 \citep{Reynolds1996}, and other low luminosity Active Galactic Nuclei (AGN), as well as a number of X-ray binary systems (see \citealt{Remillard2006} for a review).  The gas densities in these systems are low enough that the time scale for electron-ion collisions is much longer than the time scale for accretion to occur, so a one-temperature model of the gas is no longer valid (as originally recognised by \citealt{Shapiro1976}, \citealt{Ichimaru1977}, and \citealt*{Rees1982}). Instead, a better approximation is to treat the electrons and ions as two different fluids, each with its own temperature.  

Calculating the emission from accreting plasma requires predicting
the electron distribution function close to the black hole.  To date, time
dependent numerical models of RIAFs that attempt to directly connect
to observations often assume a Maxwellian distribution with a constant electron to proton temperature
ratio, $T_e/T_p$, and take the results of GRMHD simulations as the solution for the total gas temperature, $T_g = T_p + T_e$ \citep{Dibi2012, Drappeau2013,Mosci2009}. This neglects, however, several physical processes that have different effects on the electron and proton thermodynamics and that are currently only included in one-dimensional semi-analytic models. Such effects include electron thermal conduction (e.g., \citealt{Johnson2007}), electron cooling (e.g., \citealt{Narayan1995}), and non-thermal particle acceleration and emission (e.g., \citealt{Yuan2003}).  To date, extensions of the simple $T_p/T_e =$ const. prescription have been limited to post-processing models that do not self-consistently evolve the electron thermodynamics over time.  Examples include the prescription of \citet{Mosci2014} which takes $T_p/T_e = $ const. in the disc proper but sets $T_e =$ const. in the jet outflow region, as well as the model of \citet{Shcherbakov2012}, who solve a 1-D radial equation for $T_p-T_e$ at a single time-slice in the midplane to obtain a functional relationship $T_p/T_e = f(T_{g})$ that is then applied to the rest of the simulation.   To enable a more robust connection between observations of accreting black holes and numerical models of black hole accretion, it is critical to extend the detailed thermodynamic treatment of electrons used in 1D calculations to multi-dimensional models.   This is the goal of the current paper.  In particular, we describe numerical methods for separately evolving an electron energy equation in GRMHD simulations.  We focus on including heating and anisotropic thermal conduction in these models.  Future work will include electron radiative cooling and Coulomb collisions between electrons and protons.   

In a turbulent, magnetised plasma, electrons and ions are heated at different rates depending on the local plasma conditions (e.g., \citealt{Quataert1999,Cranmer2009,Howes2010,Sironi2015}). Furthermore, since the electron-to-proton mass ratio is small, electrons will both conduct and radiate their heat much more efficiently than the ions.  The combination of these effects leads to the expectation that, in general, $T_e<T_p$.  In the present paper, we thus neglect the effect of the electron thermodynamics on the overall dynamics of the accretion flow.  This allows us to treat the simulation results as a fixed background solution on top of which we independently evolve the electrons.  Even if we find that $T_e \sim T_p$ in some regions of the disc, this treatment may still be a reasonable first approximation given the uncertainties in the electron physics.

The neglect of electron cooling in the present paper is reasonable for
systems accreting at $\lesssim 10^{-5}$ of the Eddington rate, $\dot M_{\rm Edd}$, so that the synchrotron cooling time is much longer than the accretion time \citep{Mahadevan1997}.  In particular, this likely includes Sagittarius A* in the galactic center. The application of our methodology to Sgr A* is particularly important given the wealth of multi-wavelength data (e.g.,  \citealt{Serabyn1997}, \citealt{Zhao2003}, \citealt{Genzel2003}, \citealt{Baganoff2003}, \citealt{Barriere2014}) and current and forthcoming spatially resolved observations with the Event Horizon Telescope \citep{Doeleman2008} and Gravity \citep{Gillessen2010}.

The goal of this paper is to present our formalism and methodology for
evolving the electron thermodynamics and to apply the results to
2D (axisymmetric) GRMHD simulations of an accreting black hole. We show the range of possible electron temperature distributions in the inner region of the disc, which directly impacts the predicted emission. Future work will explore the impact that these results have on the emission, spectra, and images of Sagittarius A*. 

The remainder of this paper is organised as follows.  \S\ref{sec:heat}
describes our theoretical model of electron heating and anisotropic
electron conduction while \S\ref{sec:numimp} describes the numerical
implementation of this model.  \S\ref{sec:test} contains tests of the
numerical implementation, \S\ref{sec:app} applies the model to a 2D
simulation of an accretion disc around a rotating black hole, and
\S\ref{sec:conc} discusses the implications of this application and
concludes. Boltzmann's constant, $k_b$, and proton mass, $m_p$, are
taken to be 1 throughout. We use cgs units, with Lorentz-Heaviside units for the
magnetic field (e.g., magnetic pressure is $b^2/2$), and a metric signature of $({-}{+}{+}{+})$. We also assume that the gas is mostly hydrogen and ideal.  Since we also assume $n_e \approx n_p \equiv n$,  then $\rho = m_e n_e + m_p n_p \approx m_p n = n$ (setting $m_p=1$), so we use $\rho$ and $n$ interchangeably.
 
\section{Electron Thermodynamics}

\label{sec:heat}

The accreting plasmas of interest are sufficiently low density that the electron-proton Coulomb collision time is much longer than the dynamical time and so a two-temperature structure can develop, with the protons and electrons having different temperatures (and, indeed, different distribution functions).   Moreover, at the low accretion rates where radiative cooling can be neglected, the electron-electron and proton-proton Coulomb collision times are also much longer than the dynamical time \citep{Mahadevan1997}.    However, the plasma densities are high enough that the plasma is nearly charge-neutral and so we assume that $n_e \approx n_p$.  We further assume that the electron flow velocity is the same as that of the protons.\footnote{More precisely, as in standard MHD, the relative velocity between electrons and protons required to produce currents that can maintain magnetic fields near $\beta \sim 1$ is orders of magnitude less than the mean sound speed.}   This need not strictly be true (e.g., in the solar wind the relative velocities of particle species can be of order the Alfven speed; e.g. \citealt{Bourouaine2013} ), but is a reasonable first approximation. A similar approach is often used in modeling the global dynamics of the low-collisionality solar wind (e.g., \citealt{Chandran2011}).

Under these assumptions, the key difference in the electron and proton physics lies in their different thermodynamics: the protons and electrons have very different heating and cooling processes that need to be separately accounted for.   Formally, because of the low collisionality conditions we should separately solve the electron and proton Vlasov equations.  This is computationally extremely challenging, however, particularly in the global geometry required to predict the emission from accreting plasmas (even local shearing box calculations using the particle-in-cell technique to solve the Vlasov equation require an unphysical electron-proton mass ratio, thus making it difficult to reliably model the electron thermodynamics; e.g., \citealt{Riquelme2012}). As a result, we assume a fluid model in this paper.     Our fluid approximation corresponds to taking moments of the Vlasov equation and applying closures on higher moments of the distribution function.  As we shall describe, our closure corresponds to specific models for the conductive heat flux, the viscous momentum flux, and the turbulent heating rate of each particle species.

 Our basic model is thus to take a single-fluid GRMHD solution (e.g., \citealt{Komiss1999,Gammie2003,DeVilliers2003}) as an accurate description of the total fluid (composed of both the electron and proton gas) dynamics and thus the accretion flow density, magnetic field strength, and velocity field. We evolve the electrons as a second fluid on top of this background solution.  The GRMHD solution may itself include viscosity and conduction as in \citet{ManiModel}.  Our assumption that the electrons do not back react on the flow dynamics  is formally valid in the limit that $T_e \ll T_p$, but should be a reasonable approximation so long as $T_e \lesssim T_p$ in regions of large plasma $\beta \gtrsim 1$, i.e., where gas pressure forces are dynamically important.  One advantage of not coupling the electron pressure to the GRMHD solution is that we can run multiple electron models in one simulation, allowing us to explore systematic uncertainties with a minimum of computational time.

In this paper, we focus on implementing electron heating and anisotropic conduction.   Coulomb collisions are straightforward to include but are negligible for the low accretion rates at which electron cooling can be neglected.   In future work, electron cooling will be self-consistently incorporated building on the BHlight code developed by \citet{BHLIGHT}.  

\subsection{Basic Model}

The stress-energy tensors for the electron and proton fluids in our model take the form:
\begin{equation}
\begin{aligned}
T_e^{\mu \nu}=&\left( \rho_e + u_{e}  + P_e \right)u_e^\mu u_e^\nu 
+ P_e  g^{\mu \nu} + \tau_e ^{\mu \nu} +q_e^\mu u_e^\nu + u_e^\mu  q_e^\nu  \\
T_p^{\mu \nu}=&\left( \rho_p + u_{p}  + P_p \right)u_p^\mu u_p^\nu 
+ P_p  g^{\mu \nu} + \tau_p ^{\mu \nu},
\end{aligned}
\label{eq:TeandTpdef}
\end{equation}
where $\rho_k$, $u_k$, and $P_k$ are the fluid frame density, internal energy, and pressure, respectively, $u_k^\mu$ is the fluid four-velocity in the coordinate frame, $\tau_k^{\mu \nu}$ is a general stress tensor that accounts for viscous effects, and $q_e^\mu$ is the heat flux carried by the electrons.  The subscript $k$ denotes $p$ or $e$ (and will also denote the total gas quantities labeled by $g$ below). We leave $\tau_k^{\mu \nu}$ as a general tensor that will be model-specific.  For each species, ignoring electron-electron, electron-ion, and ion-ion collisions, one can take the zeroth and first moment of the Vlasov equation to show that
\begin{equation}
\nabla_\mu \left(\rho_k u_k^\mu \right) = 0
\label{eq:mass}
\end{equation}
and
\begin{equation}
  \begin{aligned}
    \nabla_\mu T_e ^{\mu \nu} = -e n u_e^\mu F_\mu^{\textrm{  }\nu}  \\
     \nabla_\mu T_p ^{\mu \nu} = e n u_p^\mu F_\mu^{\textrm{  } \nu} ,
  \end{aligned}
  \label{eq:divep}
\end{equation}
where $F^{\mu \nu} $ is the electromagnetic field tensor. In ideal, single-fluid GRMHD in the absence of shocks, the conservation of entropy equation, $\rho T_g u^\mu \p_\mu s_g = 0$, where $s_g$ is the entropy per particle, follows directly from the conservation of particle number and the stress-energy (see page 563 in \citealt{MTW}). To derive entropy equations for the electron and proton fluid used in our model, we perform the same series of manipulations; namely, contracting both equations~\eqref{eq:divep} with $u^\nu$ (the total fluid velocity, which we take to be $\approx u_p^\mu \approx u_e^\mu$) and invoking equation~\eqref{eq:mass}, which give us:
\begin{equation}
\rho T_{e} u^\mu \partial_\mu s_{p}  =Q_p ,
\label{eq:proton}
\end{equation}
and
\begin{equation}
\rho T_{e} u^\mu \partial_\mu s_{e} = Q_e - \nabla_\mu q_e^\mu  - a_\mu q_e^\mu ,
\label{eq:electron}
\end{equation}
where we have defined the heating rate per unit volume for each species as a sum of viscous and Ohmic resistance terms, $Q_e \equiv  u_\nu \nabla_\mu \tau_e^{\mu \nu} +en u_e^\mu u^\nu F_{\mu}^{\textrm{  } \nu} $ and $Q_p \equiv  u_\nu \nabla_\mu \tau_p^{\mu \nu } -en u_p^\mu u^\nu F_{\mu}^{\textrm{  } \nu} $, and where $a^\mu \equiv u^\nu \nabla_\nu u^\mu$ is the four-acceleration, which accounts for gravitational redshifting of the temperature by the metric.  We can write the heating rates in terms of the electric field four-vector, $e^\mu \equiv u_\nu F^{\nu \mu}$, and the four-currents, $J_e^\mu \equiv -ne u_e^\mu$, $J_p^\mu \equiv ne u_p^\mu$ as\footnote{ We have kept the subscripts $e$ and $p$ for the four-currents (and thus four-velocities) in equation~\eqref{eq:Qep} because the details of the Ohmic heating depend on the small but non-zero velocity difference between the proton and electron fluids (or equivalently the velocity difference between the electron/proton fluid and the total fluid). An explicit expression for these terms would require a detailed kinetic theory calculation beyond the scope of the present work (i.e., some form of ``generalized Ohm's Law,'' as in, e.g., \citealt{Koide2010}).  In our model, as described in the text, numerical resistivity provides the Ohmic heating that is then distributed to electrons and protons according to a closure model obtained from previous work in kinetic theory.  }:
   \begin{equation}
   \begin{aligned}
    &Q_e = u_\nu \nabla_\mu \tau_e^{\mu \nu} +J_e^\mu e_\mu \\ &Q_p =  u_\nu \nabla_\mu \tau_p^{\mu \nu } +   J_p^\mu e_\mu.
   \end{aligned}
   \label{eq:Qep}
\end{equation}
The intuitive understanding of the Ohmic heating terms on the right-hand side of equation~\eqref{eq:Qep} is that they are $\sim \vec{J_k}\cdot {E}$ evaluated in the rest frame of the total fluid.  To derive the entropy equation for the total fluid, we first define several total fluid variables as a sum of electron and proton terms:  $\rho = \rho_p + \rho_e \approx n$, $u_g = u_p + u_e$, $P_g = P_p + P_e$, $T_g = T_p + T_e$, $J^\mu = J_p^\mu + J_e^\mu = en (u_p^\mu - u_e^\mu)$ and $\tau_{g}^{\mu \nu} = \tau_{e}^{\mu \nu} + \tau_{e}^{\mu \nu}$, denoting total gas mass density, internal energy, pressure, temperature, current, and viscous stress.  Then, using the thermodynamic identity, $\rho T_k u^\mu \partial_\mu s_k = u^\mu \partial_\mu u_k - \left(u_k + P_k\right) u^\mu \partial_\mu \log (\rho) $, we find that the entropy per particle of the total gas, $s_g$, satisfies the relation $\rho T_g u^\mu \partial_\mu s_g = \rho T_p u^\mu \partial_\mu s_p + \rho T_e u^\mu \partial_\mu s_e$, resulting in 
\begin{equation}
   \rho T_{g} u^\mu \partial_\mu s_{g} = Q - \nabla_\mu q_e^\mu  - a_\mu q_e^\mu, 
\label{eq:total}
\end{equation}
with the total heating rate per unit volume:
\begin{equation}
Q = Q_p + Q_e = u_\nu \nabla_\mu \tau_g^{\mu \nu} +J^\mu e_\mu.
\label{eq:Qtot}
\end{equation}

In practice, we use the electron entropy equation~\eqref{eq:electron} to evolve the electron thermodynamics. To determine the overall dynamics of the electron + proton gas, we use Maxwell's equations in addition to a standard, single-fluid GRMHD evolution representing the total gas. The equations for the latter are obtained by separately summing the electron and proton parts of equation~\eqref{eq:mass} and equation~\eqref{eq:divep}, resulting in a mass conservation equation, 
\begin{equation}
\nabla_\mu \left( \rho u^\mu \right) = 0,
\label{eq:masstot}
\end{equation} 
and an energy-momentum equation,
\begin{equation}
\nabla_\mu \left(T_g^{\mu \nu} + T_{EM}^{\mu \nu} \right) = -\nabla_\mu \tau_g^{\mu \nu},
\label{eq:gem}
\end{equation}
with the total gas stress-energy tensor,
\begin{equation}
\begin{aligned}
T^{\mu \nu}_g=&\left(\rho + u_{g}  + P_g \right)u^\mu u^\nu + P_g  g^{\mu \nu},
\end{aligned}
\label{eq:Tmunugas}
\end{equation}
and the electromagnetic stress energy tensor\footnote{Here we have chosen to absorb a factor of $(4\pi)^{-1/2}$ into the definition of $F^{\mu \nu}$.}, $T_{EM}^{\mu \nu} = F^{\mu \alpha}F^{\textrm{  } \nu}_\alpha - g^{\mu \nu} F_{\alpha \beta}F^{\alpha \beta}/4$.  We have used the identity $\nabla_\mu T_{EM}^{\mu \nu} = -J^\mu F_{\mu}^{\textrm{  } \nu}$ in equation~\eqref{eq:gem}, the assumption that $u_e^\mu \approx u_p^\mu \approx u^\mu$ in equation~\eqref{eq:Tmunugas}, and the charge-neutrality assumption $n_e = n_p = n$ throughout.  Furthermore, we have dropped the electron thermal conduction terms in the evolution of the total gas properties, though we keep them in the evolution of the electron entropy (equation \ref{eq:electron}).  This is consistent if $T_e \lesssim T_p$ since electron conduction will then affect the electron thermodynamics but not the overall stress-energy of the fluid.  Finally, we take $T_{EM}^{\mu \nu}$ to be given by the ideal MHD limit (i.e., $e^\mu \rightarrow 0$): 
\begin{equation}
T_{EM}^{\mu \nu} = b^2 u^\mu u^\nu +\frac{b^2}{2}g^{\mu \nu}-b^\mu b^\nu,
\end{equation}
where $b^\mu \equiv \epsilon^{\mu \nu \kappa \lambda}u_\nu F_{\lambda \kappa}/2$ is the magnetic field four-vector defined in terms of the Levi-Civita tensor, $\epsilon^{\mu \nu \kappa \lambda}$, and $b^2 \equiv b^\mu b_\mu$ is twice the magnetic pressure.  With these assumptions, equations~\eqref{eq:masstot}, \eqref{eq:gem}, and Maxwell's equations are simply the standard single-fluid equations of ideal GRMHD except with an explicit viscosity tensor.  In standard conservative GRMHD codes (including the one used in this work), this viscosity tensor is not included explicitly but implicitly generated numerically by the Riemann solver.  Furthermore, the Riemann solver also introduces a finite numerical resistivity into Maxwell's equations, allowing for a nonzero $e^\mu$ (and thus nonzero Ohmic heating).  For further discussion of these points, see \S~\ref{sec:heatcons}.  

To summarise, we take a standard single fluid GRMHD evolution of $u^\mu, \rho, u_g$, and $P_g$ as a reasonable estimate of the total gas properties.   This corresponds to assuming that electron conduction has a negligible contribution to the dynamics of the total gas and that the adiabatic index is independent of the electron thermodynamic quantities (e.g.; $P_g/u_g \equiv [P_p+P_e]/[u_e+u_p] \equiv \gamma-1 \approx$ some function of total gas quantities only).   Formally, this assumption requires that the electron internal energy is small compared to the proton internal energy.  From this, we can calculate the heating directly from equation~\eqref{eq:total} (dropping the conduction terms) without requiring an analytic expression for $Q$. Finally, we use knowledge of the nature of heating in a collisionless plasma obtained from kinetic theory (described in \S\ref{sec:tauchi}) to relate the heating rate per unit volume of the electrons, $Q_e$, to that of the total fluid, $Q$, and directly add it to the electron entropy equation (equation~\ref{eq:electron}), as described in \S\ref{sec:electron-heating}.  This completes our model.

For simplicity, we assume that the adiabatic indices of the electron, $\gamma_e$, proton, $\gamma_p$, and total gas, $\gamma$, are constants, where $P_k = (\gamma_k -1) u_k$ for $k = e,p,$ or $g$.  This simplifies the numerical implementation of the model, as it allows us to write the entropy per particle in a simple form, $s_k = (\gamma_k-1)^{-1}\log(P_k \rho^{-\gamma_k})$, and avoids the complication of having to evaluate $T_p/T_e$ when updating the total fluid variables. This can be seen by noting that 
\begin{equation}
\frac{P_g}{u_g} \equiv \frac{P_e+P_p}{u_e + u_p} = (\gamma_e -1)(\gamma_p-1)\frac{1+T_p/T_e}{(\gamma_p-1)+ (\gamma_e -1)T_p/T_e},
\label{eq:gamma_ep}
\end{equation}
which is only constant in the limits that $T_e \ll T_p$ or $T_p \ll T_e$.   From this, we see that this simplification of $\gamma =$ const. is formally inconsistent if $\gamma_e \ne \gamma_p$, which is generally the case in the accreting systems of interest, where the electrons are typically relativistically hot ($\gamma_e \approx 4/3$) but the protons are nonrelativistic ($\gamma_p \approx 5/3)$. However, since equation~\eqref{eq:gamma_ep} is bounded between $1/3$ and $2/3$ and we expect $T_e \lesssim T_p \Rightarrow \gamma \approx \gamma_p$, we do not anticipate that this approximation will affect our results significantly.

\subsection{Electron Heating}
\label{sec:electron-heating}
We parameterise the heating term, $Q_e$ in equation~\eqref{eq:Qep} by writing $Q_e$ = $f_eQ$, where $f_e(\beta,T_e, T_p,....)\equiv Q_e/Q$ is the fraction of the total dissipation, $Q$, received by the electrons.  This function, in general, depends on the local plasma environment and our model is not limited to any particular choice of $f_e$. As knowledge in the field develops we can readily incorporate different assumptions about electron heating. A more detailed discussion of one physically-motivated prescription for $f_e$ is given in \S\ref{sec:tauchi}.  Given a GRMHD solution, the total heating rate of a fluid element moving with four-velocity $u^\mu$ in the coordinate frame can be computed (from equation~\ref{eq:total}, dropping the conduction terms):  
\begin{equation}
Q = \rho T_{g} u^\mu \partial_\mu s_{g},
\label{eq:dotq}
\end{equation}
where $s_{g}$ is the entropy per particle.  We can rewrite equation~\eqref{eq:dotq} in terms of $\kappa_{g} \equiv P_{g} \rho^{-\gamma}$, where $s_{g} = (\gamma-1)^{-1}\log (\kappa_{g})$,  as
$Q = \rho^\gamma(\gamma-1)^{-1} u^\mu \p_\mu \kappa_{g}$.
We use $\kappa_g$ to avoid the undesirable numerical properties of logarithms as the argument goes to 0. Likewise, we will often use $\kappa_e$ in place of $s_e$ in equation~\eqref{eq:electron}.

\subsection{Anisotropic Electron Conduction}

\label{sec:cond}
Some care must be taken when generalising the theory of anisotropic conduction along magnetic field lines to a relativistic and covariant formulation.  In particular, the theory must be consistent with causality in that the heat flux should not respond instantly to temperature gradients.  Our formulation of anisotropic electron conduction draws heavily on the treatment of \citet{ManiModel}, who consider a single fluid model in which the heat flux is coupled to the dynamics via the stress-energy tensor.   We give a brief summary of our approach here, highlighting those aspects of our electron-only treatment that differ from the formulation in \citet{ManiModel}. 

One can derive a perturbation solution for the heat flux, $q_e^\mu$, by expanding the entropy current in powers of $q_e^\mu$ and imposing the second law of thermodynamics.  The most straightforward relativistic generalisation of the classical, isotropic heat flux first written down by \citet{Eckart1940} is first order in this expansion and was later shown by \citet{Hiscock1985} to be unconditionally unstable, precisely because it violated causality (\citealt{ManiModel} showed the same for anisotropic conduction). \citet{Israel1979} derived a second order solution for $q_e^\mu$ which was later shown to be conditionally stable (\citealt{Hiscock1985}; \citealt{ManiModel}).  Here we use a first order reduction of that second order model that has been shown to be both stable and self-consistent \citep{Andersson2011}. We refer the reader to \citet{ManiModel} for more details.

We parameterise the heat flux as 
\begin{equation}
  q_e^\mu = \phi \hat b^\mu,
\end{equation} where $\hat b^\mu$ is a unit vector ($\hat b^\mu \hat b_\mu = 1$) along the magnetic field four-vector, $b^\mu$, and the scalar $\phi$ is given by the following evolution equation:
\begin{equation}
\nabla_\mu \left(\phi \rho u^\mu \right) = \frac{1}{\gdet}\partial_\mu \left(\gdet \phi \rho u^\mu \right) = -\rho\left[\frac{\phi -\phi^{\rm eq} }{\tau}\right],
\label{eq:phiev}
\end{equation}
where $g$ is the determinant of the metric, and we used an identity, $\nabla_\mu A^\mu = \partial_\mu(\gdet A^\mu)/\gdet$, to convert covariant derivatives into partial ones \citep[eq.~86.9 in][]{lan2}.
Here $\phi^{\rm eq}$ is the equilibrium value of the heat flux given by
\begin{equation}
\phi^{\rm eq} = -\rho \chi_e \left(\hat b^\mu \partial_\mu T_e + \hat b^\mu a_\mu T_e\right).
\label{eq:phieq}
\end{equation}
where $\chi_e$ is the thermal diffusion coefficient of the electrons and $\tau$ is the relaxation time scale for the heat flux over which it responds to temperature gradients. Note that equation~\eqref{eq:phiev} is a relaxation equation in which the heat flux relaxes on a timescale $\tau$ to the equilibrium value.

The equilibrium heat flux in equation~\eqref{eq:phieq} is the natural relativistic extension of anisotropic conduction along the magnetic field (analogous to the isotropic heat flux of \citealt{Eckart1940}).
The heat flux, $q^\mu_e$, then contributes to the electron energy equation as in equation~\eqref{eq:electron}.
Physically motivated prescriptions for the parameters $\chi_e$ and $\tau$ are all that are required to complete the model.  We discuss one choice of these in \S\ref{sec:tauchi}.

\subsubsection{Stability of Anisotropic Electron Conduction Theory}
In our formalism, we assume that the fluid velocity, $u^\mu$, and the electron number density, $n_e = \rho/m_p$, are independent of the electron thermodynamics.  Thus, in order to do a perturbative analysis we need only perturb the electron temperature, $T_e$, and the heat flux, $\phi$, in equations~\eqref{eq:electron} and~\eqref{eq:phiev}. Doing this in the fluid rest frame in Minkowski space, where $u^\mu = (1,0,0,0)$, and writing the perturbations in Fourier space as $\propto \exp(\lambda t+i\vec{k}\cdot \vec{x})$, we find the dispersion relation:
\begin{equation}
\lambda^2 + \frac{\lambda}{\tau} + (\gamma_e-1)\frac{\chi_e}{\tau}(\hat b \cdot \vec{k})^2=0
\label{eqn:disp}
\end{equation}
with the solutions:
\begin{equation}
\lambda = \frac{1}{2\tau}\left( -1 \pm \sqrt{1 -4(\gamma_e-1)\chi_e \tau (\hat b \cdot \vec{k})^2} \right).
\label{eqn:anstab}
\end{equation}
The theory is unstable if $\operatorname{Re}(\lambda)>0$, which can only occur if the term under the square root is both real and greater than unity.  However, this is impossible for any value of $k$ when $\gamma_e\ge1$, so we conclude that equations~\eqref{eq:electron} and~\eqref{eq:phiev} are unconditionally stable. This is in contrast to the case where equations~\eqref{eq:electron} and~\eqref{eq:phiev} are coupled to the ideal MHD equations, which is unstable to small perturbations unless the relaxation time is larger than a critical value (\citealt{Hiscock1985}; \citealt{ManiModel}).

\section{Numerical Implementation of Electron Heating and Conduction}
\label{sec:numimp}
The method outlined above can, in general, be applied to any GRMHD ``background'' simulation. For the rest of this work, however, we will consider only conservative codes, as the equations of ideal MHD can be naturally written in that form.  Because of this, in what follows we will seek to put all of our evolution equations in a conservative form, namely:
\begin{equation}
\frac{\p U}{\p t} + \frac{\p F^i }{\p x^i} = S,
\label{eq:cons}
\end{equation}
where $U$ is a ``conserved'' variable, $F^i$ is the corresponding flux in the $i$th direction, and $S$ is the source, which in general includes the contribution from the connection coefficients.  Equation~\eqref{eq:cons} can then be approximated in one spatial dimension by the following discretisation:
\begin{equation}
\begin{aligned}
U^{n+1} &= U^n \\ & - \Delta t  \left(\frac{F_{j+1/2}^{n+1/2} - F_{j-1/2}^{n+1/2}}{\Delta x} -S^{n+1/2}\right),
\end{aligned}
\label{eq:consnum}
\end{equation}
where the fluxes are evaluated at face centres using the chosen Riemann solver.  The generalisation to higher dimensions is straightforward.  

With that in mind, we can rewrite equation~\eqref{eq:electron}:
\begin{equation}
\p_\mu \left(\gdet \rho u^\mu \kappa_e\right) = \frac{\gdet(\gamma_e-1)}{\rho^{\gamma_e-1}}\left[f_e Q -\nabla_\mu q_e^\mu -a_\mu q_e^\mu  \right],
\label{eq:ethermcons}
\end{equation}
where we have used the definition $\kappa_e \equiv \exp[(\gamma_e-1)  s_e]$. Note that equation~\eqref{eq:ethermcons} is a quasi-conservative equation with $U_{\kappa_e} = \gdet \rho u^t \kappa_e$ and $F^i_{\kappa_e} = \gdet \rho u^i \kappa_e$ (`quasi' conservative because the standard definition of conservative equations excludes source terms with derivatives).  To solve equation~\eqref{eq:ethermcons}, we use operator splitting in the following sequence of steps:
\begin{enumerate}
\item[1.] Solve the conservative equation with $S_{\kappa_e} =0$.
\item[2.] Explicitly update $\kappa_e$ with the heating term (the first term in the brackets in eq.~\ref{eq:ethermcons}).
\item[3.] Implicitly solve a matrix equation to include the conduction source terms (the rest of the terms in square brackets in eq.~\ref{eq:ethermcons}). 
\end{enumerate}
Steps 2 and 3 are described in detail in \S\ref{sec:eheatnum} and \S\ref{sec:eCondnum}, respectively, while step 1 will be specific to the choice of the background numerical scheme.

\subsection{Heating in Conservative Codes}

\label{sec:heatcons}

Formally, the equations of ideal MHD used by conservative GRMHD simulations imply that the heating rate per unit volume, $Q$, in equation~\eqref{eq:Qtot} is identically zero.  However, conservative codes implicitly add numerical viscosity and resistivity terms to the stress-energy tensor and Maxwell's equations, respectively.  The former implies that the numerically evolved stress tensor is in fact $T_{g,\rm num}^{\mu \nu} = T_{g}^{\mu \nu} + \tau_g^{\mu \nu}  = T_{MHD}^{\mu \nu}  + \mathcal{O}\left(\textrm{truncation error}\right)$ for some numerical viscosity tensor $\tau_g^{\mu \nu}$, while the latter implies $J_\mu e^\mu= 0$ $+$ $\mathcal{O}\left(\textrm{truncation error}\right)$.  The numerical resistivity can be thought of as implicitly introducing a form of Ohm's law that allows for a nonzero electric field four-vector, $e^\mu$. Thus, even though the energy implied by $T_{MHD, \rm num}^{\mu \nu} = T_{g, \rm num}^{\mu \nu}+T_{EM, \rm num}^{\mu \nu}$ is conserved to machine precision (see below for details), $T_{MHD}^{\mu \nu}$ experiences truncation-level heating.  This manifests itself as entropy generation: truncation errors lead to dissipation of magnetic and kinetic energy close to the grid scale that is captured as internal energy.  We use this change in entropy to directly calculate the heating rate per unit volume of the gas, $Q$.   

Although $T^{\mu \nu}_{MHD,\rm num}$ is conserved to machine precision, the second law of thermodynamics is satisfied only to truncation error.  Thus there can be locally regions with $Q < 0$.  In particular, the truncation error can be positive or negative, so in places with small actual change in entropy or large truncation error the change in entropy can be negative. This is the case even in test problems in which our methods of calculating the heating give the correct, converged, answer for the fluid variables (see \S~\ref{sec:test}). Thus, while $Q$ may be instantaneously or locally negative, when integrated over a sufficient length of time and/or space in the fluid frame it will satisfy the second law of thermodynamics.

We choose this method of calculating the heating rate as opposed to introducing an explicit functional form for $Q$ because it seems reasonable to assume that for several applications, the grid-scale dissipation in conservative codes is a well-defined quantity determined by the converged large-scale physics of the problem. Turbulence, for example, takes kinetic and/or magnetic energy at the largest scales and cascades it down to a small dissipative scale where it is converted into internal energy.  For a numerical scheme with no explicit viscosity, the scale at which dissipation occurs depends entirely on the resolution of the simulation, but we expect the heating rate itself will be fixed (in an averaged sense; see, e.g., \citealt{2010ApJ...713...52D}).  We expect the same for forced reconnection at high $\beta$ values, as in the disc midplane, where the large-scale dynamics sets the rate at which the field lines of opposite sign are brought together.

The above argument relies on the conservation of energy. In an arbitrary space-time, however, conservation of energy is only well-defined if the metric is stationary (time-independent) and therefore possesses a time-like Killing vector, $K^\mu$.  If such a vector exists (as it does for the Kerr metric of interest in this work), we can construct a conserved current from the stress-energy tensor via $J^\mu \equiv  -K^\mu T^\nu_\mu$, where this current satisfies $\nabla_\mu J^\mu = 0$. This allows us to define a conserved energy in a coordinate basis: 
\begin{equation}
 E = \int  J^t  \gdet dx^1 dx^2 dx^3,
 \label{eq:Eint}
\end{equation}
where the integral is over all space (i.e. the space orthogonal to the time coordinate).  Often, the Killing vector takes the form $K = \p_t$, which simplifies equation~\eqref{eq:Eint} to
\begin{equation}
   E = \int\limits -T^t_t \sqrt{-g}dx^1 dx^2 dx^3.
\end{equation} 
Thus, $-T^t_t$ can be thought of as the conserved energy per unit volume for a particular choice of coordinates.  The total energy, $E$, is conserved to machine precision, modulo fluxes of energy through the boundaries, so entropy can only be generated by conversion of one form of energy to another.  

\subsection{Calculating the Total Heating Rate}

To calculate the heating generated at each time step, we introduce an entropy-conserving equation as a reference to compare with the energy conservation equation. The entropy-conserving equation is simply a conservation equation like equation~\eqref{eq:cons} with $U_{\kappa_{g}}= \gdet \rho u^t \kappa_{g}$, $F^i_{\kappa_{g} } = \gdet \rho u^i \kappa_{g}$, and $S_{\kappa_{g}}=0$.  If we call the solution to this equation $\hat \kappa_{g}$ (the $\hat{\phantom{\kappa_g}}$ denotes the solution corresponding to entropy conservation), then we show in Appendix~\ref{app:heat} that the total heating rate of the fluid (measured in the fluid rest frame) incurred over an interval $\Delta t$ (measured in the coordinate frame),
\begin{equation}
Q =  \left(\frac{\rho^{\gamma-1}}{\gamma-1}\right)^{n+1/2}\left[\frac{\rho u^t(\kappa_{g} - \hat \kappa_{g})}{\Delta t}\right]^{n+1},
\label{eq:qdottot}
\end{equation}
where $u^t$ accounts for the transformation of $\Delta t$ from the coordinate frame to the fluid rest frame, $n$ and $n+1$ denote the values at the beginning and the end of the time step, respectively, so that $t_{n+1} = t_n + \Delta t$, and $n+1/2$ denotes the values when calculating the fluxes. To compute the dissipation rate via eq.~\eqref{eq:qdottot}, we set $\hat \kappa_{g}^{n} =  \kappa_{g}^n$ at the beginning of each time step and use $\hat \kappa_{g}^{n+1/2} =  \kappa_{g}^{n+1/2}$ when calculating the fluxes. Physically, equation~\eqref{eq:qdottot} means that the Lagrangian heating rate is set by the difference between the entropy implied by the total energy conserving solution ($\kappa_g$) and the entropy implied by the entropy conserving solution ($\hat \kappa_g$).

\label{sec:heatcalc}

\subsection{Electron Heating Update}
\label{sec:eheatnum}

Let us call $\hat \kappa_e$ the solution to equation~\eqref{eq:ethermcons} without any source terms.  On top of this adiabatic evolution, electrons receive a fraction, $f_e$, of the heating of the gas, $Q_e = f_e Q$. In discrete form, this can be written as follows, 
\begin{equation}
\label{eq:dq}
  \frac{(\rho^{\gamma_e})^{n+1/2}}{\gamma_e-1}(\kappa_e-\hat \kappa_e)^{n+1} = f_e^{n+1/2} \frac{(\rho^{\gamma})^{n+1/2}}{\gamma-1} (\kappa_{g}-\hat \kappa_{g})^{n+1}.
\end{equation}
Therefore, the heating update to the electrons, $\hat\kappa_e^{n+1}\to\kappa_e^{n+1}$, takes the following form:
\begin{equation}
\label{eq:kupdate}
\kappa_e^{n+1} = \hat \kappa_e^{n+1}+ \frac{\gamma_e-1}{\gamma-1} \left(\rho^{\gamma-\gamma_e}f_e\right)^{n+1/2} (\kappa_{g}-\hat \kappa_{g})^{n+1}.
\end{equation}

\subsection{Electron Conduction Update}
\label{sec:eCondnum}
We note that the evolution equation for the heat flux $\phi$ (equation~\ref{eq:phiev}) is already in a quasi-conservative form if we define $U_\phi = \gdet \rho u^\mu \phi$ and $F^i_\phi = \gdet \rho u^i \phi$.  We treat the evolution of $\phi$ in an operator split way similar to the evolution of $\kappa_e$ with the following series of steps:
\begin{enumerate}
\item[1] Solve the conservative equation with $S_{\phi} =0$.
\item[2] Implicitly solve a matrix equation to include the source terms. 
\end{enumerate}
The source terms in the electron entropy and $\phi$ equation due to conduction are given by:
\begin{equation}
\begin{aligned}
&S_{\kappa_e,\rm cond} = \gdet(\gamma_e-1)\rho^{1-\gamma_e} \left(-\nabla_\mu q_e^\mu - a_\mu q_e^\mu\right)\\
&S_{\phi} = -\rho\gdet\left[\frac{\phi + \rho \chi_e \left(\hat b^\mu \partial_\mu T_e + T_e \hat b^\mu a_\mu \right)}{\tau}\right],
\end{aligned}
\end{equation}
which are discretised in space by using slope-limited derivatives across three grid cells and discretised in time by centring the time derivatives at $t_{n+1/2}$.  The latter discretisation gives us an implicit equation for the variables $\kappa_e$ and $\phi$ at time $t_{n+1}$. If we call $\hat \phi^{n+1}$ the heat flux after being updated by step 1, and $\kappa_{e,\rm \rm H}$ the electron entropy after being updated by heating, then this matrix equation takes the form:
\begin{equation}
\begin{pmatrix}
a_{11} & a_{12} \\
a_{21} & a_{22}
\end{pmatrix}
\begin{pmatrix}
\kappa_e^{n+1} \\ \phi^{n+1}
\end{pmatrix}
 =\begin{pmatrix}  
b_1 \\b_2
\end{pmatrix},
\end{equation}
with components
\begin{equation}
\begin{aligned}
&a_{11} = \left(\frac{\gdet \rho u^t}{\Delta t}\right)^{n+1} \\
&a_{12} = \left[\gdet (\gamma_e-1) \rho^{1-\gamma_e}\right]^{n+1/2} \left[\frac{\hat b^t}{\Delta t}\right]^{n+1} \\
&a_{21} =\left[ \frac{\gdet \rho^2 \hat b^t \chi_e}{\tau}\right]^{n+1/2}  \left[\frac{ \rho^{\gamma_e-1} }{\Delta t}\right]^{n+1}\\
&a_{22} =\left( \frac{\gdet \rho u^t }{\Delta t}\right)^{n+1}\\
\end{aligned}
\end{equation}
and
\begin{equation}
\begin{aligned}
b_1  = &\left(\frac{\gdet \kappa_{e,\rm H}\rho u^t}{\Delta t}\right)^{n+1} + \left[(\gamma_e-1) \rho^{1-\gamma_e}\right]^{n+1/2} \\ & \times \left[\gdet \left(\frac{q_e^t}{\Delta t}\right)^n - \left(\p_i \left(\gdet q_e^i\right)- \gdet q_e^\mu a_\mu\right)^{n+1/2}\right]\\
b_2 = &\left(\frac{\gdet \hat \phi\rho u^t}{\Delta t}\right)^{n+1} - \gdet\left(\frac{ \rho \phi}{\tau}\right)^{n+1/2}\\ &+ \gdet \left(\frac{\rho^2 \chi_e}{\tau} \right)^{n+1/2} \\
& \times  \left[ \left(\hat b^t\right)^{n+1/2} \left(\frac{T_e}{\Delta t}\right)^n - \left(\hat b^\mu \p_\mu T_e + \hat b^\mu a_\mu T_e\right)^{n+1/2} \right].
\end{aligned}
\end{equation}
The system of equations has a straightforward solution, 
\begin{equation}
\begin{pmatrix}
\kappa_e^{n+1} \\ \phi^{n+1}
\end{pmatrix}
 =
\frac{1}{a_{11}a_{22}-a_{21}a_{12}}\begin{pmatrix}  
b_2 a_{11}-b_1 a_{21} \\b_1 a_{22} - b_2 a_{12}
\end{pmatrix}.
\end{equation}

To ensure that the heat flux, $\phi$, does not reach unphysically large values, we apply a limiting scheme to keep $|\phi|\lesssim \left(u_{e} + \rho_e c^2\right) v_{t,e} \equiv \phi_{\rm max}$, where $\rho_e = \rho m_e/m_p$ and $v_{t,e}$ is the electron thermal speed. Since we are considering physical systems in which the electrons are always at least mildly relativistic, this limit effectively reduces to $|\phi|\lesssim u_{e} c/\sqrt{3}$, which corresponds to a `saturated' heat flux in which the heat is redistributed at the electron thermal speed.  The numerical implementation of this limit is to replace the values of the thermal diffusivity, $\chi_e$, and the relaxation time-scale, $\tau$, with `effective' values \citep{ManiModel}:
\begin{equation}
  \chi_{\rm eff} = \chi_e f\left(\frac{|\phi|}{\phi_{\rm max}}\right),
\end{equation}
and
\begin{equation}
  \tau_{\rm eff} = \tau f\left(\frac{|\phi|}{\phi_{\rm max}}\right),
\end{equation}
where 
\begin{equation}
  f(x) = 1-\frac{1}{1+\exp\left(-\displaystyle\frac{x-1}{0.1}\right)} + \epsilon,
\end{equation}
which sharply transitions from $1 \rightarrow \epsilon$ for some small $\epsilon$ as $|\phi|\rightarrow \phi_{\rm max}$. Thus, according to equation~\eqref{eq:phiev}, when $|\phi|>\phi_{\rm max}$, $|\phi|$ decays exponentially on a timescale $\sim \epsilon \tau$ until it drops below $\phi_{\rm max}$. The parameter $\epsilon$ is chosen such that the criterion for numerical stability is always satisfied (see \S\ref{sec:numstab} and Appendix~\ref{app:stab}).

\subsubsection{Numerical Stability of Electron Conduction}
\label{sec:numstab}
A detailed derivation of the criteria for numerical stability is in Appendix~\ref{app:stab}.  The basic result is that for a Courant-Friedrichs-Lewy (CFL) number, $\mathcal{C}$, reasonably chosen between $0$ and $1$, the relaxation time, $\tau$, must satisfy
\begin{equation}
\tau >  f(\mathcal{C})\left( \frac{\Delta t}{\Delta x}\right)^2 \chi_e,
\label{eq:taulim}
\end{equation}
where $f(\mathcal{C})$ is a function of the CFL number. This can be understood as a requirement that the relaxation time $\tau$ (which we are free to choose as arbitrarily large, though which should correspond to a physical time scale), must be larger than the time step $\Delta t$ (which is limited by computational expense) by the ratio between $\Delta t$ and the standard Courant limit for a diffusive process $\Delta t_{\rm diff} = \Delta x^2/\chi_e$.

\subsection{Treatment of the Floors}
Conservative codes deal poorly with vacua of internal energy and density.  Because of this, many schemes employ floors on internal energy and density to ensure that the errors in solving for the primitive variables from the conservative variables do not produce unphysically small or negative values.  The nature of the model outlined above requires special care to be taken when these floors are activated, as they introduce artificial changes in internal energy, which act as a source of heat, and density, which change the conversion between entropy and internal energy.  

\subsubsection{Electron Energy Floors}
Though the second law of thermodynamics states that the heating term $u^\mu \partial_\mu \kappa_{g}$ should be positive definite, numerically we find that $u^\mu \partial_\mu \kappa_{g}$ can be locally negative because of truncation errors. This introduces the possibility of the electron internal energy going to zero (or even becoming negative) due to truncation error fluctuations in our heating term.  To correct for this, we implement a floor on the electron internal energy that is $1\%$ of the floor on the total gas internal energy.  That is, if $u_{e}$ drops below $0.01 u_g$, we reset $u_{e}$ to $0.01 u_g$.

\label{sec:efloors}

\subsubsection{Total Gas Internal Energy Floors}
When the floor on internal energy of the total gas is activated, there is an artificial increase in $u_g$ which then shows up in our heating term.   We treat this addition of energy as if it were a physical, isochoric addition to the energy of the gas and add it to the electrons as described above.  We emphasise that the internal energy floor does not affect the system dynamics in any significant way because it is only activated in magnetically-dominated regions where the value of the internal energy is dynamically irrelevant.

\subsubsection{Density Floors}
When the floor on density is activated, the total gas internal energy remains unchanged. However, the value of $\hat u_g \equiv \hat \kappa_{g} \rho^\gamma /(\gamma-1)$ increases by a factor of $(\rho_{\rm floor}/\rho_{\rm init} )^{\gamma}$, where $\rho_{\rm init}$ is the pre-floor density.  To correct for this, we require conservation of the evolved gas entropy when the density floor is activated by decreasing $\hat \kappa_{g}$ by a factor of $(\rho_{\rm floor}/\rho_{\rm init})^{\gamma}$. Furthermore, we enforce that the evolved electron entropy remains unchanged by the density floor in the same manner by decreasing $\kappa_e$ by a factor of $(\rho_{\rm floor}/\rho_{\rm init} )^{\gamma_e}$.  Similar to the internal energy floors, the density floors do not affect the dynamics of the system. 

\section{Tests of Numerical implementation}
\label{sec:test}

In this section we describe a series of tests that demonstrate the robustness and accuracy of our method of evolving the electron internal energy. We implemented the model described in \S\ref{sec:heat} and \S\ref{sec:numimp} into the conservative GRMHD code, HARM2D (High-Accuracy Relativistic Magnetohydrodynamics; \citealt{Gammie2003,2006ApJ...641..626N}). To speed up the computations, we parallelised the code using OpenMP and MPI via domain decomposition.

\subsection{Tests of Electron Heating}

In what follows we demonstrate the validity and convergence of our implementation of electron heating using a number of tests.  The 2nd order convergence of HARM in smooth flows and 1st order convergence in discontinuous flows is well documented in \citet{Gammie2003} and we will not reproduce it here.

\label{sec:testheat}

\subsubsection{Explicit Heating in a Hubble-Type Flow}
To test whether our discretizations of the heating is correctly time centred, i.e., converges at the expected 2nd order in time, we focus here on solving the electron equation when we introduce an explicit heating term to the total energy equation.  We do this in an unmagnetised, 1D Hubble-type flow with $v \propto x$ (restricting ourselves to non-relativistic velocities).  In the local rest frame of a fluid element, this velocity field gives an outflow in both directions that is homogenous and isotropic,  causing the density to uniformly decrease with time as matter leaves the computational domain.  The velocity profile also scales with time to satisfy the momentum equation $\left(\p v /\p t + v \p v /\p x = 0\right)$.   In the absence of heating, the internal energy and pressure evolve according to entropy conservation ($P \propto \rho^\gamma$), so that the solution at later times is given by \citep{Sasha2007}:
\begin{equation}
\begin{aligned}
v &= \frac{v_0 x }{1+v_0 t} \\
u_g &=  \frac{u_{g,0}}{\left(1+v_0 t\right)^\gamma}\\
\rho &= \frac{\rho_0}{1+v_0 t} .
\end{aligned}
\end{equation}
If we now add a cooling term to the energy equation, of the form:
\begin{equation}
Q = -\frac{u_{g,0} v_{0}\left(\gamma-2 \right)}{\left(1+v_0t\right)^3},
\end{equation}
the internal energy should evolve as
\begin{equation}
u_g = \frac{u_{g,0}}{\left(1+v_0 t\right)^2}.
\end{equation}
Plugging these solutions in the electron entropy equation (\ref{eq:ethermcons}) for $f_e = 1$ and $u_{e}(t=0 ) = u_0$, we obtain:
\begin{equation}
\kappa_e =  \frac{\left(\gamma-2\right)\left(\gamma_e-1\right)}{\gamma_e-2}\frac{u_{0}}{\rho_0^{\gamma_e}}\frac{1}{\left(1+v_0 t\right)^{2-\gamma_e}}.
\end{equation}

\begin{figure}
\centering
\includegraphics[scale = .3]{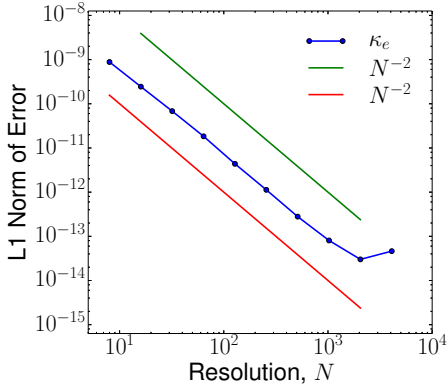}
\caption{L1 norm of the error in the electron entropy for heating in a 1D Hubble-type flow (see \S\ref{sec:hubble}). Above a resolution of $\sim 1000$ the relativistic errors in the analytic result are comparable to the numerical truncation errors, so convergence is no longer seen.}
\label{fig:hubblecon}
\end{figure}

For the numerical test, we set these analytic solutions as the boundary and initial conditions in a one-dimensional grid and check if we maintain this solution after a dynamical time of $L/\max[v(t=0)]$.  We set $\gamma = 5/3$, $\gamma_e = 4/3$, $\max (v_0x) = 10^{-3}c$, and $\max(\rho v_0 x/u_g) =1$, on a computational domain of $0\le x \le 1$.  Formally, since $\Theta_e \equiv kT_e/m_e c^2 \ll 1$, the choice of $\gamma_e =4/3$ is unphysical.  However the motivation for this choice stems from the fact that our primary application is to the inner regions of an accretion disc around a black hole, where we expect $\gamma_e \approx 4/3 \ne \gamma \approx 5/3$.  We find that our calculation converges at second order (see Figure~\ref{fig:hubblecon}), up until the point at which the errors in the analytic solution due to relativistic effects become important (which, for $\max (v_0x) = 10^{-3}$ is $\delta \kappa_e /\kappa_e \sim v^2/c^2 \sim 10^{-6}$).

\label{sec:hubble}

\subsubsection{1D (Noh) Shock Test}

\label{sec:testshock}

In Appendix~\ref{App:shock}, we show that for a high Mach number shock in which the electrons are assumed to receive a constant fraction $f_e$ of the `viscous' heating in the shock, the post shock electron internal energy $u_{e}$ is given by:
\begin{equation}
\frac{u_{e}^f}{u_{g}^f} =\frac{f_e}{2} \left[ {\left(\frac{\gamma + 1}{\gamma - 1}\right)^{\gamma_{e}}\left(1-\frac{\gamma}{\gamma_e}\right)+ 1 + \frac{\gamma}{\gamma_e} } \right]\frac{\gamma^2-1}{\gamma_e^2-1},
\label{eq:shockrat}
\end{equation}
where $u_g^f$ and $\gamma$ are the post-shock internal energy and the adiabatic index of the fluid.   Equation~\eqref{eq:shockrat} assumes that the electrons do not back react on the shock structure, consistent with the model developed in this paper. When $\gamma = \gamma_e$, equation~\eqref{eq:shockrat} is equal to $f_e$, while for $\gamma = 5/3$ and $\gamma_e = 4/3$ it is $\sim 0.76 f_e$. Here we check whether our numerical implementation of electron heating is consistent with this result.
 
The initial conditions for this test are an unmagnetised, non-relativistic ($\gamma=5/3$), uniform density and internal energy fluid.  The velocity profile is discontinuous at the center of the grid, with a left and a right state given by $v_l = -v_r  = {\rm constant} >0$.  The resulting solution is two shocks propagating outwards with a static region in between.  We focus on a non-relativistic ($|v| = 10^{-3}c)$ flow of initially cold gas so that the Mach number of the flow satisfies $M\gg1$. For this test we fix $f_e = 0.5$ and show the results for both $\gamma_e = 4/3$ and $\gamma_e = 5/3$.

\begin{figure*}
\includegraphics[scale = .4]{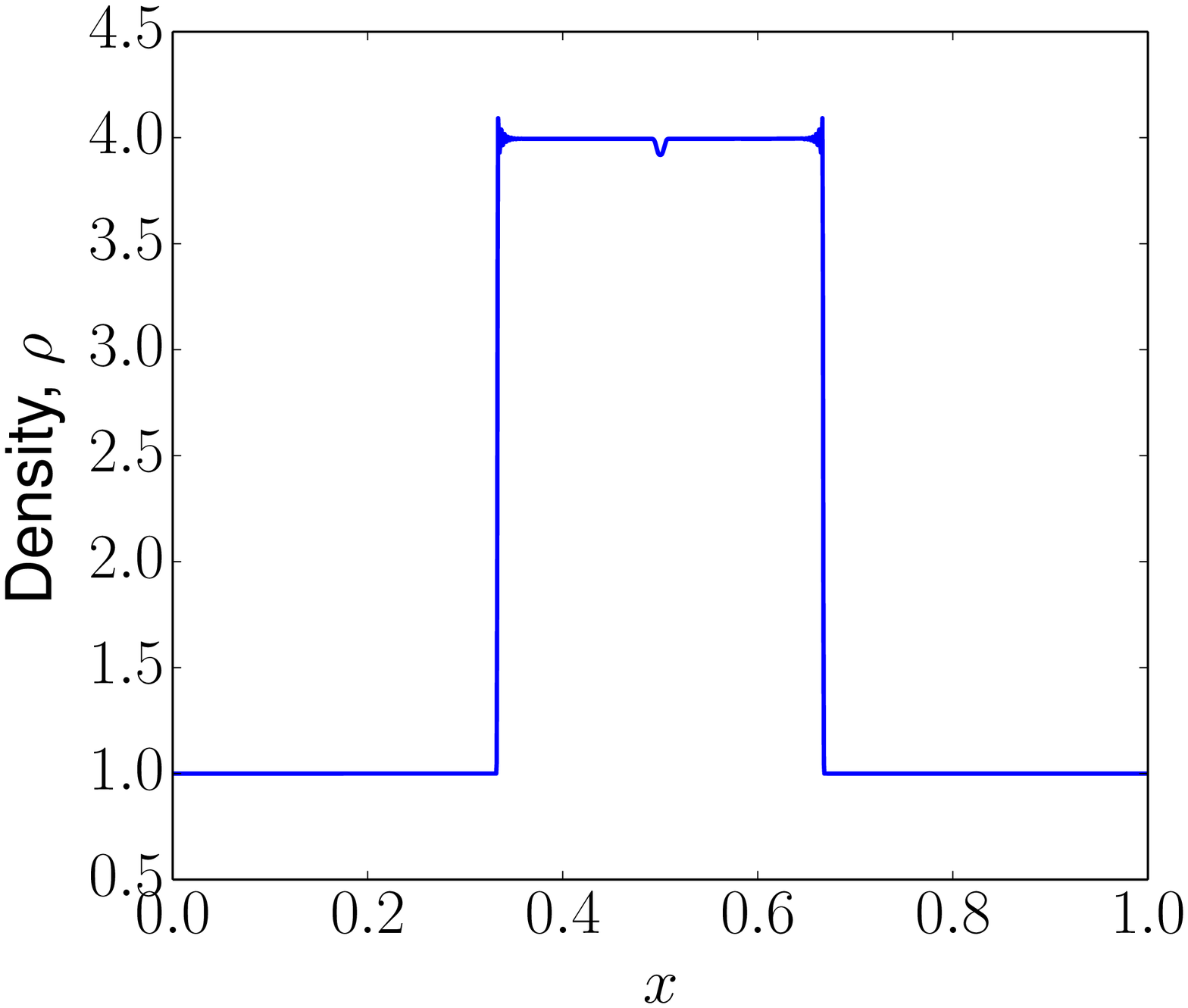}\\
\includegraphics[scale = .4]{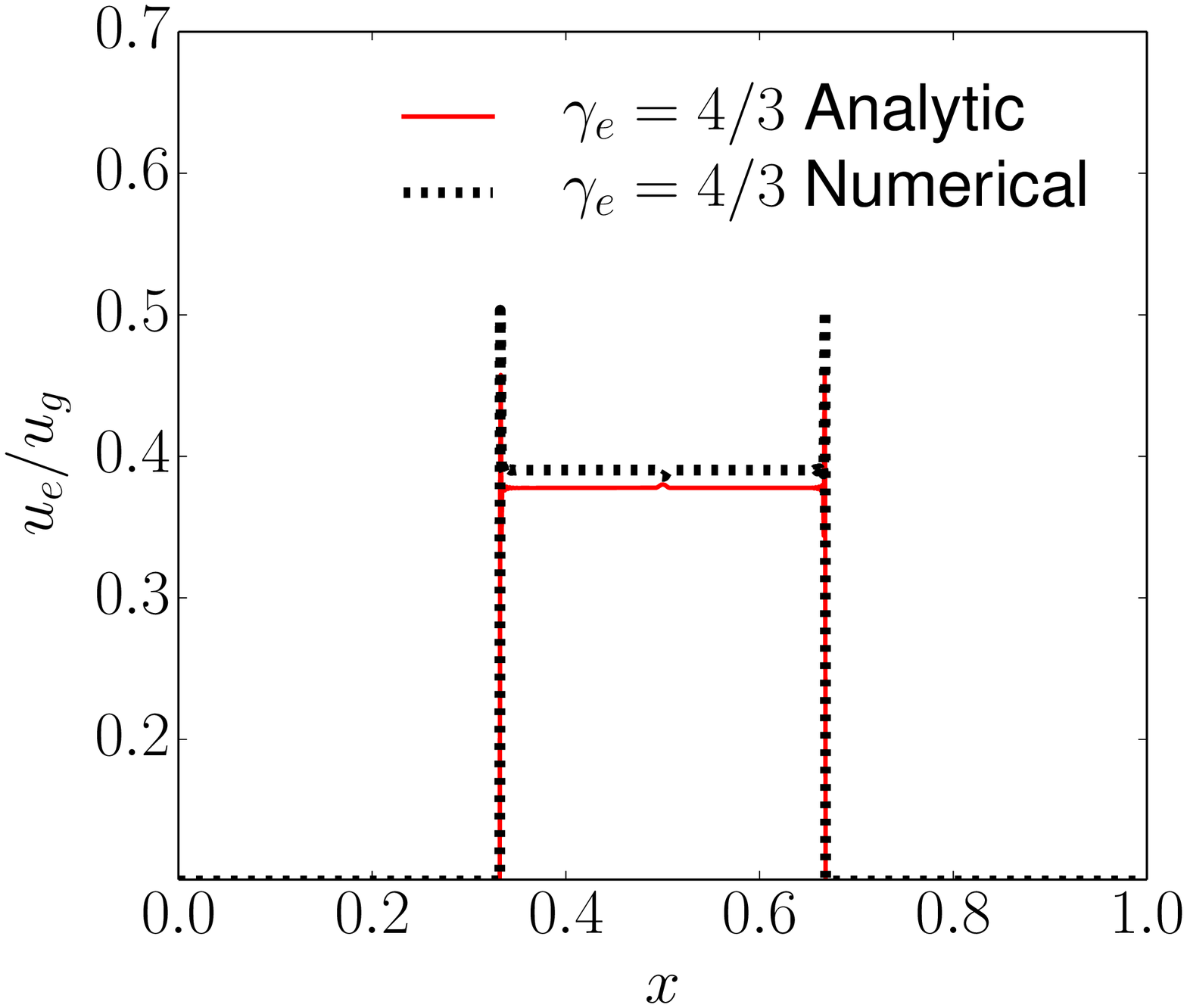}
\includegraphics[scale = .4]{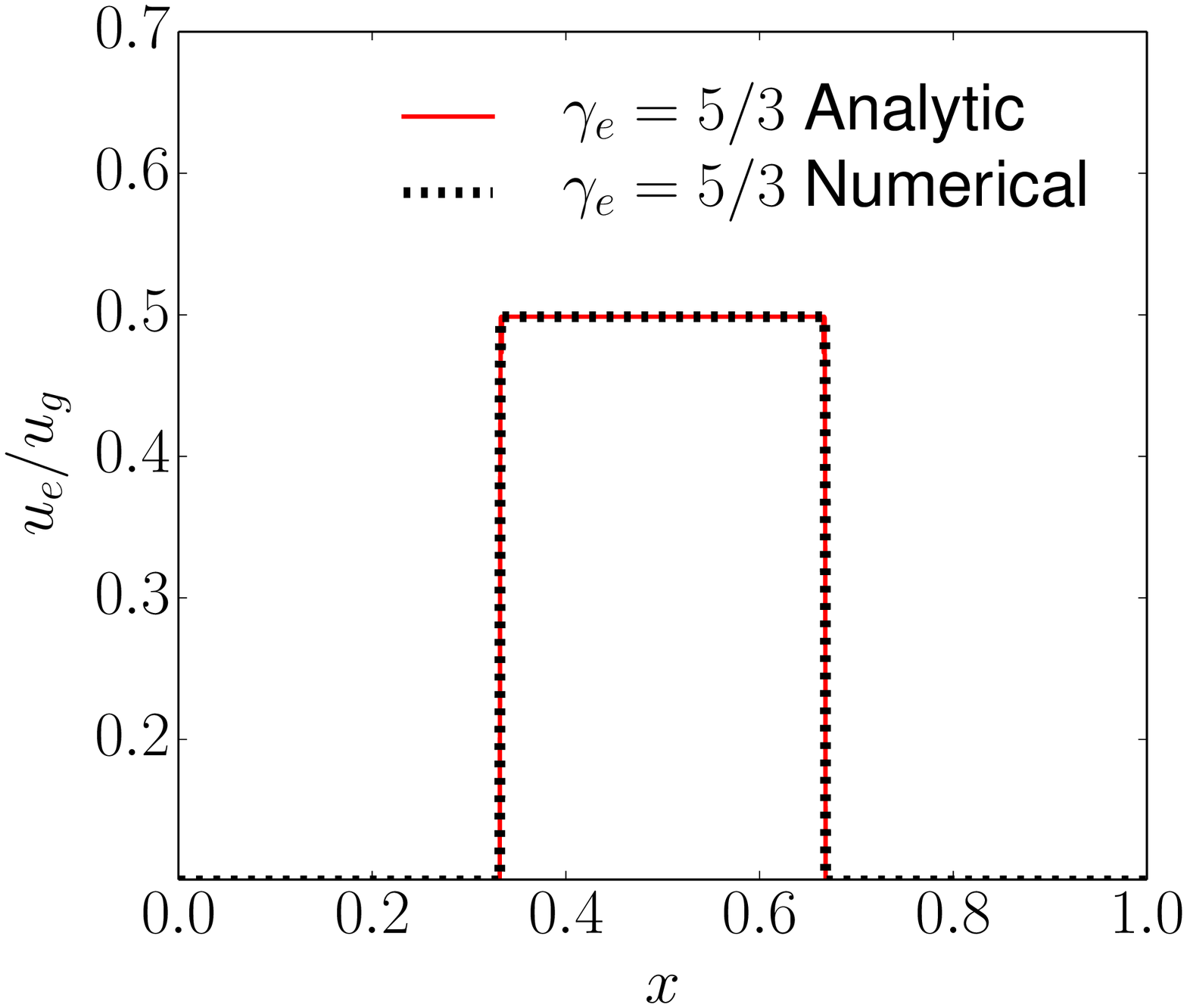}
\caption{High Mach number shock results for an electron heating fraction $f_e = 0.5$ at a resolution of 2000 cells. Top: solid blue line shows the fluid density in a numerical simulation. Density undergoes a jump of $\rho_2/\rho_1 = (\gamma+1)/(\gamma-1)=4$ at the two shocks, located at $x\approx 0.35$ and $x = 0.65$. Left: electron internal energy relative to total fluid internal energy for $\gamma_e = 4/3$. The analytic solution is shown with the solid red line and the numerical solution with the dotted black line. Right: the same for $\gamma_e = 5/3$. The analytic solution uses the functional form for $u_e/u_g (\rho)$ (see Appendices~\ref{App:shock} and \ref{App:viscshock} for details) and applies it to the density returned by the simulation.   At this resolution all the fluid variables are essentially converged.  The $\gamma_e = 5/3$ electrons show convergence to the expected result of $u_{e} = f_e u_g$ (the numerical and analytical lines are essentially on top of each other) while the $\gamma_e = 4/3$ electrons converge to a value that is greater than the analytic result ($u_e = 0.379 u_g$ for $f_e = 0.5$; equation~\ref{eq:shockrat}) by $\sim 3\%$. This is because the internal shock structure is never well resolved without an explicit bulk viscosity (see \S\ref{sec:testshock} and Appendix~\ref{App:viscshock} for details). }
\label{fig:shock}
\end{figure*}

\begin{figure}
\includegraphics[scale = .35]{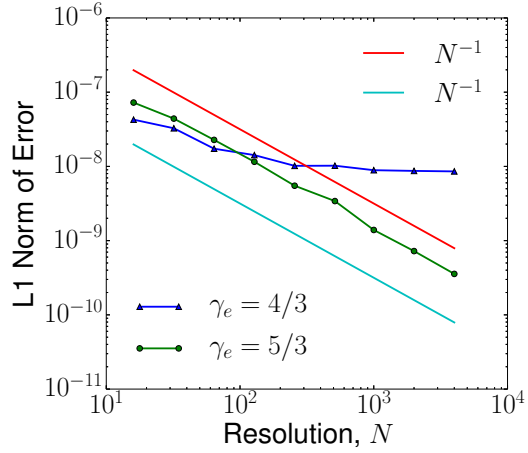}
\caption{Convergence of the post-shock electron internal energy in the 1D shock test to the analytic solution (equation~\ref{eq:shockrat}). The shock's Mach number is $\sim49$ (see Sec.~\ref{sec:testshock}). The $\gamma_e =\gamma = 5/3$ electrons converge at 1st order, as expected, but the $\gamma_e = 4/3$ electrons do not converge to the correct solution to better than $\sim 3 \%$ (see Figure~\ref{fig:shock}).  This is because our calculation of the heating requires a well-resolved shock structure, which is not the case for modern shock-capturing conservative codes (see \S\ref{sec:testshock} for details).  Introducing an explicit bulk viscosity to resolve the shock structure leads to convergence for $\gamma_e \ne \gamma$ (see Appendix~\ref{App:viscshock}). For $\gamma_e = \gamma$, a convenient cancellation makes the evolution of the electron entropy independent of the shock structure. }
\label{fig:shockcon}
\end{figure}

\begin{figure}
\includegraphics[scale = .35]{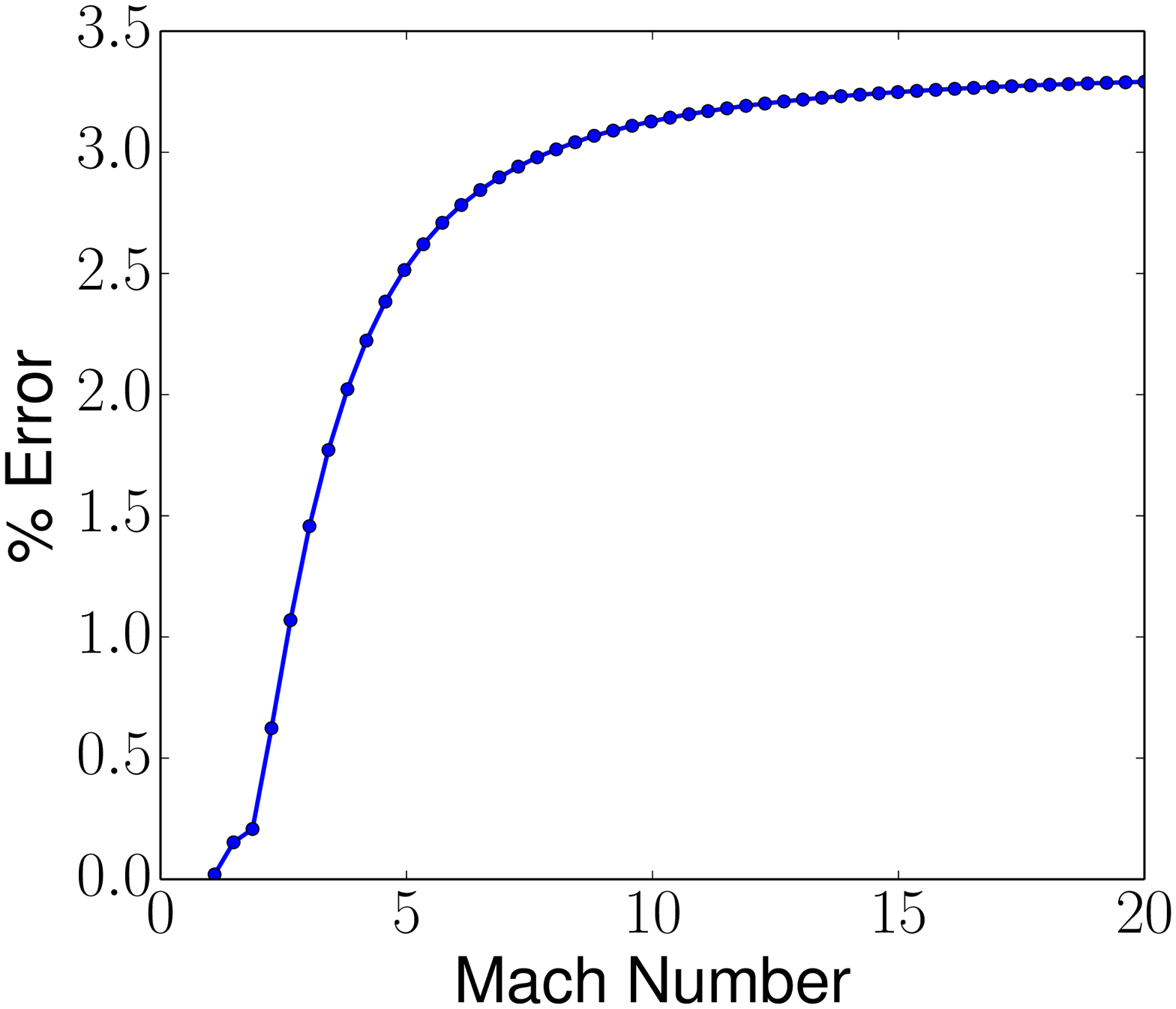}
\caption{Percent error in the post-shock electron internal energy for $\gamma_e = 4/3$ and $\gamma = 5/3$ as a function of Mach number in the 1D shock test as computed by HARM at a resolution of 2000 for a fractional heat given to the electrons of $f_e = 1/2$ and an initial $u_{e}/u_g = 0.1$.  The analytic solution is given by equation~\eqref{eq:Mache}. The final time was fixed such that the two shocks were located at  $x=0.25$ and $x = 0.75$ in a $0\le x\le 1$ domain. Note that the fractional errors are always $\lesssim 3.3 \%$. The change in the percent error as Mach number goes to 1 is because the flow becomes increasingly smooth and the electron internal energy is no longer converged at a resolution of 2000. }
\label{fig:Mach}
\end{figure}

Figure~\ref{fig:shock} shows the density and electron internal energy as a function of position for a shock with $M\sim 49$ at $t = 0.6 L/|v_l|$, where $L$ is the size of the computational domain.  Figure~\ref{fig:shockcon} shows that our simulation converges at 1st order to the analytic result for the post-shock electron internal energy when $\gamma=\gamma_e$ but to a value differing from the expected result by $\sim 3\%$ when $\gamma_e \ne \gamma$ (Figure~\ref{fig:shock}).  This difference is smaller at lower Mach number, as shown explicitly in Figure~\ref{fig:Mach}.   The modest discrepancy between the analytic post-shock electron temperature and the HARM solution is because an accurate, converged calculation of the heating term requires a well-resolved shock structure that gets better resolved at higher resolution. This is not the case for modern shock capturing techniques, for which the numerical width of the shock is always a few grid points and our heating calculation is never able to resolve the shock.    This is not an issue when $\gamma = \gamma_e$ because the factors of density in equation~\eqref{eq:kupdate} cancel, removing the dependence on the shock structure. We show in Appendix~\ref{App:viscshock} that introducing an explicit bulk viscosity leads to convergence to the analytical result at 2nd order for $\gamma\ne \gamma_e$.  However, the $\lesssim 3\%$ error as seen in Figures~\ref{fig:shock} and \ref{fig:Mach} is sufficient for our purposes so we do not include a bulk viscosity in our calculations.

\subsubsection{2D Forced MHD Turbulence Test}

\label{sec:turb}
Another test problem with a known, converged heating rate is driven turbulence in a periodic box. If we inject the fluid with a constant energy input rate of $\dot E_{\rm in}$ at large-scales, we should find that $\int Q dV= \dot E_{\rm in}$ after saturation of kinetic and magnetic energies has been reached.   Thus, for the electron heating model outlined above, after this saturation point the electrons should receive a fraction $f_e$ of $\dot E_{\rm in}$. Furthermore, if we have a periodic box in which the particle number is fixed, then the total internal energy change from adiabatic expansion/compression will sum to zero.  Thus the analytic result we expect for our model of electron heating is that $\int \rho T_e \dot{s}_{e} dV = f_e \int Q dV = f_e \dot E_{\rm in}$.  This test checks whether our model satisfies this result numerically.

We start with a static, uniform density fluid with $\beta=6$ and sound speed $c_{s,0} = 8.6 \times 10^{-4} c$ in a 2D periodic box.  The initial magnetic field is uniform: the magnetic field lines are straight and lie in the plane of the simulation.  Then, at each time step, we give Gaussian random kicks to the velocity such that the wave number satisfies $\vec{k}\cdot \delta\vec{v} = 0$ (i.e. the driving is incompressible), and $\sigma_v^2 \propto k^6 \exp(-8 k/k_{\rm peak})$ (compare to \citealt{2009Lemaster}).  The normalisation is fixed such that the rate of energy injection is equal to $\dot E_{\rm in} = 0.5 \bar{\rho} c_{s,0}^3$. This leads to a rms turbulent velocity that is $\sim 0.8 c_{s,0} \sim 1.8 v_A$, so that the turbulence is subsonic and roughly Alfv\'enic. The peak driving wave number is set to half the box size: $k_{\rm peak} = 4\pi/L$. Furthermore, we ensure that no net momentum is added to the box by subtracting off (from the kicks) any average velocity that would have been generated by the kicks. For this test we fix $f_e = 0.5$, $\gamma=5/3$ and $\gamma_e = 4/3$.

\begin{figure}
\includegraphics[scale = .4]{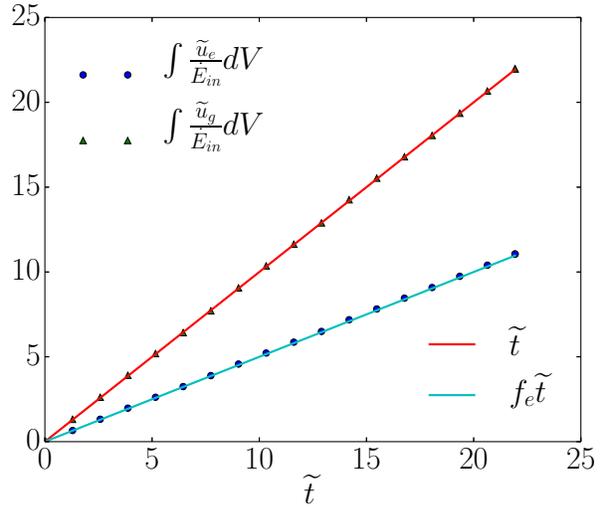}
\caption{Electron and total gas internal energies summed over the grid as a function of time at a resolution of $512^2$ for the forced subsonic MHD turbulence simulation with an assumed electron heating fraction of $f_e = 0.5$ and initial $\beta = 10$.  We define $\widetilde{t} = (t-t_i)c_{s,0}/L$ and $\widetilde{u}_{g,(e)} = u_{g,(e)}-u_{g,(e)}(t=t_i)$, where $t_i\sim 4 L/c_{s,0}$ is the time at which kinetic and magnetic energy roughly saturate.  We normalised the integrated internal energy by the energy injection rate, $\dot E_{\rm in} = 0.5 \rho c_{s,0,}^3$, so that the $y$-axis has dimensions of time which we measure in units of $L/c_{s,0}$.  In these variables, the analytic solutions for the total gas and electron internal energies are lines with slopes of 1 and $f_e=0.5$, which are plotted as solid lines, to be compared to the simulation results which are represented by points. We find that the electron heating rate is $0.5$ of the total heating rate, consistent with the analytic solution given the input value of $f_e = 0.5$.  For a numerical comparison at different resolutions, see Table~\ref{table:turbfit}, which shows the results of applying a linear regression fit to the internal energies. }
\label{fig:turbe}
\end{figure}

Figure~\ref{fig:turbe} shows our results for the electron and total internal energies as a function of time in the box at $512^2$. We see that once approximate saturation of the turbulence is reached at $t \sim L/c_{s,0}$ (or $\widetilde{t} \sim 0$ in the figure), the internal energies are in very good agreement with a linear fit, as expected given the constant rate of energy injection. For a parameterization of  $\int u_{e} dV = g_e t +b_e$ and $\int u_{g} dV = g_{g} t +b_{g}$, $g_e$ and $g_{g}$ represent the electron and total heating rates, respectively. These can be compared to the energy injection rate, $\dot{E}_{in}$, which is a fixed constant of $0.5 \bar{\rho} c_{s,0}^3$.  At a resolution of $512^2$, we find the total heating rate differs from $\dot{E}_{in}$ by $\sim 4 \%$, while the electron heating rate differs from $f_e\dot{E}_{in}$ by $\sim 2\%$.  Unfortunately, a rigorous convergence study of these quantities is not possible because of the nature of turbulence in 2D.  Due to the inverse energy cascade, the kinetic and magnetic energies never truly saturate and convergence of \emph{any} of the fluid variables is never achieved.  This can be seen from Table~\ref{table:turbfit}, where the values of $g_e$ and $g_{g}$ are quoted at various resolutions, neither of which display significant convergence to the expected $0.5 \dot{E}_{in}$ and $\dot{E}_{in}$, respectively.  Nevertheless, we find the percent level error found at all resolutions to be sufficiently small to satisfy our error tolerance in the full accretion disc simulation. 

\begin{table}
\begin{minipage}{6cm}
  \caption{Turbulence Test Linear Fits (\S\ref{sec:turb})}
\centering
\begin{tabular}{c c c c c  }
\hline\hline \\
 Resolution: & 128& 256 & 512  \\ [0.5ex] %
\hline
\\
$g_e-0.5\dot{E}_{in}$\footnote{Fractional error in the electron heating rate relative to the analytic solution.}&$0.0027 \dot{E}_{in}$&$0.0042 \dot{E}_{in}$&$0.0024  \dot{E}_{in}$ \\
$g_{g}-\dot{E}_{in}$\footnote{Fractional error in the total heating rate relative to the analytic solution.}&$-0.00017\dot{E}_{in}$&$0.0012\dot{E}_{in}$&$-0.0016 \dot{E}_{in}$ \\
[1ex]
\hline
\end{tabular}
\label{table:turbfit}
\end{minipage}
\end{table}

\subsubsection{Shadow Solution}

For our two-temperature model, we can seek a solution in which the electron fluid simply `shadows' the total gas, in that $u_{e} \propto u_g$.  Such a solution is found by setting $u_{e}(t=0) = f_e u_{g}(t=0)$ at some initial time, since we can solve the first law of thermodynamics for the electrons with $u_{e} = f_e u_g$ for all time, assuming that $\gamma_e = \gamma$ and $f_e$ is a constant. This can be seen from the first law:
\begin{equation}
u^\mu\partial_\mu u_{e} = f_e \rho T_{g} u^\mu\partial_\mu s_{g} - \frac{u_{e} + P_e}{\rho}u^\mu\partial_\mu \rho,
\label{eq:etherm}
\end{equation}
because the electrons will always get a fraction, $f_e$ of the entropy-generated heat (the first term on the RHS of equation~\ref{eq:etherm}), while the compression term is directly proportional to $u_{e} \propto f_e u_g$.  This solution is valid regardless of the details of the overall fluid evolution, so we can apply it to an arbitrarily complicated system.

For this test, we evolve the electron internal energy in the full accretion disc simulation around a rotating black hole as outlined in \S\ref{sec:app}. We initially apply small ($\delta u_{e}/u_{e} \sim 0.04$) perturbations to the electron internal energy about the average value of $u_{g,e,0} = 0.5 u_{g,0}$, and set $f_e =0.5$ and $\gamma_e = \gamma =5/3$.  For this test alone, we set the floor on electron internal energy to be a fraction $f_e$ of the floor on the total fluid (as opposed to our usual choice of $1\%$).  If this latter step were neglected, the floors would cause the polar regions to differ significantly from the expected result, though leaving the disc and corona unaffected (i.e. they still satisfy the analytic result).   The test is whether or not our simulation can maintain this result over the run time of $2000 M$.  

Running this test at a resolution of $512^2$ gives an average fractional error of 
\begin{equation}
\frac{1}{N^2}\sum^{N-1}_{j=0}\sum^{N-1}_{i=0} \left|\frac{\left(\left[{u_{e}}/{u_g}\right]_{ij}-f_e\right)}{f_e}\right| \sim 0.8 \%,
\end{equation}
which is smaller than our initial perturbations and shows that our numerical solution correctly evolves equation~\eqref{eq:etherm} even in a complex problem with MHD turbulence, weak shocks, and other heating processes in the presence of a curved metric.

\subsection{Tests of Electron Conduction}

Our model and testing suite for conduction closely resembles that of \citet{grim}, so we leave the details to Appendix~\ref{app:condtests}.   In summary, we have found second order convergence for linear modes, for a static, 1D atmosphere in the Schwarzschild metric, and for a relativistic, spherically symmetric Bondi accretion flow.  We also show that the electrons properly conduct along field lines in a 2D test.

\section{Application to an Accreting Black Hole in 2D GRMHD Simulations}
We apply the new methods discussed in \S\ref{sec:heat} and the numerical implementation described in \S\ref{sec:numimp} to the astrophysical environment of an accretion disc surrounding a spinning black hole as described by the Kerr Metric with a spin parameter of $a=0.9375$.  For this spin the last stable circular orbit is $\approx 2.04 r_g$ and the thin disc radiative efficiency is $\approx 0.18$  \citep{Novikov1973}.  We use the conservative code HARM \citep{Gammie2003} as our background GRMHD solution.  Our initial conditions for the total fluid are the \citet{Fishbone1976} equilibrium torus solution (see Appendix~\ref{app:fishbone}) with inner radius $r_{\rm in} = 6 r_g$ and with the maximum density of the disc occurring at $r_{\rm max} = 12 r_g$. Note that here and throughout $r$ and $\theta$ refer to the Boyer-Lindquist coordinates.  This equilibrium solution has a temperature maximum of $\approx 7.5 \times 10^{10} K$ and a thickness\footnote{Here we define $h/r \equiv\left.{\iint \rho u^t |\theta-\pi/2| \gdet \dd\theta \dd\phi }\middle/{\iint \rho u^t \gdet \dd\theta \dd\phi}\right.$.}  of $h/r \sim 0.18 $ at $r_{\rm max}$.  %
We normalise the torus density distribution such that the maximum value of density in the torus is $\rho_{\rm max}c^2 = 1$ and 
perturb the internal energy of the gas with random kicks on the order of $\delta u_{g}/u_{g} \sim 0.04$ to provide the perturbations for the magnetorotational instability (MRI, \citealt{1991ApJ...376..214B}) to develop.\footnote{In addition to the electron specific floor described in \S\ref{sec:efloors}, there are also floors on the density and internal energy of the HARM single fluid GRMHD solution.   These are $\rho_{\rm floor}c^2=\max\left[b^2/50,10^{-4}(r/r_g)^{-3/2}\rho_{\rm max}c^2\right]$ and $u_{\rm floor} = \max\left[b^2/250,10^{-6}(r/r_g)^{-5/2}\rho_{\rm max}c^2\right]$.  Note that the unit choice for the background "atmosphere" is such that the initial torus maximum density is $\rho_{\rm max}c^2=1$ and the initial torus internal energy is $u_{\rm max} \approx 0.01$.}
  We overlay this equilibrium solution with an initial magnetic field with $2P_{\rm max}/b^2_{\rm max} = 100$ (where max refers to the maximum value inside the torus), defined by the scalar vector potential: 
\begin{equation}
A_\varphi \propto \left(\rho / \rho_{\rm max} -0.2\right) \cos\theta,
\label{eq:Aphi}
\end{equation}
if $\rho > 0.2\rho_{\rm max}$ and 0 otherwise. This vector potential defines two meridional loops contained in the torus that are antisymmetric about the equator.  This choice ensures that the field lines are not along constant density.  Since constant density implies constant temperature when entropy is constant, field lines along constant density would be isothermal in the initial condition (as would happen if we dropped the factor of $\cos\theta$ in eq.~\ref{eq:Aphi}).  2D MHD torus simulations are unable to reach a statistical steady state in which the initial conditions are forgotten, so initially isothermal field lines could artificially suppress electron conduction even at later times. We choose the 2-loop initial condition to avoid this.

For the electrons, we start with $u_{e}/u_g = 0.1$, and run two different models for $f_e$, described below. For conduction runs, we set the initial heat flux to zero.

Our conduction runs are all in $256^2$ grids with a physical size of the domain in spherical polar coordinates of $(R_{\rm in},R_{\rm out})\times(\theta_{\rm in},\theta_{\rm out}) = (0.8r_{\rm H},1000 r_g)\times(0,\pi)$, where $r_{\rm H} = r_g(1+\sqrt{1-a^2})$ is the black hole event horizon radius. For $a = 0.9375$, $r_{\rm H} \approx 1.35 r_g$. In the regions with $r< 50 r_g$, the code uses modified Kerr-Schild coordinates ($t$, $x^1$, $x^2$, and $\varphi$) of \citet{Gammie2003}, so that the regions with the highest resolution are near the mid-plane close to the horizon. For $r>50 r_g$, we use hyper-exponential coordinates to move out the outer radial boundary $r = R_{\rm out}$ and limit unphysical reflection effects by defining the internal code coordinate $x^1$ implicitly by the equation \citep{Sasha2011}:
\begin{displaymath}
r/r_g =  \left\{
\begin{array}{ll}
   \exp({x^1}) &:\quad r \le50 r_g\\
\exp\left\{x^1 + [x^1-x^1(r=50 r_g)]^4\right\} &:\quad r>50 r_g.
 \end{array}
\right.
\end{displaymath}
The electron heating-only (i.e. without conduction) runs have the same parameters but a higher resolution of $512^2$. At the inner and outer radial boundaries we apply the standard outflow (copy) boundary conditions, at the polar boundaries we apply the standard antisymmetric boundary conditions (with all quantities symmetric across the polar axis except $u^\theta$ and $B^\theta$, whose signs are reversed).

\begin{figure*}
    \includegraphics[scale = .105]{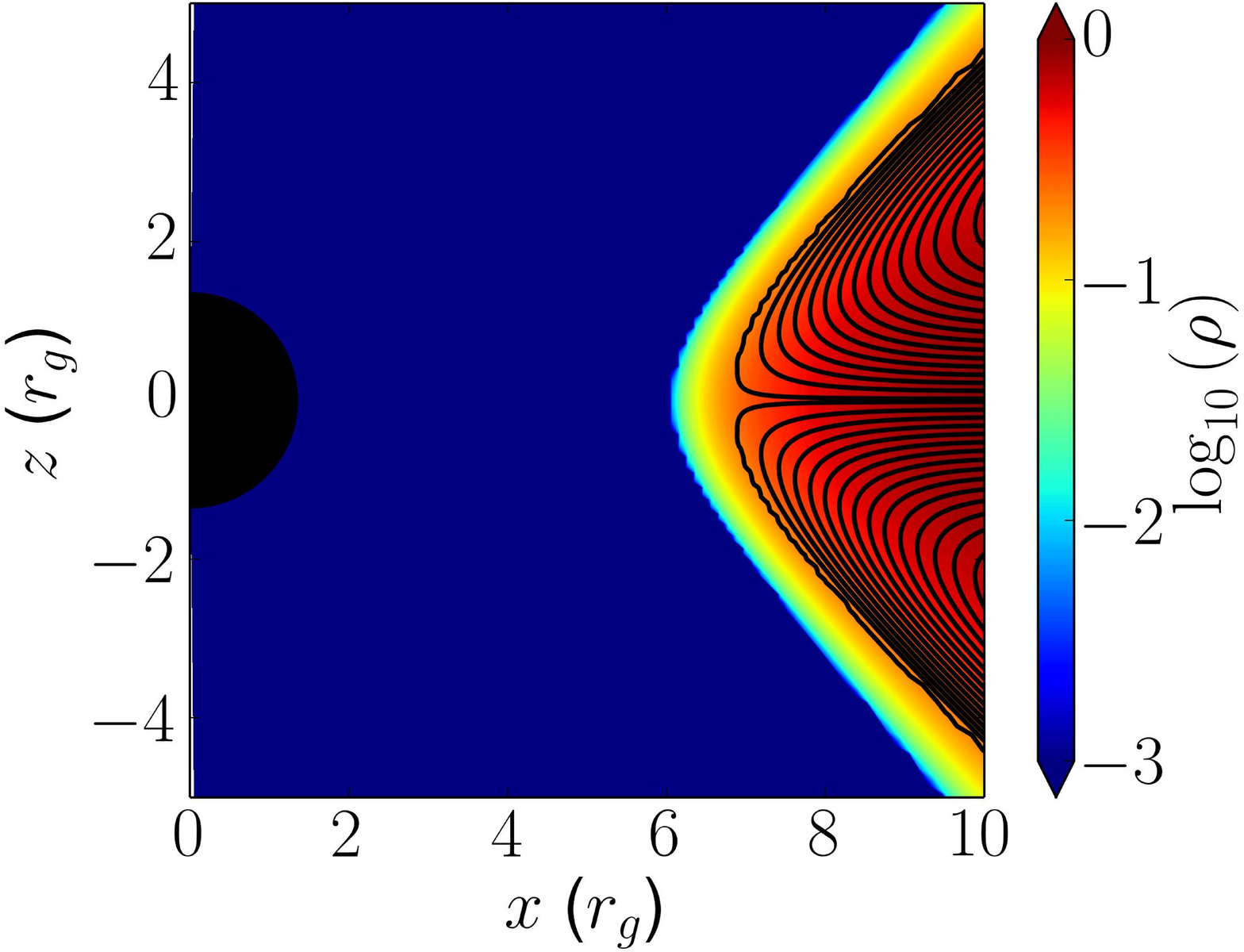} 
     \includegraphics[scale = .105]{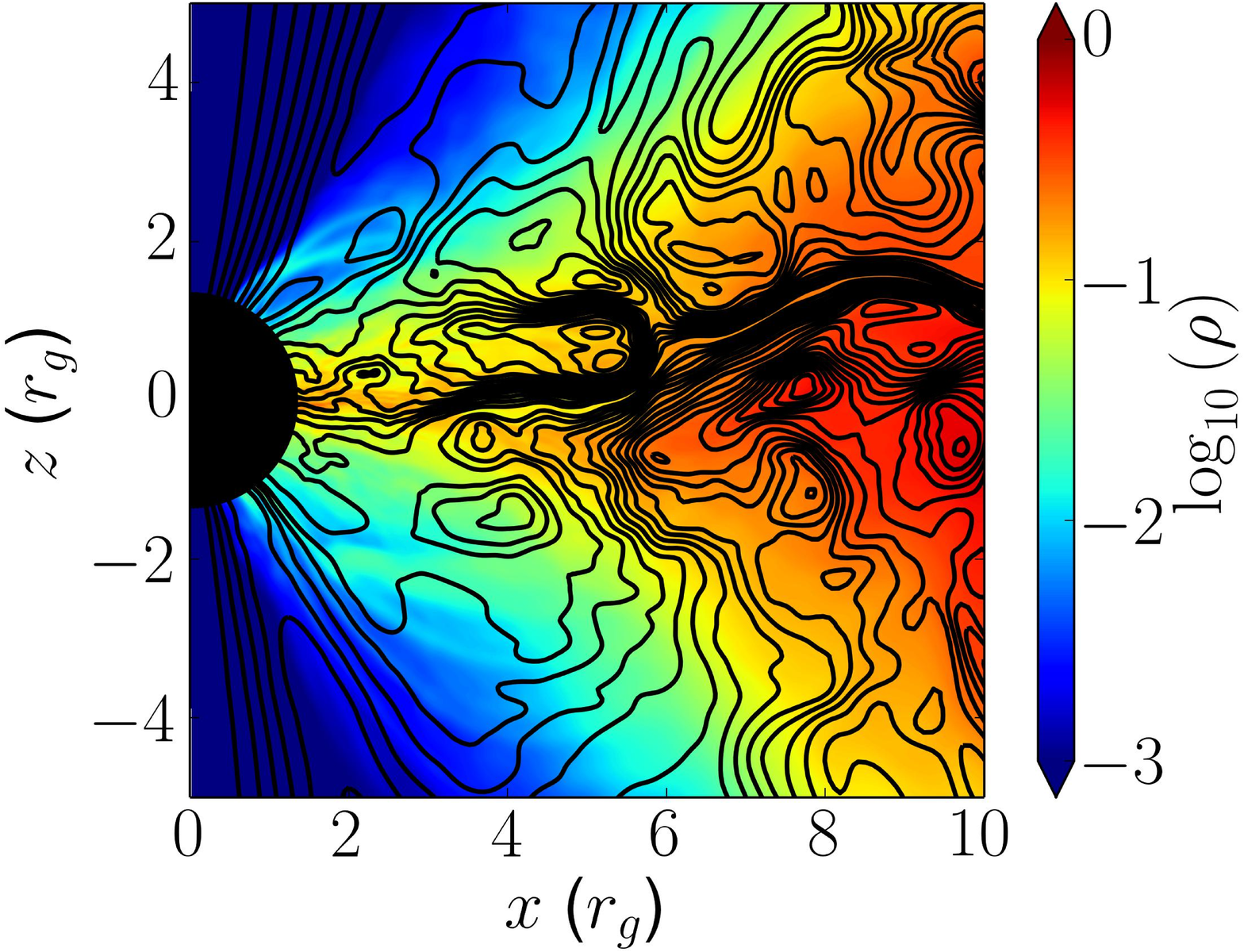}
\includegraphics[scale = .105]{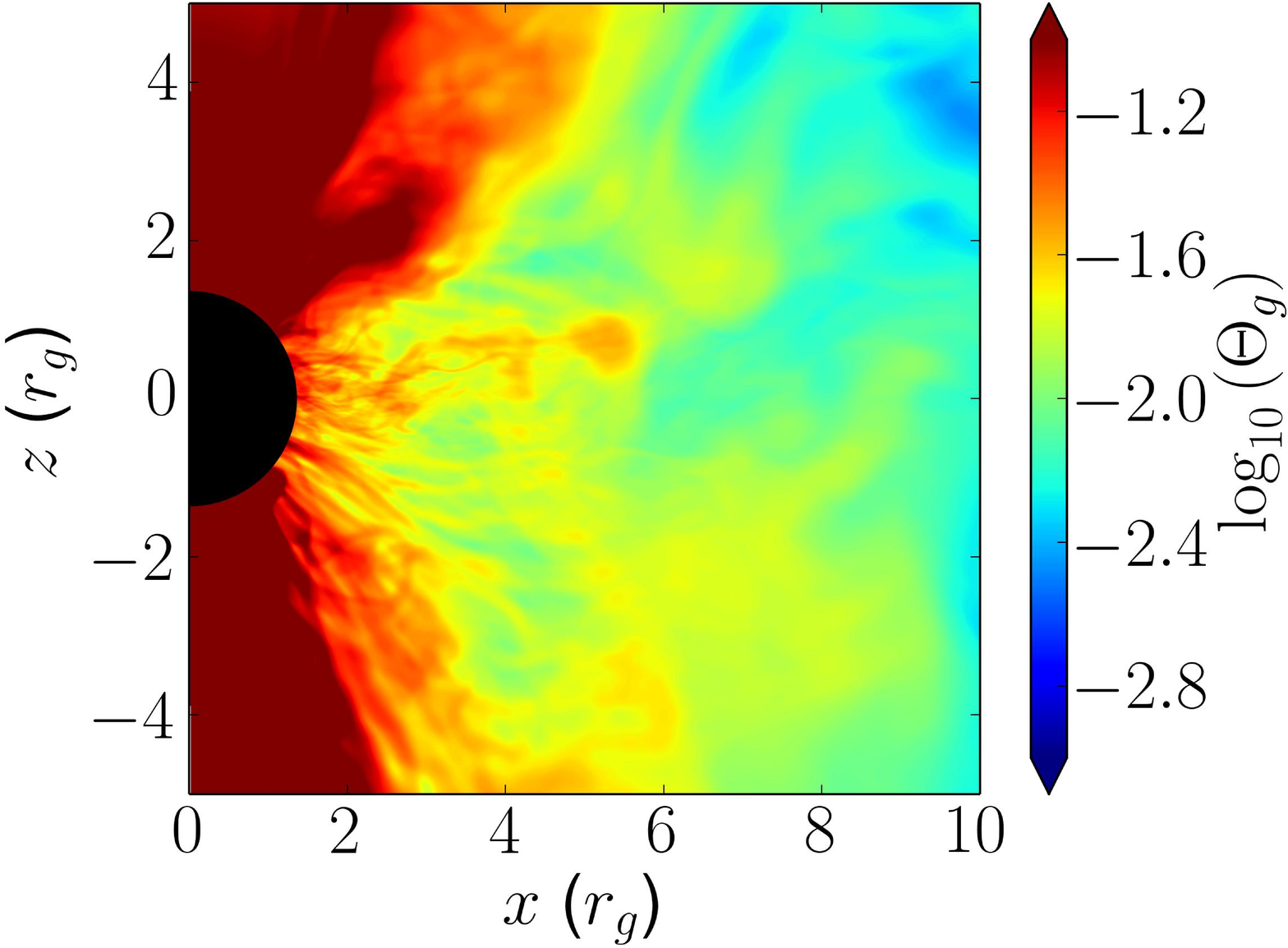}
\includegraphics[scale = .105]{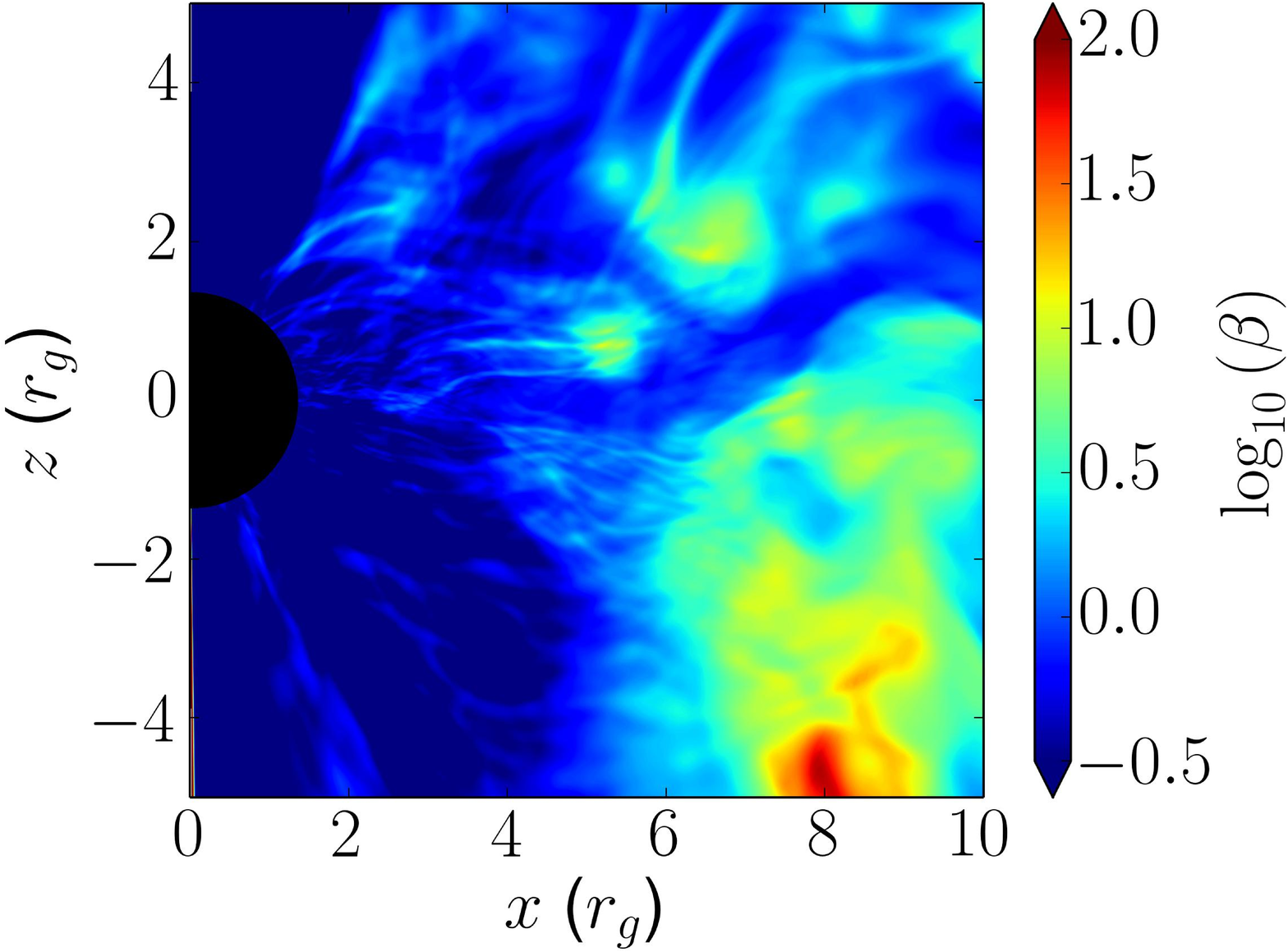}
\includegraphics[scale = .105]{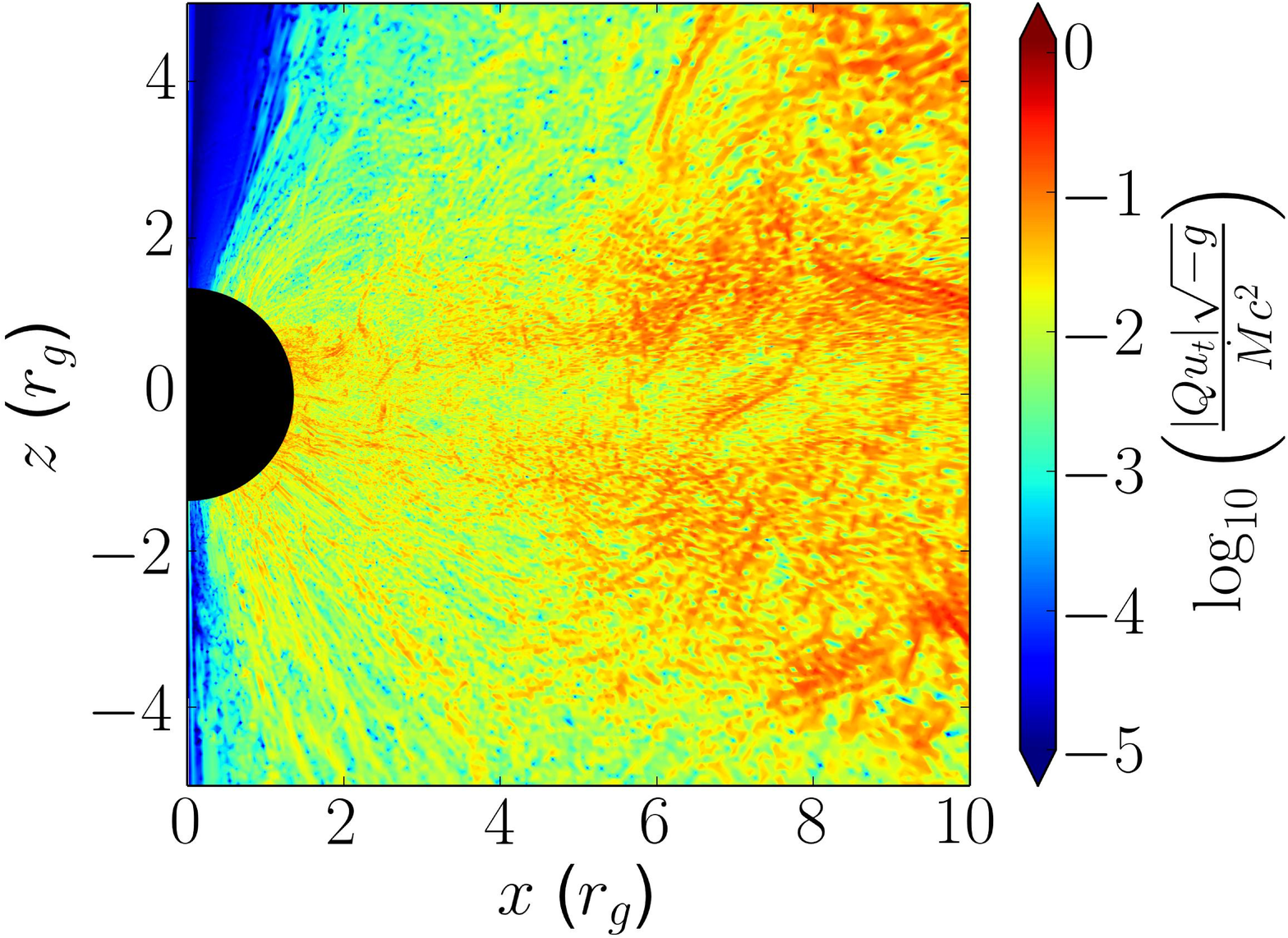}
\caption{Properties of our 2D black hole accretion simulations. The top panel shows the density over-plotted with magnetic field lines in the initial conditions (left) and averaged over $900-1100$ $r_g/c$ (right).  The remaining panels are the total gas temperature in units of $m_p c^2$ (middle left), the plasma parameter, $\beta \equiv 2P_{g}/b^2$ (middle right), and the absolute value of the heating rate per unit volume in the coordinate frame, $|Q u_t|$, in units of $\dot M c^2/(\sqrt{-g})$ (bottom), all averaged over time in the interval $900-1100$ $r_g/c$.  Note that for calculating the average $\beta$, we use $2\langle P_g\rangle/\langle b^2\rangle$, where $\langle \rangle$ denotes an average over time.  These plots represent the background GRMHD solution on top of which we separately solve the electron entropy equation.}%
\label{fig:2loopbg}
\end{figure*}

Figure~\ref{fig:2loopbg} shows the background HARM solution for the density, magnetic field, temperature, plasma $\beta\equiv 2P_g/b^2$, and the heating rate per unit volume in the coordinate frame, $-Qu_t$, averaged over the time interval $900-1100$ $r_g/c$, as well as the initial field configuration. After $\sim 1200 r_g/c$ the turbulence starts to decay, an artefact of 2D simulations in which MRI turbulence is not sustainable.  %

\begin{figure}
\includegraphics[scale = .35]{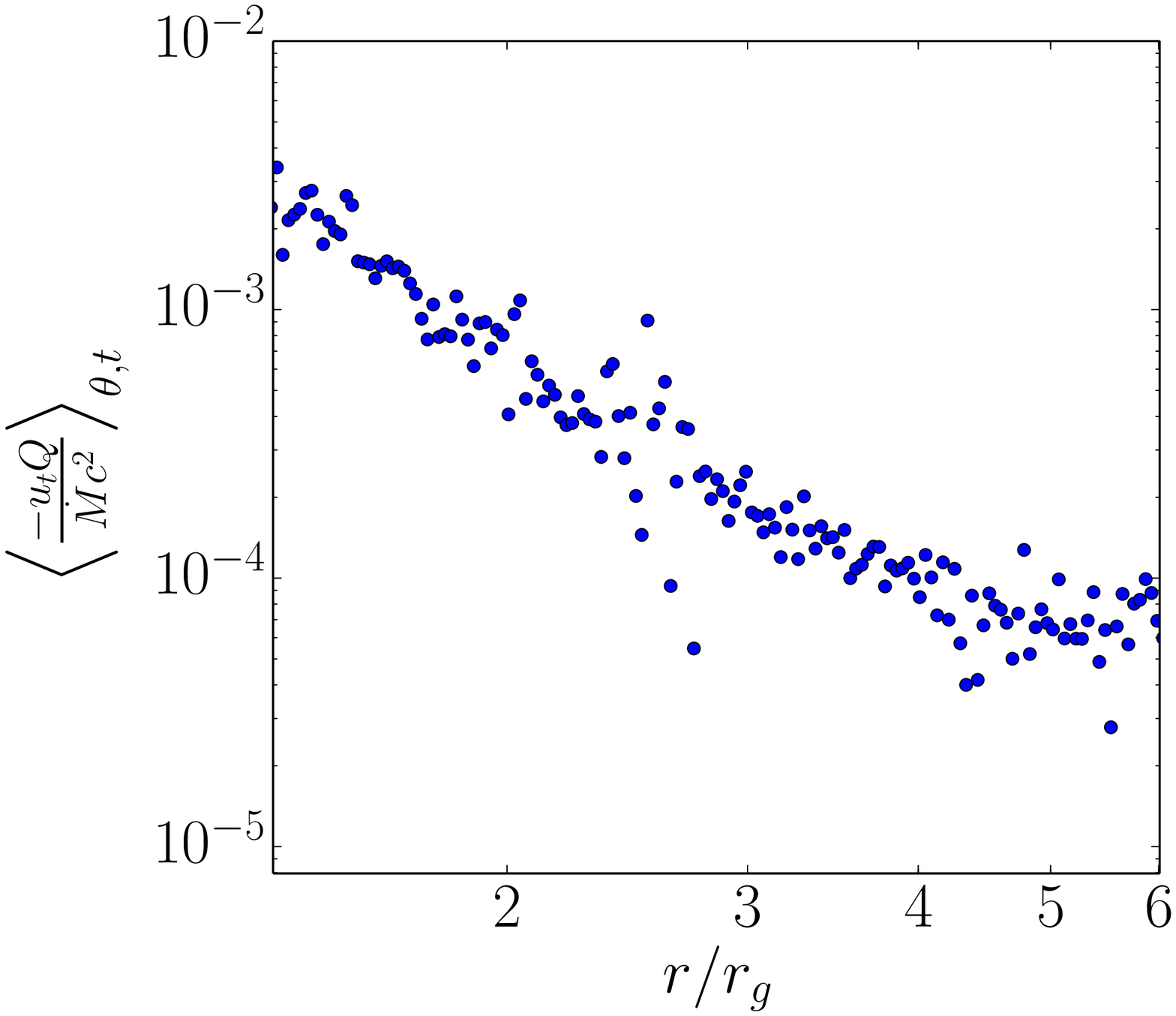}
\caption{Mass-weighted average (see eq.~\ref{eq:massaveragingdef}) of the heating rate per unit volume in the coordinate frame, averaged over time in the interval $900-1100$ $r_g/c$ and over $\theta$ from 0 to $\pi$. Note that for our metric sign convention, $u_t \le 0$. The total volume integrating heating out to $\sim 6 r_g$ is $\sim 0.17 \dot M c^2$ (equation~\ref{eq:qdotint}), comparable to the \citet{Novikov1973} heating rate for this black hole spin.}%
\label{fig:qdotphi}
\end{figure}

As noted in \S\ref{sec:heatcons}, we find locally that $Q < 0$ (violating the second law of thermodynamics) in many regions due to truncation errors. This is because HARM satisfies the total energy equation to machine precision but only satisfies the second law of thermodynamics to truncation error.  However, while $Q$ may be instantaneously or locally negative, when integrated over a sufficient length of time and/or space in the fluid frame it will satisfy the second law of thermodynamics.  In our torus simulation, for instance, Figure~\ref{fig:qdotphi} shows that when averaged over $\theta$ and time ($900-1100$ $r_g/c$), the heating rate is entirely positive definite within the region of interest.  Furthermore, when integrated over the volume enclosed between the event horizon, $r_H$, and $r= 6 r_g$ (roughly the radius at which the accretion time $\sim 1000$ $r_g/c$), we find
\begin{equation}
\int\limits_0^{2\pi}\int\limits_0^{\pi} \int \limits_{r_H}^{6 r_g} -u_t Q (r^2+a^2\cos^2\theta)\sin\theta\,  {\rm d}r\, {\rm d}\theta\, {\rm d}\varphi \approx 0.17 \dot M c^2,
\label{eq:qdotint}
\end{equation}
where the factor of $-u_t$ converts $Q$ to the coordinate frame. In equation~\eqref{eq:qdotint}, $\dot M$ is the accretion rate of the black hole in terms of coordinate time (corresponding to time measured by a distant observer) at the event horizon radius, $r=r_{\rm H}$, 
\begin{equation}
  \dot M  =  \int\limits_{r=r_{\rm in}} \rho u^r (r^2+a^2\cos^2\theta)\sin\theta\,  {\rm d}\theta\, {\rm d}\varphi.
\end{equation}
The heating rate in equation~\eqref{eq:qdotint} is in excellent agreement with that expected for a rapidly spinning black hole (e.g., the \citealt{Novikov1973} model predicts a radiative efficiency of $\approx 0.18$ for $a = 0.9375$).

In this work, all mass-weighted averages are computed using the weighting function: $\rho u^t \gdet$, which represents the conserved mass per unit coordinate volume.  For example, a radial average of a function $f(x^1, x^2, x^3, t)$ is computed as:
\begin{equation}
  \frac{\int \limits ^{x^1_{\rm max}} _{x^1_{\rm min}} f(x^1, x^2, x^3, t) \rho u^t \gdet dx^1}{\int \limits ^{x^1_{\rm max}} _{x^1_{\rm min}} \rho u^t \gdet dx^1}.
\label{eq:massaveragingdef}
\end{equation}

 \label{sec:app}

\subsection{Electron Parameter Choices}
\label{sec:tauchi}

Here we describe physically motivated estimates of the electron heating fraction, $f_e$, and the electron thermal diffusivity, $\chi_e$, appropriate for low-collisionality accretion flows such as that of Sagittarius A*. A more comprehensive exploration of physical models will be explored in future work.

We consider two simple models for the electron heating fraction $f_e$.   The first sets $ f_e = 1/8$, a constant.   Because the electron adiabatic index is not the same as the proton (total) adiabatic index, and the heating is not spatially uniform, a constant $f_e$ model does {\em not} necessarily lead to a constant $T_p/T_e$.  The second, more physical model, sets $f_e$ based on theoretical models of the dissipation of MHD turbulence in low-collisionality plasmas.  These generically predict that electrons receive most of the turbulent heating at low $\beta$ while protons receive most of the turbulent heating at high $\beta$. This is true both for reconnection \citep{Numata2014} and collisionless damping of turbulent fluctuations \citep{Quataert1999}.  This dependence on $\beta$ is the key qualitative feature of our chosen model of $f_e$.   For concreteness, we use the specific calculations of \citet{Howes2010} who provided a simple fitting function for the electron to proton heating rate as a function of plasma parameters in calculations of the collisionless damping of turbulent fluctuations in weakly compressible MHD turbulence like that expected in accretion discs.   These models do a reasonable job of explaining the measured proton and electron heating rates in the near-Earth solar wind \citep{Howes2011}. The functional form of $f_e$ is derived from the relations:
\begin{equation}
\frac{Q_p}{Q_e} = c_1 \frac{c_2^2 + \beta_p^{2-0.2\log_{10}(T_p/T_e)}}{c_3^2 +  \beta_p^{2-0.2\log_{10}(T_p/T_e)}}\sqrt{\frac{m_p T_p}{m_e T_e}}e^{-1/\beta_p},
\label{eq:qrat}
\end{equation}
with $c_1 = 0.92$, $c_2 = 1.6/(T_p/T_e)$, and $c_3 = 18+5\log_{10}(T_p/T_e)$ for $T_p/T_e>1$, while $c_2 = 1.2/ (T_p/T_e)$ and $c_3 = 18$ for $T_p/T_e <1$. The corresponding result for $f_e$ is simply 
\begin{equation}
f_e \equiv \frac{Q_e}{Q_p+Q_e} = \frac{1}{1+Q_p/Q_e}.
\end{equation}
The critical assumption used in deriving equation~\eqref{eq:qrat} is that the turbulent fluctuations on the scale of the proton Larmor radius have frequencies much lower than the proton cyclotron frequency.   This is believed to be well-satisfied for weakly compressible MHD turbulence in accretion disks (e.g.,  \citealt{Quataert1998}).  For concreteness, we note that for $T_p/T_e = 1$ and $\beta_p = (0.1,0.3,1,10)$, we have $Q_p/Q_e = (0,0.01,0.16,8.6)$, while for $T_p/T_e=10$ and $\beta_p = (0.1,0.3,1,10)$, $Q_p/Q_e = (0,0.001,0.09,12)$, respectively.  This demonstrates the strong transition from predominantly electron to predominantly proton heating with increasing $\beta_p$, with the transition happening at a value of $\beta_p$ that depends weakly on the proton to electron temperature ratio. This implies that we expect strong electron heating in the corona and jet regions but suppressed electron heating in the bulk of the disc. 

 We reiterate that the key feature of equation~\eqref{eq:qrat} is not the precise value of the predicted $Q_p/Q_e$, but rather the transition from $Q_p \gtrsim Q_e$ for $\beta_p \gg 1$ to $Q_p \ll Q_e$ for $\beta_e \ll 1$.  This qualitative transition is much more robust than the specific functional form in equation~\eqref{eq:qrat} (e.g., \citealt{Quataert1999}; \citealt{Numata2014}).

For the electron thermal diffusion parameters, since $\chi_e$ is a diffusion coefficient, we assume that it has the form 
\begin{equation}
\chi_e = \alpha_e c r,
\end{equation}
where $\alpha_e$ is a dimensionless thermal diffusivity, and $r$ is the radial distance from the center of the black hole, which is comparable to the density scale height of the disc, $H$. Since we are interested in fairly relativistic electrons, we choose the relevant velocity to be $c$ in our diffusivity estimate.  In what follows, we consider a range of dimensionless diffusivities, $\alpha_e \sim 0.1-10$.    A typical value of $\alpha_e \sim 1$ is motivated by the idea that particles scatter roughly after moving a distance comparable to the length-scale over which the magnetic field strength, density, etc. change.  In fact, for high beta plasmas, the mean free path due to wave-particle scattering can be significantly lower, reducing the thermal diffusivity significantly.  In Appendix~\ref{app:whist} we discuss the specific limits imposed by electron temperature anisotropy instabilities present in a turbulent plasma.  In particular, the whistler and firehose instabilities lead to limits on $\Delta T_e/T_e$  (eq.~\ref{eq:dPe-limits}) and thus the electron viscosity and thermal diffusivity,  where the temperature anisotropy is defined with respect to the local magnetic field. In terms of the electron thermal diffusivity, this becomes $\chi_e = \min\left(\alpha_e r c , \chi_{\rm max}\right)$, where $\chi_{\rm max}$ is set by velocity space instabilities and is estimated in Appendix~\ref{app:whist}. Finally, we choose the relaxation time scale, $\tau$ to be given by the thermal time scale: 
\begin{equation}
\tau \sim  \frac{\chi_e}{v_{th}^2} \sim \frac{\chi_e}{c^2}  .
\label{eq:tau}
\end{equation}
Comparing this to the stability condition given by equation~\eqref{eq:taulim}, we see that stability is ensured if 
\begin{equation}
v_{th} \lesssim \frac{\Delta x}{\Delta t}.
\label{eq:vthlim}
\end{equation}
Since we have a non-uniform grid, the time step $\Delta t$ is essentially set by the light crossing time of the smallest grid cell (i.e. that nearest the horizon), meaning that $\Delta t \lesssim c \Delta x$ near the horizon and $\Delta t \ll c \Delta x$ further from the horizon.  For a reasonable choice of a CFL number of 0.5, we find that equation~\eqref{eq:vthlim} is satisfied everywhere and is not a limiting factor in our simulation. Moreover, we find that the exact value of $\tau$ is not critical as long as it satisfies numerical stability and is not too long (e.g. is less than a local dynamical time).

\subsection{Electron Heating Only}
In this section we focus solely on the effects of separately evolving the electron internal energy equation without conduction in our black hole torus simulation and compare the results for different electron heating models.
\subsubsection{Constant Electron Heating Fraction}
\begin{figure}
\includegraphics[scale = .105]{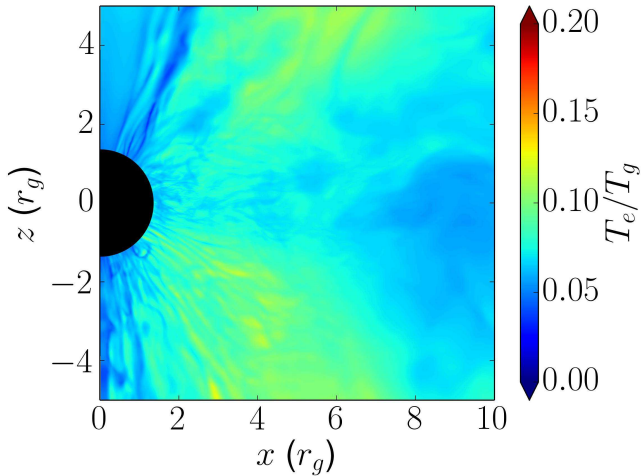}
\caption{Ratio of $\left<T_e\right>/ \langle T_{g}\rangle$ in our black hole accretion simulation, where $\langle\rangle$ denotes an average over time in the interval $900-1100$ $r_g/c$ (where $T_g$ is the temperature of the single fluid GRMHD simulations and $T_e$ is the electron temperature). These results assume a constant fraction of dissipated heat is given to the electrons ($f_e = 1/8$).  Compare with the more physical $\beta$-dependent heating results in Figure~\ref{fig:Tratbeta}. }%
\label{fig:Tratfe3}
\end{figure}

Figure~\ref{fig:Tratfe3} shows the temperature ratio, $T_e/T_g$, averaged over the interval $900-1100r_g/c$ for $f_e = 1/8$. We reiterate that $T_g$ here is the temperature inferred from the underlying single fluid GRMHD solution (approximately the proton temperature in our model) while $T_e$ is the electron temperature determined from our separate electron entropy equation.  We include this constant $f_e$ result primarily because it is conceptually similar (although quantitatively different) to the constant $T_p/T_e$ assumption often used in the literature.  Notice that the resulting $T_e/T_{g}$ ratio, seen in Fig.~\ref{fig:Tratfe3}, is non-uniform despite the constant $f_e$. Also note that due to the fact that MHD turbulence is unsustainable in 2D simulations, the heating dies off after $\sim 1200 r_g/c$ of evolution and prevents the outer $r \gtrsim 10 r_g$ region of the disc from ever being heated substantially. However, as we will see later, conduction can occur at a much faster (electron thermal) speed along the magnetic field lines and can affect the solution at somewhat larger radii.

\subsubsection{$\beta$-Dependent Electron Heating}

\label{sec:appbeta}
\begin{figure}
\includegraphics[width=0.97\columnwidth]{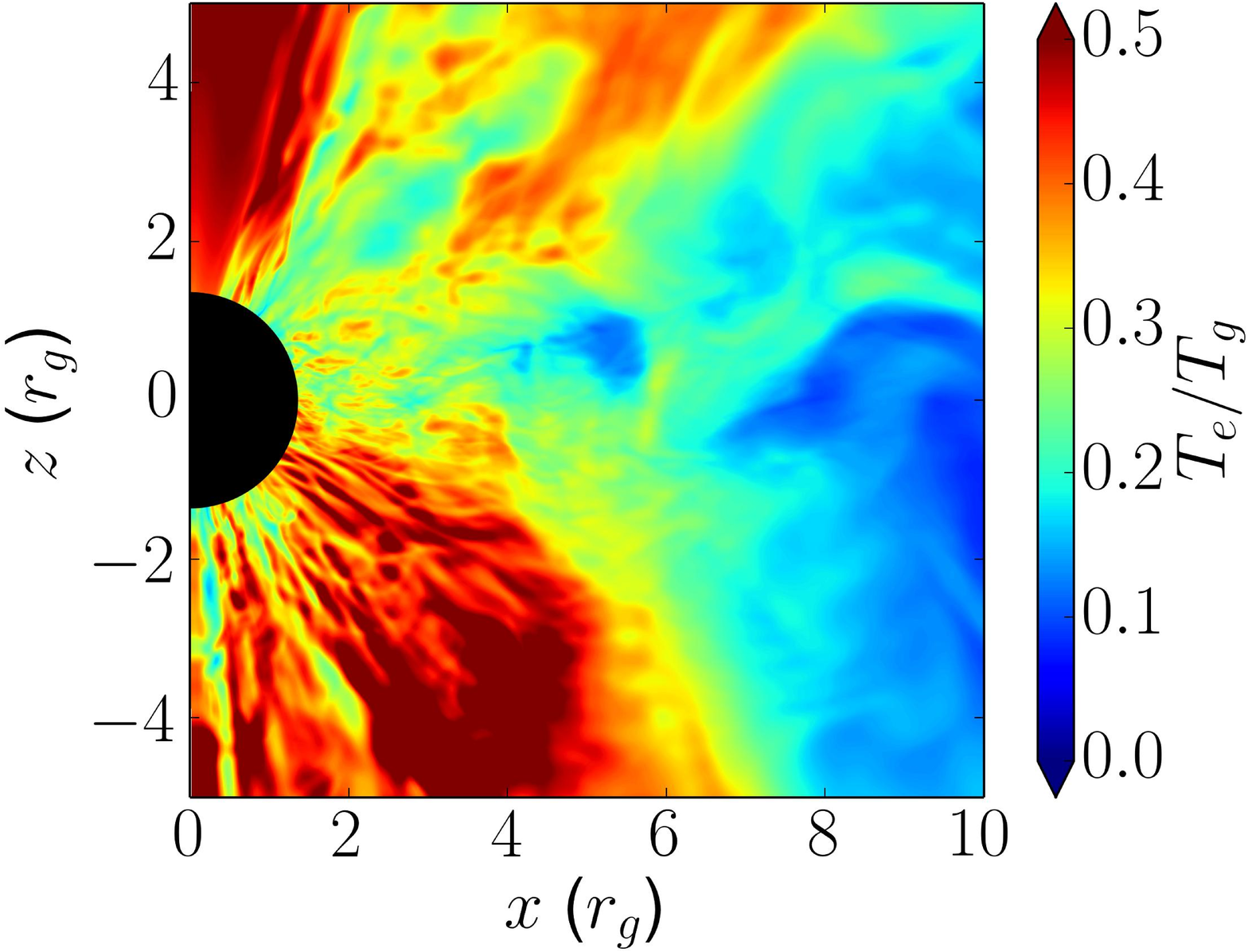}
\includegraphics[width=\columnwidth]{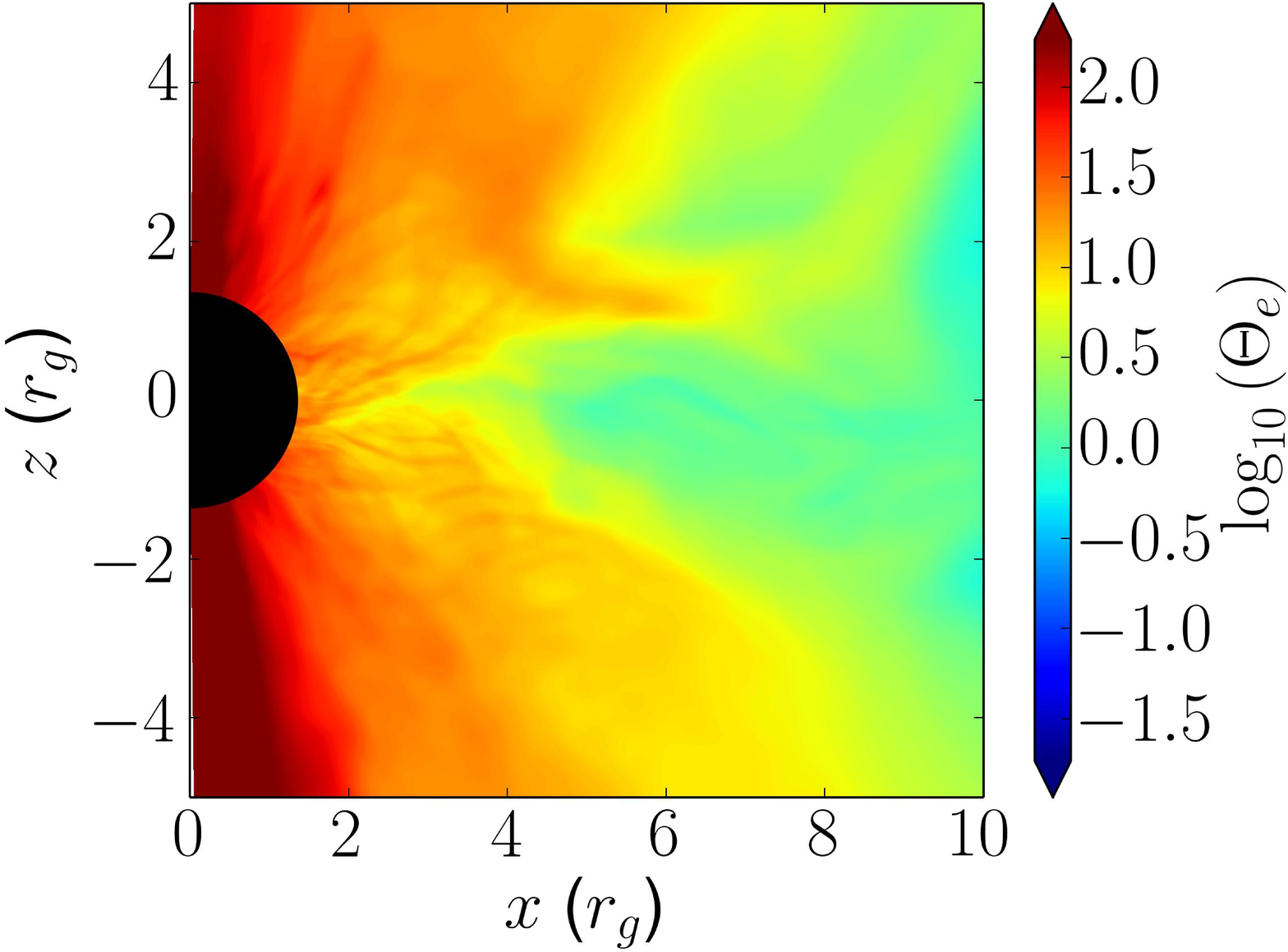}
\includegraphics[width=0.97\columnwidth]{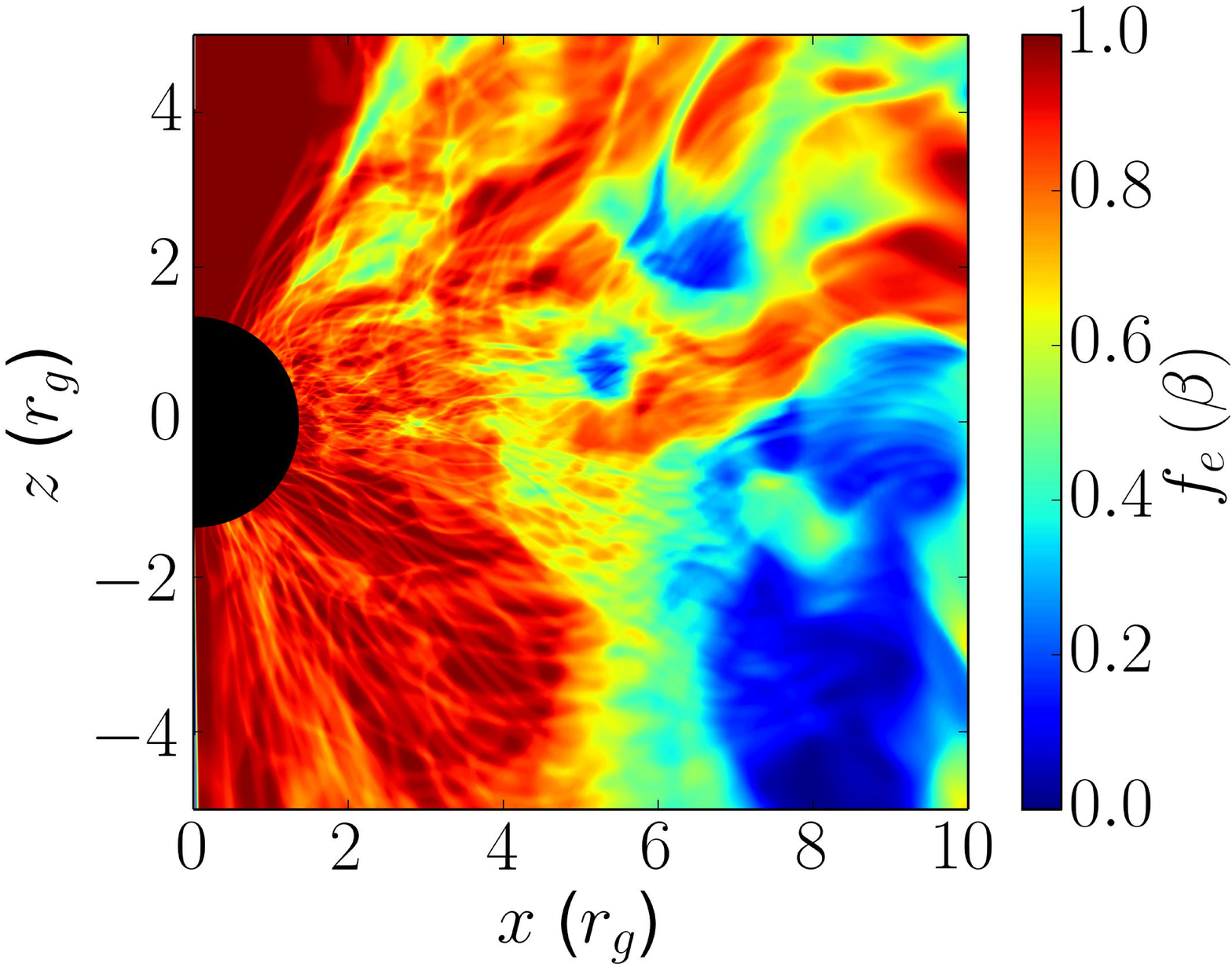}

\caption{Ratio of $\left<T_e\right>/\langle T_{g}\rangle$ (top), electron temperature, $\left<T_e\right>$, in units of $m_e c^2$ (middle), and electron heating fraction, $\left<f_e\right>$ (bottom), where $\left<\right>$ denotes an average over time in the interval $900-1100$ $r_g/c$.  These results are for $\beta$-dependent heating (see \S\ref{sec:appbeta}).  Compare to Figure~\ref{fig:Tratfe3} for a constant electron to proton heating ratio.  The highly non-uniform distribution of $\beta$ (see Figure~\ref{fig:2loopbg}) and the strong $\beta$ dependence of the electron-to-total heating ratio (equation~\ref{eq:qrat}) lead to a strong angular dependence of $T_e/T_{g}$. }
\label{fig:Tratbeta}
\end{figure}

\begin{figure}
\includegraphics[scale = .35]{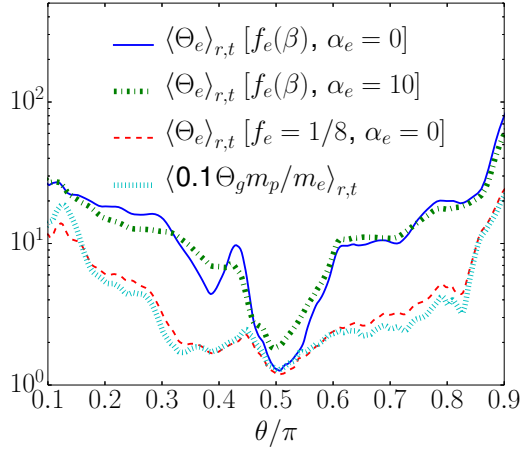}
\caption{Mass-weighted average of total gas and electron temperature (in units of $m_e c^2$) as a function of the polar angle, $\theta$.  We show the electron temperature with and without conduction for a $\beta$-dependent electron heating fraction, $f_e$, as well as without conduction for a constant electron heating fraction $f_e = 1/8$.  The results are averaged over time from $900-1100$ $r_g/c$ and averaged over $r$ from $5-7$ $r_g$. Note that the total gas temperature has been multiplied by a constant fraction to more clearly compare to the electron temperatures.  The electron temperature with $\beta$-dependent heating displays much stronger $\theta$ variation because the electron heating fraction itself varies with $\theta$ (see Figure~\ref{fig:Tratbeta}).  Conduction has only a modest effect on redistributing heat in $\theta$ due to the geometry of the field.}
\label{fig:Ttheta}
\end{figure}

Figure~\ref{fig:Tratbeta} shows the temperature ratio, $T_e/T_{g}$, the electron temperature itself, $\Theta_e$, and the electron heating fraction, $f_e$, averaged over the interval $900-1100$ $r_g/c$ for the $\beta$-dependent heating model of equation~\eqref{eq:qrat}, which we regard as a more physical electron heating model than $f_e =$ const.  We note that this leads to hot electrons being strongly concentrated in the corona of the torus in between the disc and the jet, where $\beta$ is the smallest (and $f_e \sim 1$ from equation ~\ref{eq:qrat}). This is also clear from the 1D profiles of electron temperature as a function of polar angle in Figure~\ref{fig:Ttheta}.

Figure~\ref{fig:Ttheta} shows the mass-weighted average over radius ($r=5{-}7 r_g$) of the electron and gas temperatures, plotted versus the polar angle, $\theta$.  The $f_e=1/8$ electrons and total gas temperatures have mild variation in $T$ with $\theta$, while the $f_e = f_e(\beta)$ electrons have significantly higher temperatures in the polar regions. This demonstrates that the non-uniformity of the electron temperature in the $f_e(\beta)$ model is primarily caused by the strong $\beta$-dependence of our model of $f_e$ as opposed to any non-uniformity of the heating rate itself (Figure~\ref{fig:2loopbg}).

\subsection{Conduction and Electron Heating}

\begin{figure}
\includegraphics[scale = .35]{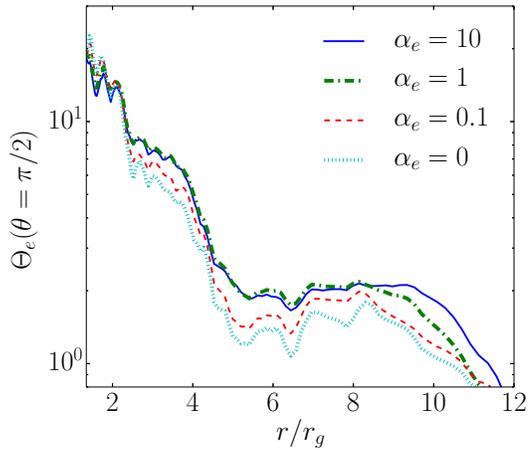}  
\caption{Electron temperature in the mid-plane ($\theta=\pi/2$) in units of $m_e c^2$ for black hole accretion simulations with $\beta$-dependent heating and for electron conduction with dimensionless conductivity $\alpha_e = 0,0.1,1,10$ (where the electron thermal diffusivity is $\chi_e = \alpha_e r c$; see \S\ref{sec:tauchi}). The results are time averaged over the interval $900-1100 r_g/c$. For $alpha_e \gtrsim  1$, conduction redistributes energy from small to large radii, increasing the electron temperature at larger radii.  Compare to Figure~\ref{fig:Ttheta}, which shows that redistribution of heat in the polar direction is less efficient. }
\label{fig:logTcon}
\end{figure}

\begin{figure}
\includegraphics[scale = .105]{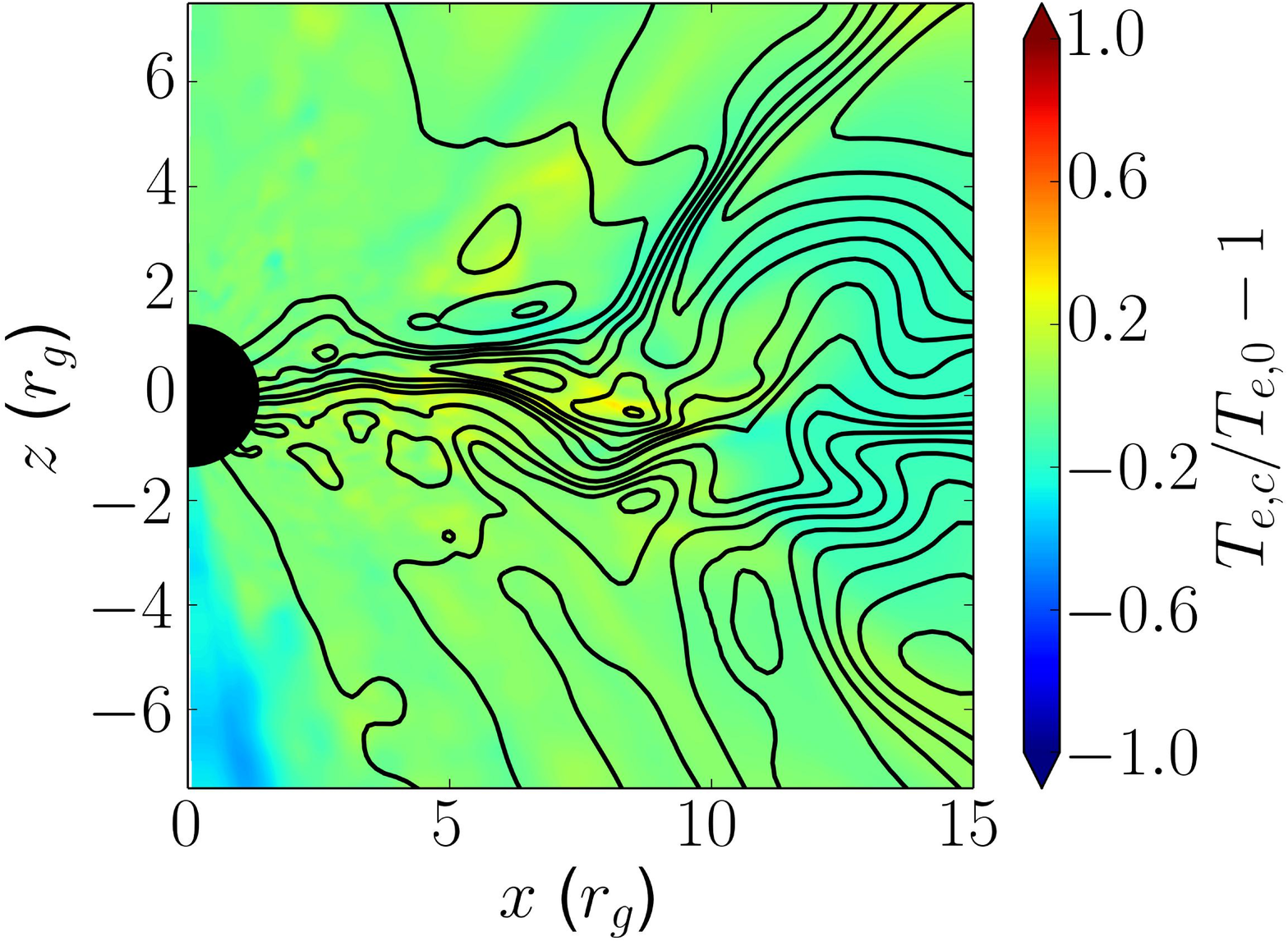}
\includegraphics[scale = .105]{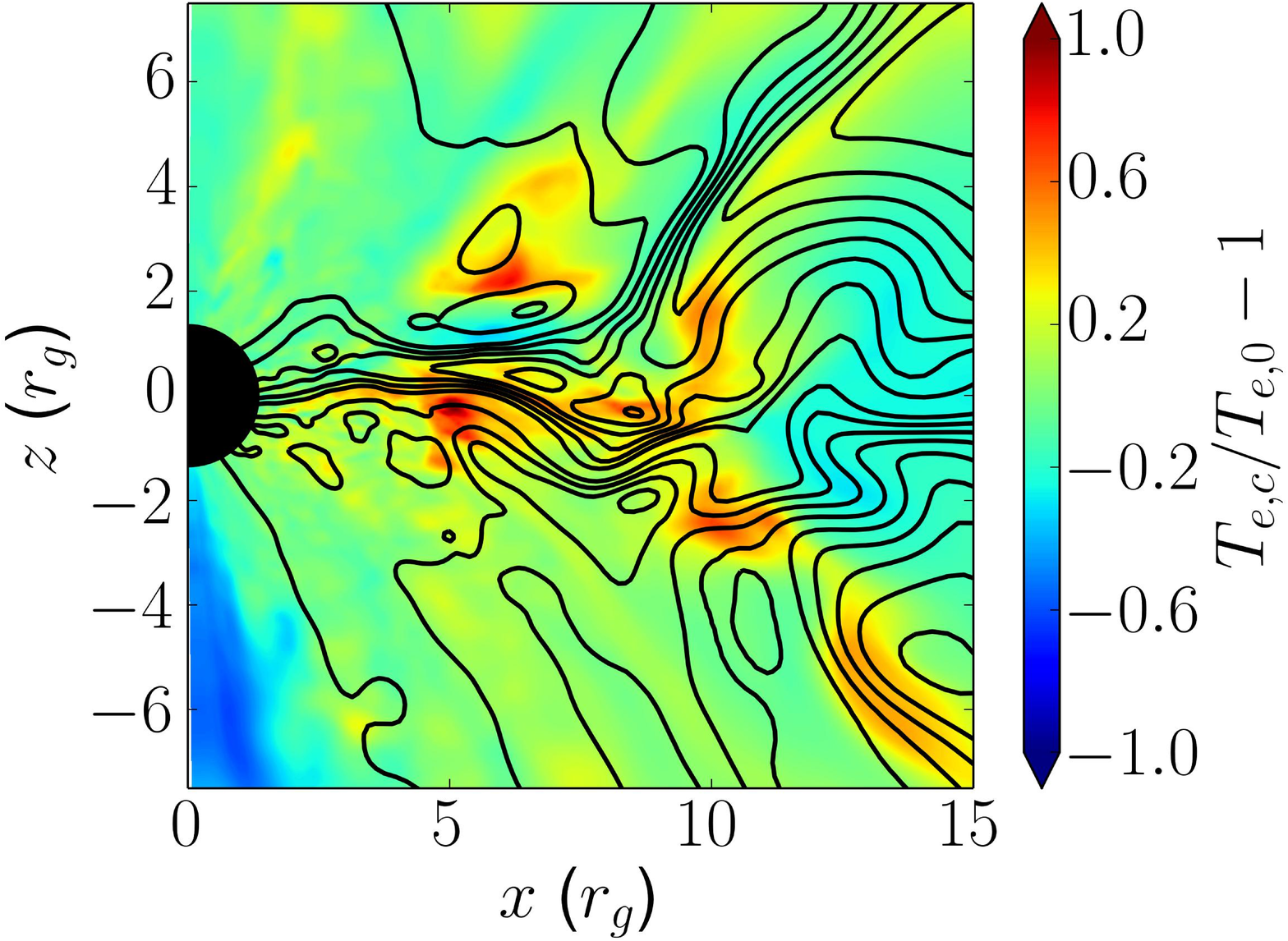}
\includegraphics[scale = .105]{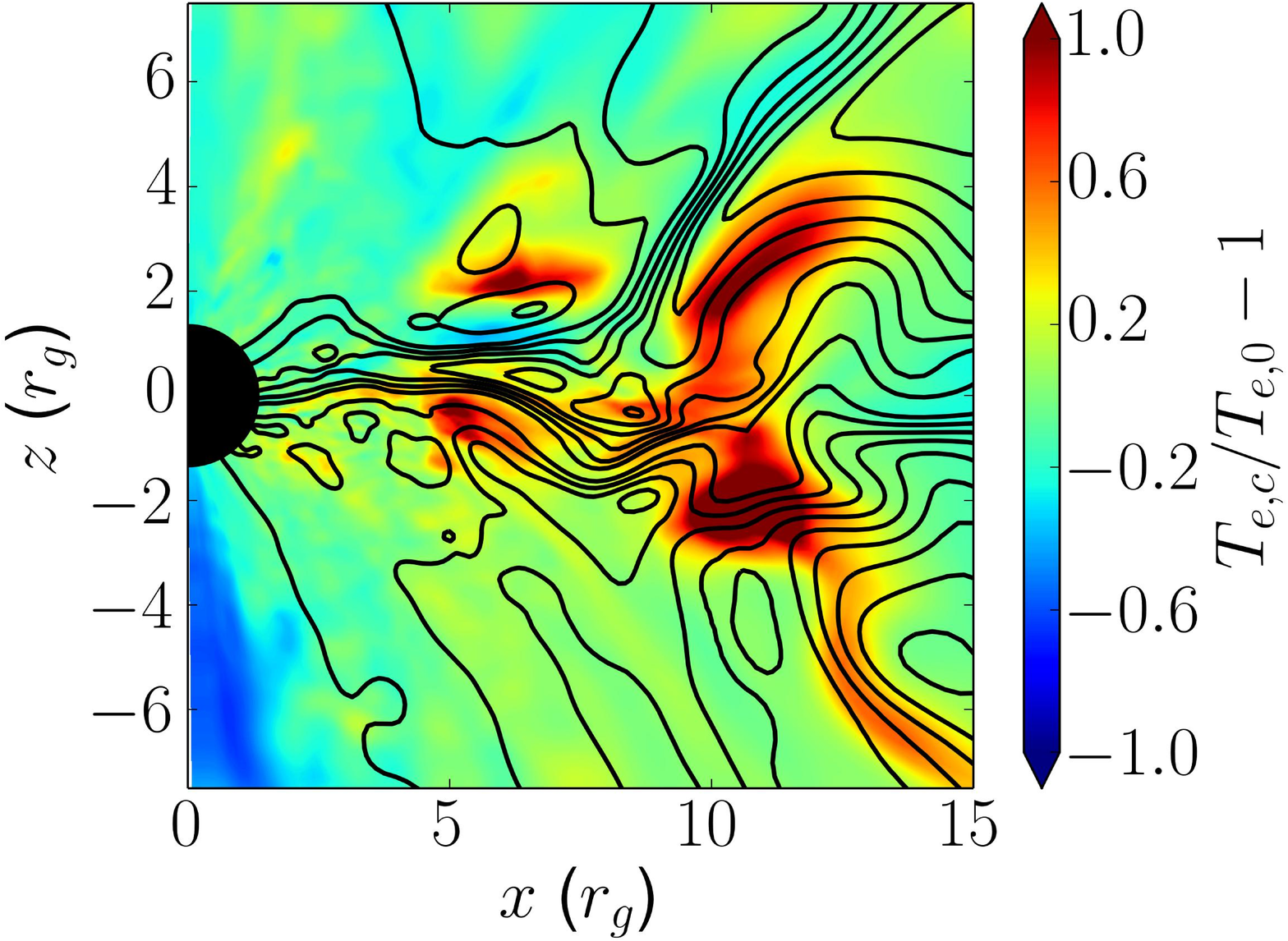}
\caption{Fractional difference in electron temperature between solutions with and without electron conduction shown in colour (see colour bar for details) over-plotted with magnetic field lines shown as solid black lines.  The fractional difference is calculated as $\left<T_{e,c}\right>/\left<T_{e,0}\right>-1$, where $\left<\right>$ denotes an average over time from $900-1100$ $r_g/c$.   The results include $\beta$-dependent electron heating for $\alpha_e=0.1$ (top), $\alpha_e = 1$ (middle), and $\alpha_e = 10$ (bottom panel), where the electron thermal diffusivity is $\chi_e = \alpha_e r c$ (\S\ref{sec:tauchi}).  Higher $\alpha_e$ allows more heat to flow from the inner regions to larger radii. For $\alpha_e = 0.1$ conduction has a negligible effect on the electron temperature, while for $\alpha_e \gtrsim 1$ conduction leads to order unity changes in $T_e$.}
\label{fig:TcoTdis1}
\end{figure}

We now consider the effects of electron conduction on the electron temperature structure of black hole accretion discs.   We focus on the more physical model of $\beta-$dependent heating described in \S\ref{sec:tauchi}.  In all of our calculations, we include the velocity space instability limit on the electron thermal conductivity (Appendix~\ref{app:whist}), although runs without this limit produce similar results because $\beta$ is modest ($\lesssim 1-10$) in the inner regions of these simulations (Figure~\ref{fig:2loopbg}).  Figure~\ref{fig:logTcon} shows the electron temperature as a function of radius at the mid-plane in the simulations with and without conduction. Figure~\ref{fig:TcoTdis1} shows the effects of conduction more quantitatively via the fractional change in temperature between the electron temperature solution with conduction and that without.

\begin{figure}
\includegraphics[scale = .105]{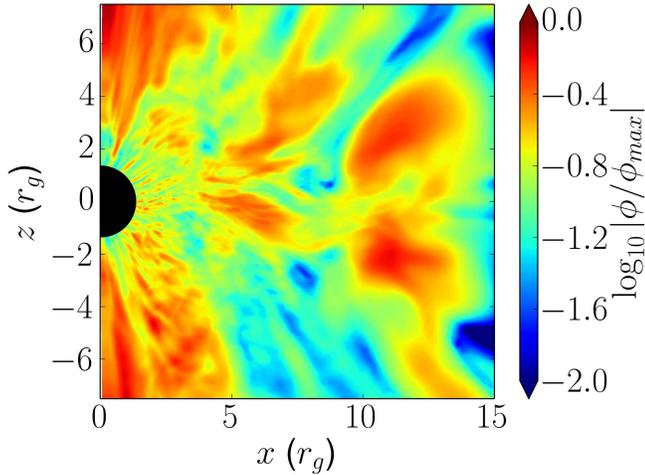}
\caption{$\langle |\phi|\rangle /\langle\phi_{\rm max}\rangle$, the ratio of the electron heat flux to the maximum value $\phi_{\rm max} = (u_{e} + \rho_e c^2) v_{t,e}$, where $\langle \rangle$ denotes an average over time from $900-1100$ $r_g/c$.  This is calculated based on the results of a black hole accretion simulation with $\beta$-dependent electron heating and a dimensionless electron thermal conductivity of $\alpha_e = 10$ (where the electron thermal diffusivity is $\chi_e = \alpha_e r c$; see \S\ref{sec:tauchi}).  Comparison with Figure~\ref{fig:TcoTdis1} shows that conduction has a significant effect on redistributing heat only in the regions where the heat flux is saturated or nearly saturated.  However, even for a high electron thermal conductivity of $\alpha_e=10$, the heat flux is still well below the saturated value over much of the domain.}%
\label{fig:phirat}
\end{figure}

To summarise Figures~\ref{fig:logTcon} and~\ref{fig:TcoTdis1}, conduction has little effect on the electron temperature for $\alpha_e \lesssim 1$.  However, for $\alpha_e \gtrsim 1$,  conduction leads to a significant radial redistribution of heat such that the electron temperature is factors of a few larger at large radii.  Even for $\alpha_e>1$, however, the angular redistribution of heat is much less efficient, as seen in the radially and time-averaged electron temperatures in Figure~\ref{fig:Ttheta} for $\alpha_e = 10$. This is primarily because of the structure of the magnetic field, as can be seen by noting that the regions where conduction modifies the temperature in Figure~\ref{fig:TcoTdis1} largely follow magnetic field contours which do not efficiently connect the polar and equatorial regions. To aid the interpretation of these results, Figure~\ref{fig:phirat} shows the heat flux $\phi$ normalised to the maximum value $\phi_{\rm max} = (u_{e} +\rho_e c^2) v_{t,e}$; even for $\alpha_e =10$ the heat flux is still well below the saturated value in significant parts of the domain.  We now summarise and interpret these results in more detail.

\begin{figure}
\includegraphics[scale = .105]{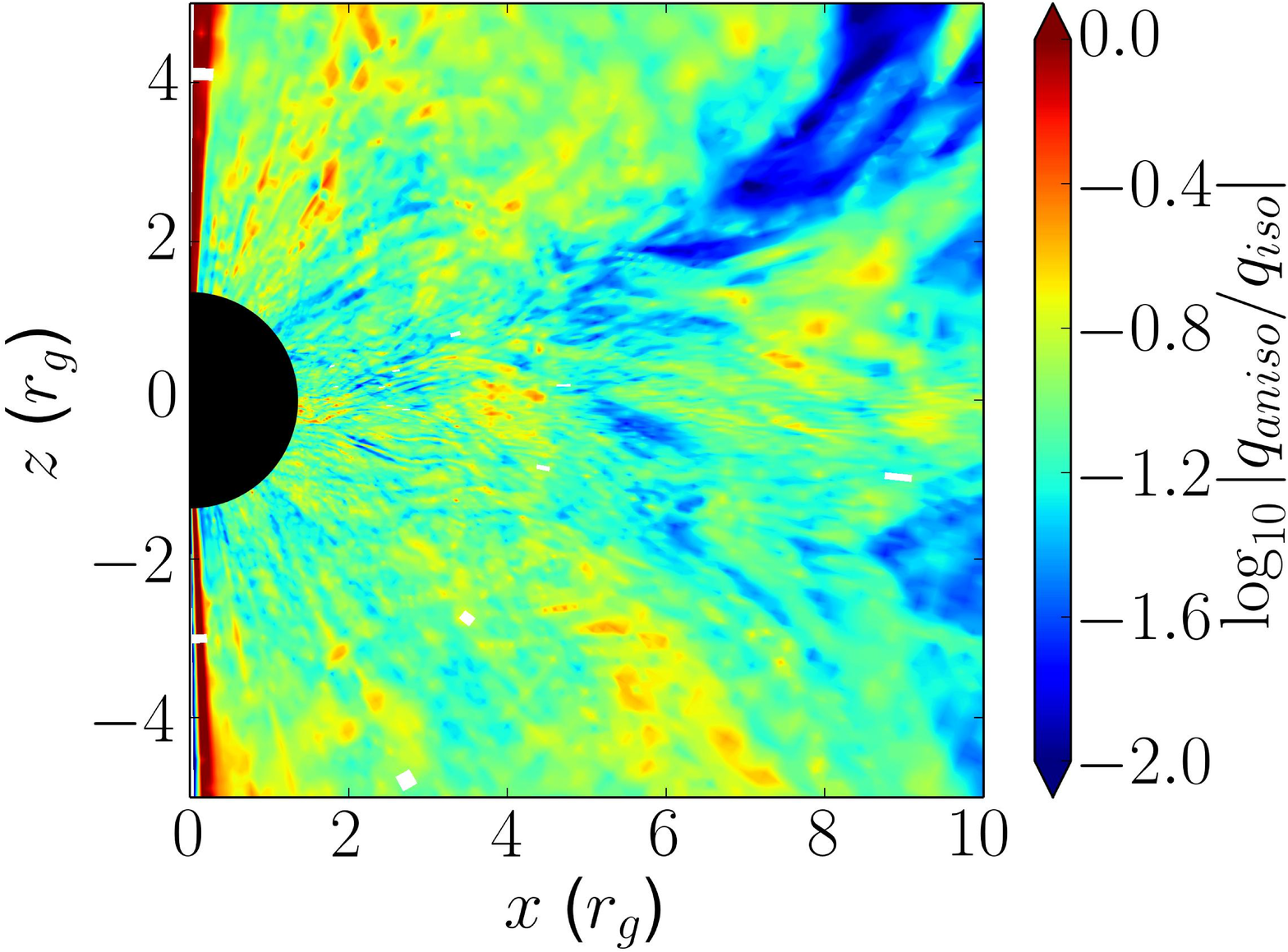}
\caption{$\langle |q_{\rm aniso}| \rangle /\langle |q_{\rm iso}|\rangle$, the ratio of the anisotropic (field-aligned) heat flux to the isotropic heat flux , where $\langle \rangle$ denotes an average over time from $900-1100$ $r_g/c$.  This is calculated based on the results of a black hole accretion simulation with $\beta$-dependent electron heating but without conduction.  The factor of $\sim 5-10$ suppression of the field aligned heat flux is roughly consistent with that expected from local shearing box calculations of MRI turbulence, where $\vec{B}$ is aligned with the $\hat \varphi$ direction (e.g. \citealt{Guan2009}).}%
\label{fig:qrat}
\end{figure}

For $\alpha_e \lesssim 1$ we find conduction to have only a small effect on the electron thermodynamics in the accretion disc,  despite the relatively high conductivity.  We can understand this result as being due to the suppression of the isotropic heat flux by being projected along field lines, quantified by the ratio,
\begin{equation}
\epsilon^2 \equiv \frac{\left( q_\mu q^\mu\right)_{\rm aniso}}{ \left(q_\mu q^\mu\right)_{\rm iso}},
\label{eq:isorat}
\end{equation}
where $q^\mu_{\rm iso}$ and $q^\mu_{\rm aniso}$ are evaluated using the electron temperature as evolved \emph{without} conduction and which we now define.  For this diagnostic, we use 
\begin{equation}
q^\nu _{\rm iso} =  -\rho \chi_e h^{\mu\nu}\left(\partial_\mu T_e  +T_e a_\mu\right),
\end{equation}
where $h^{\mu \nu} = u^\mu u^\nu + g^{\mu \nu}$ is the projection tensor that projects along a space-like direction perpendicular to the fluid velocity $u^\mu$.  This projection ensures that the heat flux in the fluid frame has a zero time-component. Likewise, for $q^\nu_{\rm aniso}$, we use the first order anisotropic heat flux: $q^\nu_{\rm aniso} = \left(\hat b^\mu q_\mu^{\rm iso}\right) \hat b^\nu$. Note that in equation~\eqref{eq:isorat}, $\epsilon$ is always $\le 1$ because both heat fluxes are mutually orthogonal to $u^\mu$. Figure~\ref{fig:qrat} shows $\left|q_{\rm aniso}\right|/\left| q_{\rm iso}\right|$ in our torus simulation, where we find the suppression of the isotropic heat flux to be around $\epsilon\sim 0.2$.  The simplest explanation for this small number is that the field is predominantly in the $\varphi$ direction, where the temperature gradient is identically 0 in 2D simulations.  For instance, in local shearing box calculations, \citet{Guan2009} found that the typical angle between $\vec{B}$ and $\hat \varphi$ was $\sim 10-15^{\circ}$, corresponding to a suppression of the heat flux with $\epsilon\sim 0.25$.


Contrary to the $\alpha_e < 1$ cases, setting $\alpha_e \ge 1$ causes conduction to have a significant effect by redistributing the electron heat from the coronal regions to the bulk of the torus at larger radii. This redistribution of heat causes the electron temperature to actually exceed the total gas temperature in certain regions, which formally violates our assumption that $T_e\ll T_p$.

While the calculation with $\alpha_e=10$, or with $\chi_e= 10 r c$, might seem to use an unphysically large conductivity, roughly corresponding to a length scale for conduction of $\sim 10 H$, where $H$ is the disc density scale height, the heat flux in these calculations is limited to be smaller than the value set by the physically motivated whistler criterion in equation~\eqref{eq:chimax} and to be less than the saturated heat flux $\sim u_{e}c$. As Figure~\ref{fig:phirat} shows, the heat flux is saturated in only part of the domain. Furthermore, the appropriate length scale for conduction should be the scale height \emph{along field lines}, which could be significantly greater than the overall density scale height if the field has a large toroidal component. For these reasons, we believe that the larger $\alpha_e$ solutions may in fact be physical because they correspond to a heat flux closer to the saturated value $\sim u_{e}c$ expected in low-collisionality plasmas.

\section{Conclusions}

\label{sec:conc}
We have presented a method for evolving a separate electron entropy equation in parallel to the standard equations of ideal General Relativistic MHD.  Our motivation is the study of two-temperature radiatively inefficient accretion flows (RIAFs) onto black holes, in which the electron-proton Coulomb collision time is sufficiently long that the proton and electron thermodynamics decouple (e.g., \citealt{Rees1982}).    Understanding the electron temperature distribution close to the black hole is necessary for robustly predicting the radiation from the numerical simulations of black hole accretion (and outflows) in the sub-Eddington regime. 

The long-term goal of the present work is to incorporate the key processes that influence the electron thermodynamics in RIAFs into GRMHD simulations:   heating, thermal conduction, radiative cooling, and electron-proton Coulomb collisions.  In the present paper we have focused on the first two of these processes.  Specifically, we have developed, implemented, and tested a model that quantifies the rate of heating in a conservative GRMHD simulation (\S\ref{sec:heat}).   We then assign a fraction $f_e$ of this heating to the electrons based on a microphysical model of the key heating processes (e.g., turbulence, reconnection, shocks; see, e.g., \S\ref{sec:tauchi}).   In addition, we have implemented and tested a model of relativistic anisotropic conduction of heat (by electrons) along magnetic field lines, based on the \citet{ManiModel} formulation of anisotropic relativistic conduction (\S\ref{sec:cond}).   The electron thermal diffusivity is a free parameter in this calculation.   For the black hole accretion disc applications of interest, we advocate a `saturated' heat flux in which the thermal diffusivity is $\sim r c$, subject to additional constraints imposed by velocity space instabilities and scattering by wave-particle interactions (Appendix~\ref{app:whist}).

We implemented our electron energy model in a conservative GRMHD code HARM2D \citep{Gammie2003}, though the model we have developed can be applied to any underlying GRMHD scheme.   For simplicity, the implementation in this paper neglects the back reaction of the electron pressure on the dynamics of the accretion flow.  We believe that this is a reasonable first approximation given some of the uncertainties in the electron physics.  Formally, this approximation is is valid only when $T_e \ll T_p$ though we expect it to be a reasonable first approximation when $T_e \lesssim T_p$ in regions with plasma $\beta \gtrsim 1$, i.e., in the regions where gas thermal pressure forces are dynamically important.

We have demonstrated that our implementations of electron heating and conduction are accurate and second order convergent in several smooth test problems (\S\ref{sec:test} and Appendix~\ref{app:condtests}).   For shocks, the heating converges at first order but to a post shock temperature that differs from the analytic solution by $\lesssim 3\%$ when the electron adiabatic index differs from the adiabatic index of the fluid in the GRMHD solution (e.g., Figure~\ref{fig:shock}).  This discrepancy arises because standard Riemann solvers `resolve' the shock structure with only a few grid points.  Including an explicit bulk viscosity to broaden and resolve the shock leads to a converged numerical solution for the post-shock electron energy that agrees with the analytic solution (Appendix~\ref{App:viscshock}).    In practice, the $\lesssim 3\%$ discrepancy between the numerical and analytic solutions for standard Riemann solvers is sufficiently accurate given other uncertainties in the electron physics. For this reason, we do not use bulk viscosity in our calculations.   Moreover, strong shocks are rare and account for a negligible fraction of the dissipation in accretion disc simulations with aligned black hole and accretion disc angular momentum.

In addition to formulating and testing our electron energy equation model, we have also presented a preliminary application of these new methods to simulations of black hole accretion.  Specifically, we have studied the impact of realistic electron heating and electron thermal conduction on the spatial distribution of the electron temperature in 2D (axisymmetric) simulations of black hole accretion onto a rotating black hole.   We find that the resulting electron temperatures differ significantly from the assumption of a constant electron to proton temperature ratio used in previous work to predict the emission from GRMHD simulations \citep{Mosci2009,Dibi2012,Drappeau2013}; see, e.g. Figures~\ref{fig:Tratbeta}-\ref{fig:logTcon}. This is due to the strong $\beta$-dependence of the electron heating fraction, $f_e$, described in \S\ref{sec:tauchi}: electrons are preferentially heated in regions of lower $\beta$, causing $T_e/T_p$ to be larger in the coronal regions compared to the midplane.   In addition, we find that the effect of thermal conduction on the electron temperatures is suppressed by the fact that the heat flux must travel along field lines, which are predominantly toroidal and thus not aligned with the temperature gradient.    Specifically, we find that electron conduction modifies the temperature distribution only if the effective electron mean free path along the magnetic field is $\gtrsim$ the local radius in the flow (see Figure~\ref{fig:TcoTdis1}).   In this case, there is a net transfer of heat from the corona to the bulk of the disc. This increases the electron temperature at larger radii by a factor of $\sim 2$.

It is important to stress that the unsustainability of MHD turbulence in 2D simulations (e.g., \citealt{Guan2008}) limits how thoroughly we can interpret the accretion disc results presented in this paper. Since a steady state is never truly reached, the bulk of the disc retains memory of the initial conditions and only the innermost regions ($r \lesssim 10 r_g$) develop significant turbulence. This could artificially limit the effects of electron conduction because the thermal time for relativistic electrons is $\sim r/c$ and is thus substantially shorter than the local dynamical time only at large radii.    Future work will use the methods developed here in 3D simulations.

It is also important to stress that, as in previous work, our results for both the gas and electron temperature are not reliable when $b^2 \gg \rho c^2$. In these regions the ratios of $b^2/\rho c^2$ and $b^2/u_g$ are so large that the evolution of the density and internal energy are dominated by truncation errors in the magnetic field, to which they are nonlinearly coupled by the total energy
equation.  This requires the use of density and internal energy floors.  Because our calculation of the electron heating rate relies on quantifying the entropy changes in the underlying GRMHD solution, our predicted electron temperatures also become unreliable when $b^2 \gg \rho c^2$.   In the accretion disc simulations, this only affects the regions close to the pole where there is very little matter, not the evolution of the electrons in the bulk of the accretion disc or corona.    We have specifically tested several treatments of the internal energy and density floors which produce dramatically different results in the poles but are all consistent in the higher density regions for both the fluid variables and the electron temperature.

Future applications of the methods developed in this paper will center on using our electron temperature calculations to predict the emission from accreting black holes.  In particular, we hope to produce more accurate images of the radio and IR emission of Sagittarius A* (and M87) that can be used to interpret the forthcoming spatially resolved observations by the Event Horizon Telescope \citep{Doeleman2009} and Gravity \citep{Gillessen2010}.

\section*{Acknowledgements}
We thank F. Foucart for useful discussions, as well as all the members of the horizon collaboration, horizon.astro.illinois.edu, for their advice and encouragement. We also thank Dmitri Uzdensky for a useful and thorough referee report.   This work was supported by NSF grant AST 13-33612 and NASA grant NNX10AD03G, and a Romano Professorial Scholar appointment to CFG.   EQ is supported in part by a Simons Investigator Award from the Simons Foundation and the David and Lucile Packard Foundation.   MC is supported by the Illinois Distinguished Fellowship from the University of Illinois.   Support for AT was provided by NASA through Einstein Postdoctoral Fellowship grant number PF3-140131 awarded by the Chandra X-ray Center, which is operated by the Smithsonian Astrophysical Observatory for NASA under contract NAS8-03060, and by NSF through an XSEDE computational time allocation TG-AST100040 on TACC Stampede. This work was made possible by computing time granted by UCB on the Savio cluster.

\bibliographystyle{mn2efix}
\bibliography{Kappa}
\appendix

\section{Tests of Electron Conduction}
\label{app:condtests}

This Appendix outlines tests of our numerical implementation of electron conduction that demonstrate that our calculations are robust and second-order accurate.  The tests are taken directly from \citet{grim}, to which we refer the reader for more details.

\subsection{Conduction Along Field Lines}
\label{app:statcond}

This test is simply to check whether the electrons conduct heat along field lines properly.  The initial conditions are a 2D, periodic box of physical size $1\times1$ with uniform pressure and a small, density variation (and hence temperature variation) of the form:
\begin{equation}
\rho = \rho_0 \left(1- e^{- \frac{(x-0.5)^2+(y-0.5)^2}{0.005}   }\right).
\end{equation}
The field lines are sinusoidal and given by
\begin{equation}
\begin{aligned}
&B_x = B_0  \\
&B_y = B_0 \sin(8\pi x),
\end{aligned}
\end{equation}  
derived from a scalar potential of
\begin{equation}
A_z = B_0\left(y + \frac{1}{8\pi}\cos(8\pi x)\right).
\end{equation}
For the conduction parameters, we choose $\chi_e = 0.5/\rho$ and $\tau = 1$ and run the simulation for $10 \tau$.

\begin{figure}
\includegraphics[scale = .1]{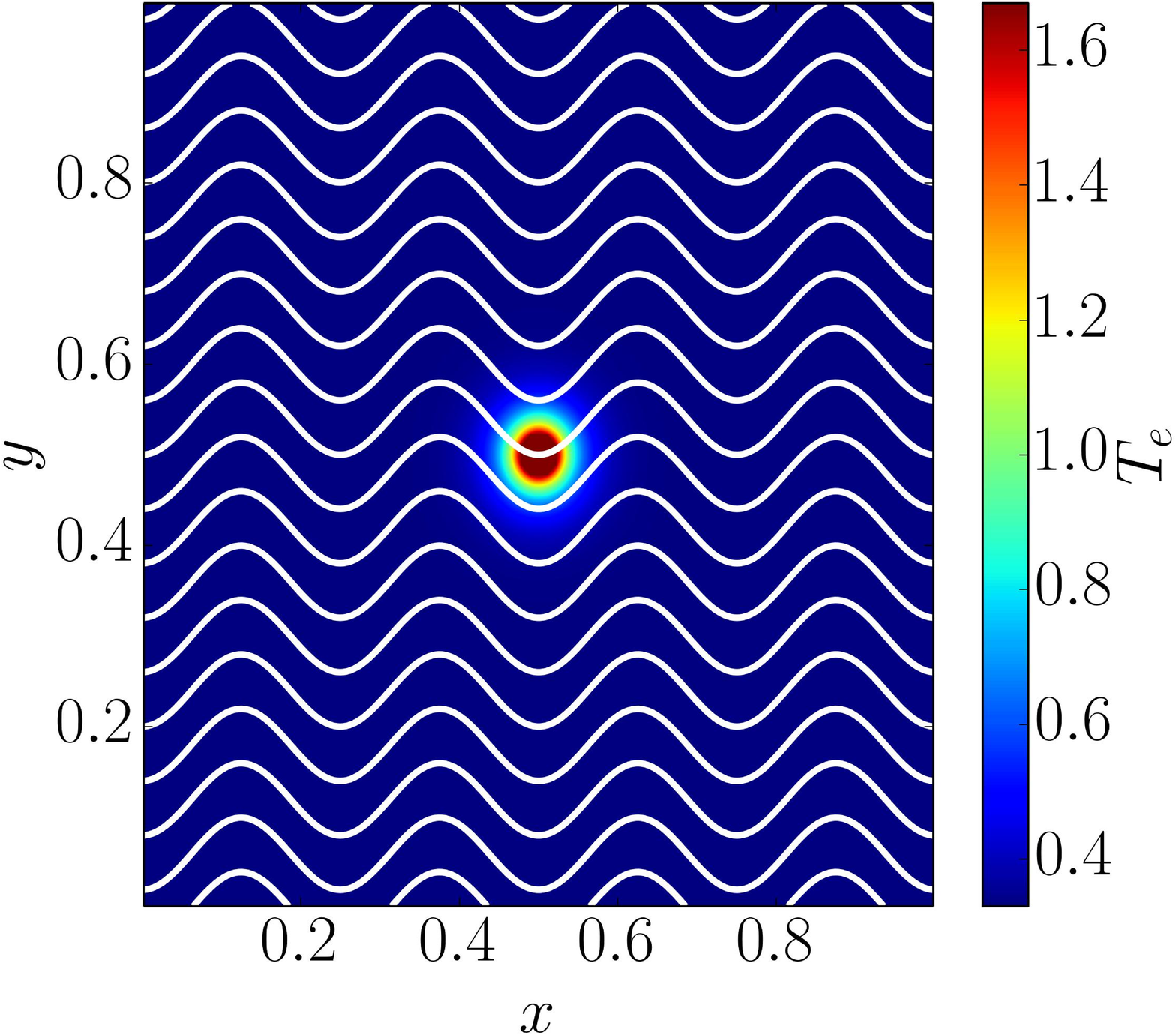}
\includegraphics[scale = .1]{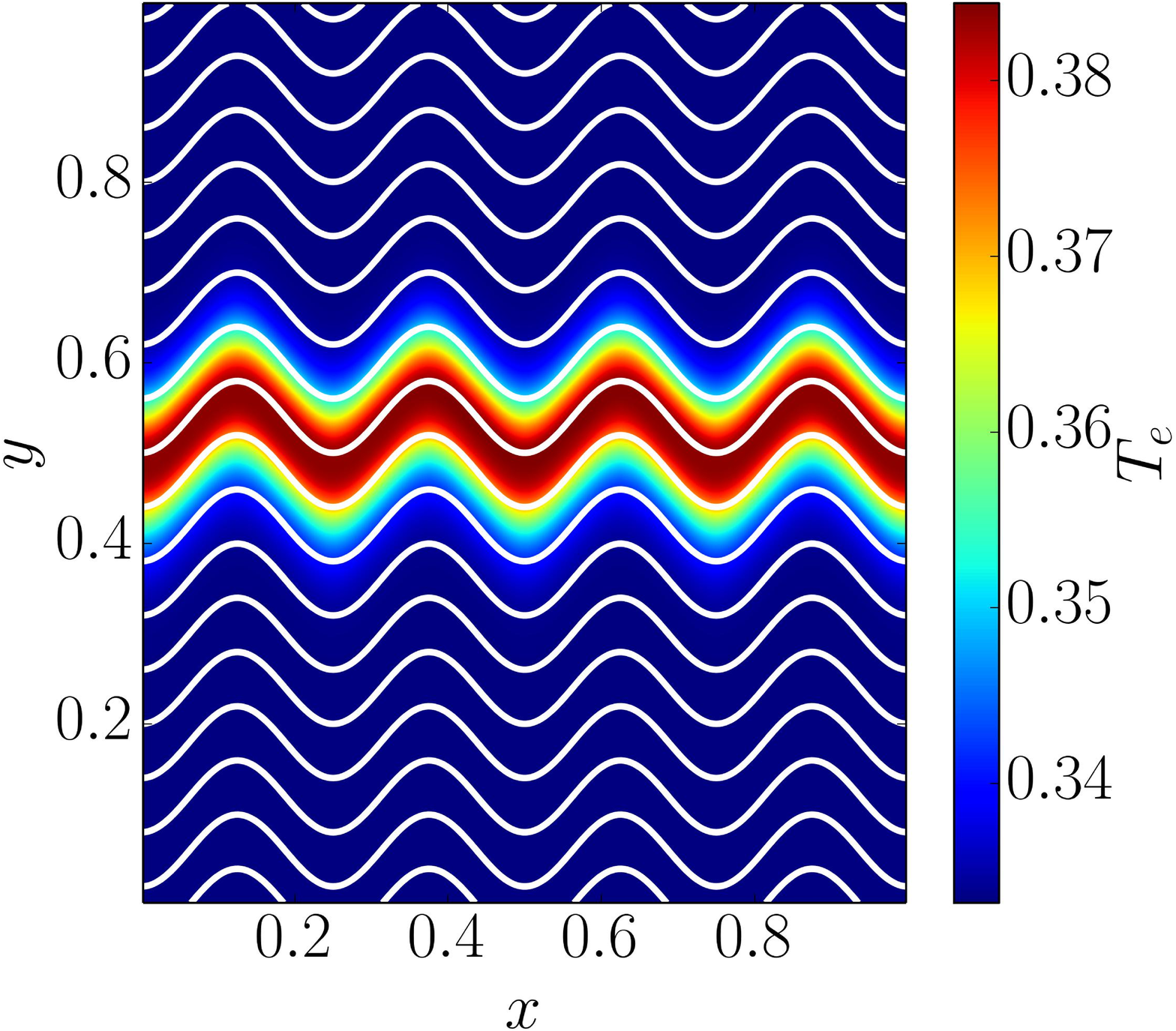}
\caption{Temperature profiles over-plotted with magnetic field lines in the 2D anisotropic conduction test from \citet{grim}, adapted for electron conduction (see \S\ref{app:statcond}). The top panel is at the initial time while the bottom panel is at the end of the run ($t = 10\tau$).  The field lines become isothermal, consistent with heat conduction only along the magnetic field.  }
\label{fig:Tstat}
\end{figure}

Figure~\ref{fig:Tstat} shows that the final state of the fluid is that of isothermal field lines, exactly as expected, with heat flux equilibrating the temperature along the magnetic field lines.  This shows that our implementation of conduction properly limits the heat flux to be parallel to the magnetic field.

\subsection{Linear Modes Test}
\label{sec:modes}
\begin{figure}
\includegraphics[scale = .4]{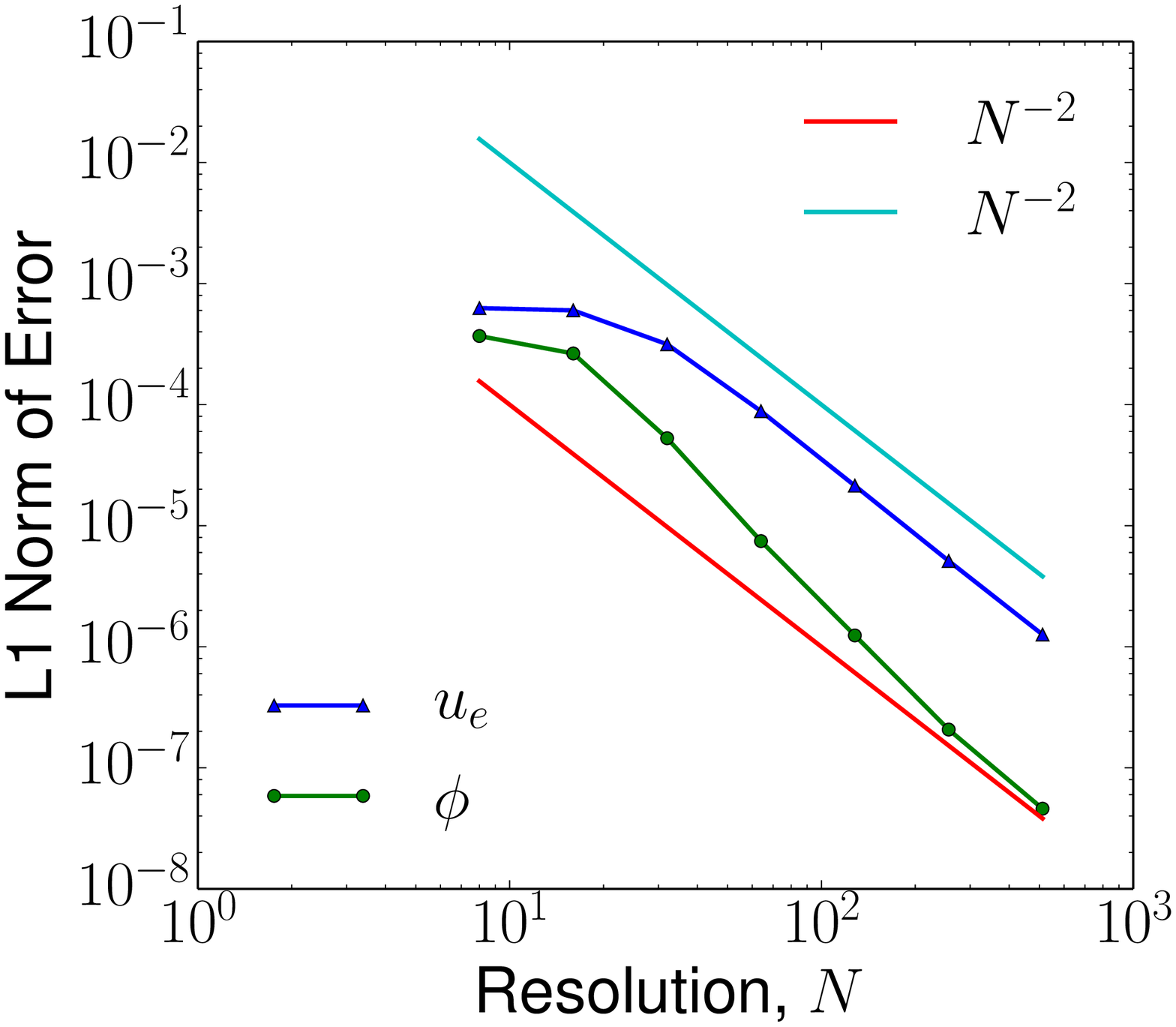}
\caption{L1 Norm of errors in the 2D linear modes test after one period as computed from the eigenfrequencies given in equation~\eqref{eqn:anstab}. See \S\ref{sec:modes}.}
\label{fig:Linmodescon2D}
\end{figure}

This test checks whether our implementation of conduction gives the correct eigenmodes corresponding to Equation~\eqref{eqn:anstab}. Writing $\lambda = -\alpha \pm i \omega$, we initialise perturbations in an otherwise uniform box about the equilibrium solution with wave number $k = 2\sqrt{2} \pi$ and run the simulation for one period:  $t = 2\pi/\omega$. The analytic solution is that the perturbations, $\delta$, should obey $\delta(t =2\pi/\omega) = \delta(t= 0) e^{-2\pi \alpha/\omega} $. We choose $\hat b = 1/\sqrt{3} \hat x + \sqrt{2}/\sqrt{3} \hat y $ and $\vec{k}= 2\pi \hat x + 2\pi \hat y$. We find that both $\phi$ and $u_{e}$ converge at second order to the analytical solution as shown in Figure~\ref{fig:Linmodescon2D}.

\subsection{1D Atmosphere in a Schwarzschild Metric}
\label{app:atm}

\begin{figure}
\includegraphics[scale = .4]{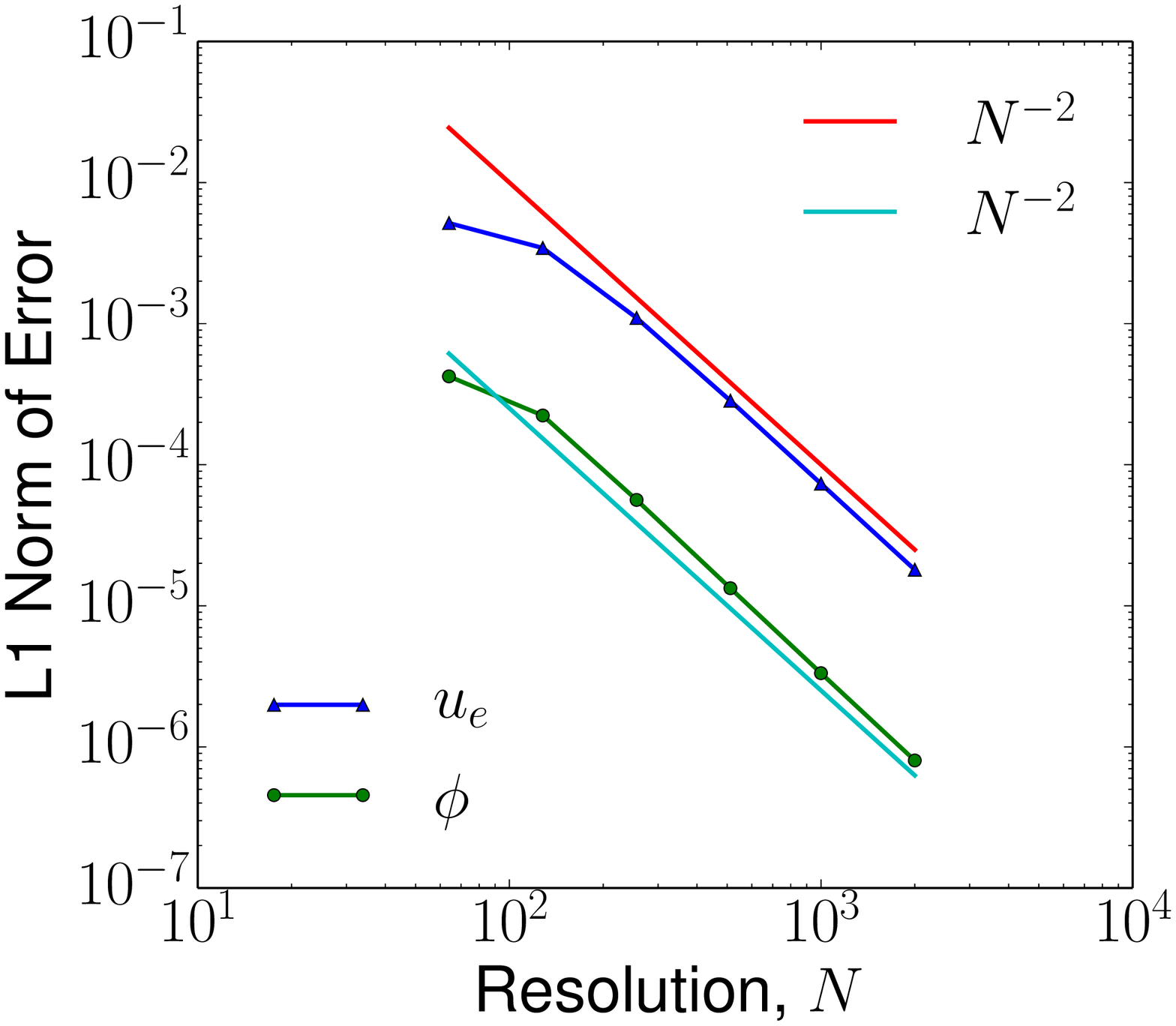}
\caption{L1 norms of the error in both the heat flux and electron internal energy for the 1D atmosphere test in the Schwarzschild metric (\S\ref{app:atm}).  }
\label{fig:atmcon}
\end{figure}

This test checks whether our implementation of the electron conduction gives the correct analytic result in a non-trivial space-time.  In the Schwarzschild metric, the solution for a fluid in hydro-static equilibrium reduces to a system of two ordinary differential equations, which can be solved for any given temperature profile (see \citealt{grim} for details).  For this test, we initialise the temperature and heat flux of the electrons to be this equilibrium solution for a purely radial field and see if the code can maintain it over a time of $100$ $r_g/c$ in a computational domain of $1.4 r_g \le r \le 90 r_g$.   To compute the error, we again use the L1 norm and find 2nd order convergence for both $\phi$ and $u_{e}$, as shown in Figure~\ref{fig:atmcon}. 

\subsection{Relativistic Bondi Accretion}
\label{app:bondi}

\begin{figure}
\includegraphics[scale = .4]{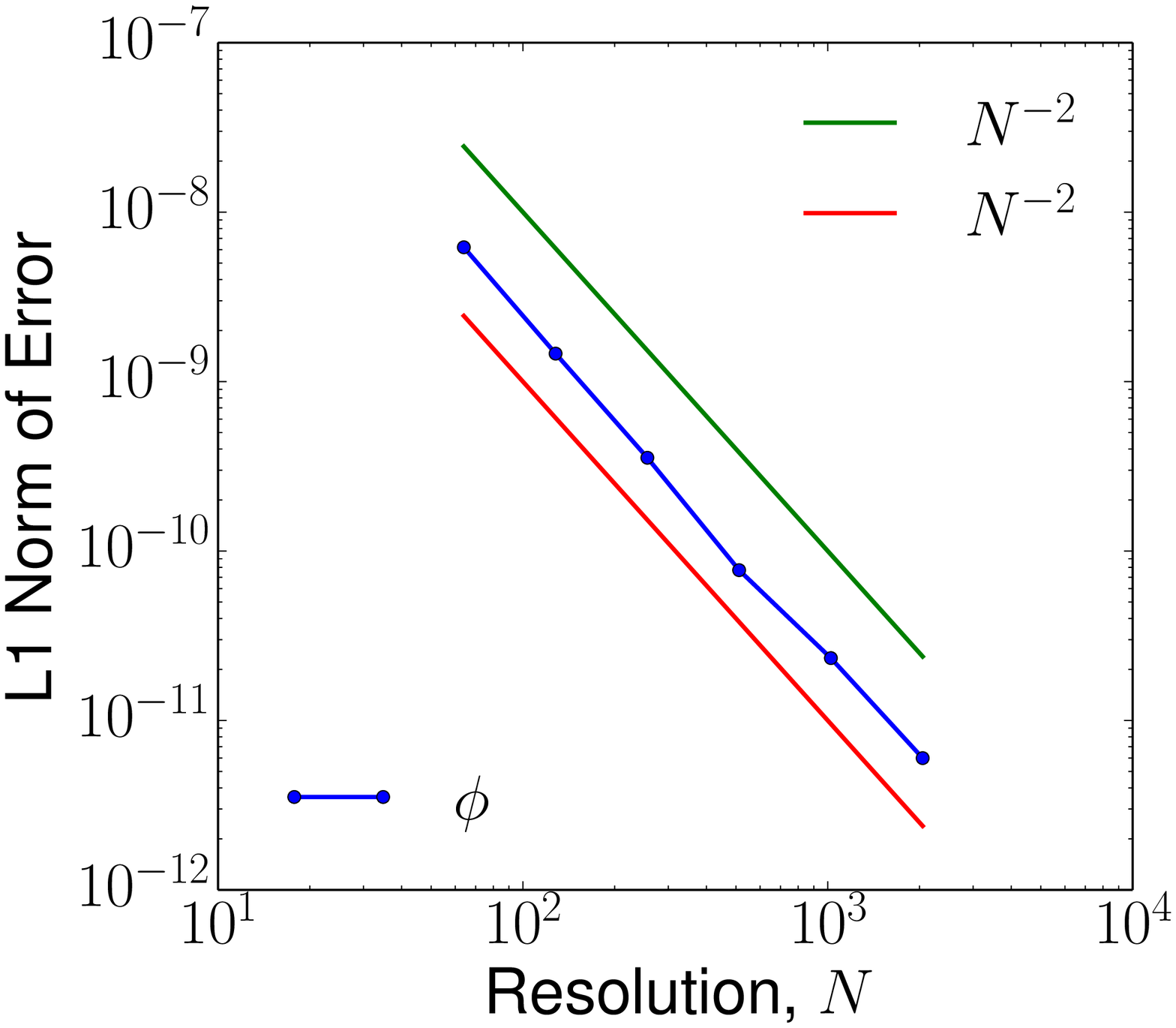}
\caption{L1 norms of the error in the magnitude of the heat flux for the relativistic Bondi accretion test (\S\ref{app:bondi}).  }
\label{fig:bondicon}
\end{figure}

This test checks whether our implementation of the electron conduction gives the correct analytic result in a fluid with $u^i \ne 0$, which activates terms that were not present in the 1D atmosphere test.  For the standard, spherically symmetric, steady-state Bondi solution for an accreting black hole \citep{Hawley1984}, we can solve equation~\eqref{eq:phiev} by numerical integration if we assume that the heat flux does not back-react on the electron temperature.  For this test, we set the initial condition of the fluid variables to be the Bondi solution and the initial conditions of $\phi$ to be given by the solution to equation~\eqref{eq:phiev} with Dirichlet boundary conditions.  We choose the sonic point to occur at $r_c = 20 M$ and fix the outer boundary at a spherical radius of $R_{\rm out} = 40 M$ to have $\phi(r=40 M) = 0$. The inner radius of the grid is inside the event horizon at $r = 1.6 M$.   The test is whether or not the code can maintain this state over a period of $t = 200 M$. We find second order convergence of the heat flux to the analytical solution, as shown in Figure~\ref{fig:bondicon}.

\section{Derivations}
\subsection{Total Heating Rate}

This section derives the result quoted in equation~\eqref{eq:qdottot}.

First, we introduce the variable $\hat \kappa_{g}$, which is equivalent to $\kappa_{g} \equiv P_g\rho^{-\gamma}$ at the beginning of the time step and at the $n+1/2$ ``predictor'' step, but which is evolved over a time step according to:
\begin{equation}
\p_\mu(\gdet\rho \hat \kappa_{g} u^\mu) =0.
\label{eq:seom}
\end{equation}
We discretise equation \eqref{eq:seom} in a standard way (i.e. equation~\ref{eq:consnum}):
\begin{equation}
\begin{aligned}
& \frac{\left(\gdet\rho \hat \kappa_{g} u^t\right)^{n+1}-\left(\gdet\rho \kappa_{g} u^t\right)^{n}}{\Delta t} \\ &   + \frac{[\gdet\rho \kappa_{g} u^x]^{n+1/2}_{i+1}-[\gdet\rho \kappa_{g} u^x]^{n+1/2}_{i}}{\Delta x}=0,
\end{aligned}
\label{eq:eomsn}
\end{equation}
where the square brackets indicate fluxes computed via the Riemann solver at cell interfaces and the generalisations to higher dimensions is straightforward. Note that we have dropped the $\hat{\phantom{\kappa}}$ in the $n+1/2$ and $n$ terms because $\hat\kappa_g = \kappa_{g}$ at the beginning of the time step and at the $n+1/2$ step. We obtain the new value of entropy, $\hat \kappa_{g}^{n+1}$, at $t_{n+1}\equiv t_n+\Delta t$ via solving equation~\eqref{eq:eomsn}.  We emphasise that $\hat \kappa_{g}$ is not the true entropy at $t^{n+1}$ but the entropy evolved according to equation~\eqref{eq:seom} [or its discretised equivalent equation~\ref{eq:eomsn}] and thus does \emph{not} include any heating.

At the end of the time step (i.e. at $t = t_{n+1}$), we compute the ``true'' value of the entropy due to
the full GRMHD evolution, according to the definition of $\kappa_{g}$:
\begin{equation}
\kappa_{g}^{n+1} =\left( \frac{P_g}{\rho^\gamma}\right)^{n+1}.
\label{eq:snp1}
\end{equation}
Unlike $\hat \kappa_{g}^{n+1}$, which does not include any heating, $\kappa_{g}^{n+1}$ accounts for the heating as implied by the conservative evolution of the underlying GRMHD scheme. The difference $(\kappa_{g}-\hat \kappa_{g})^{n+1}$
is related to the heating incurred during time step $n$, and we will use it below.

To compute the heating rate we evaluate the quantity:
\begin{equation}
  \label{eq:qdot}
\begin{aligned}
Q &\equiv \rho T u^\mu \p_\mu s_g  = \frac{\rho^{\gamma}}{\gamma-1} u^\mu \p_\mu \kappa_g
\\ &\equiv \frac{\rho^{\gamma-1}}{\gamma-1} (\rho \kappa_g u^\mu)_{;\mu}  \\ &\equiv \frac{1}{\gamma-1}\frac{\rho^{\gamma-1}}{\gdet}\partial_\mu \left(\gdet\rho  \kappa_g u^\mu\right),
\end{aligned}
\end{equation}
where the third equality holds because \[ \begin{aligned}
(\rho \kappa_g u^\mu)_{;\mu} &= \partial_\mu (\gdet\rho  \kappa_g
u^\mu)/\gdet \\ &\equiv \kappa_g \partial_\mu (\gdet\rho  u^\mu)/\gdet + \rho
u^\mu \partial_\mu \kappa_g
\end{aligned}
 \]  and the first term vanishes due to conservation of mass.
We evaluate eq.~\eqref{eq:qdot} at the $n+1/2$ time step in a discretised form by centring the time derivatives at $n+1/2$ but evaluating the prefactor at the $n+1$ time step:
\begin{equation}
  \label{eq:qdotn}
\begin{aligned}
  Q^{n+1/2} &= \left(\frac{1}{\gamma-1}  \frac{\rho^{\gamma-1}}{{\gdet}}\right)^{n+1/2} \\ &\times \left\{\frac{(\gdet\rho \kappa_g u^t)^{n+1}-(\gdet\rho \kappa_g u^t)^{n}}{\Delta t}\right. \\&  + \left. \frac{[\gdet\rho \kappa_g u^x]^{n+1/2}_{i+1}-[\gdet\rho \kappa_g u^x]^{n+1/2}_{i}}{\Delta x}\right\}.
\end{aligned}
\end{equation}
Now, multiplying eq.~\eqref{eq:eomsn} by $\left(\frac{1}{\gamma-1}\rho^{\gamma-1}/\gdet\right)^{n+1/2}$ and adding the result  to eq.~\eqref{eq:qdotn}, we obtain equation~\eqref{eq:qdottot} of the main text:
\begin{equation}
  \label{eq:qdotn1}
  Q^{n+1/2} = \left(\frac{\rho^{\gamma-1}}{\gamma-1}\right)^{n+1/2}\frac{\{\rho u^t (\kappa_g-\hat \kappa_g)\}}{\Delta t}^{n+1}.
\end{equation}

\label{app:heat}

\subsection{Whistler Instability Limit on Conduction}

We assume that the electrons are relativistic with $\Theta_e = k T_e/m_e c^2 \gtrsim 1$.    If the electrons relax to thermal equilibrium with a scattering rate $\nu_e$, relativistic kinetic theory implies that the electron viscosity $\eta_e$ and thermal  diffusivity $\chi_e$ satisfy \citep{anderson1971}
\begin{equation}
\eta_e \simeq \Theta_e \frac{c^2}{\nu_e} \hspace{0.5in} \chi_e \simeq 1.6 \frac{c^2}{\nu_e}
\label{eq:e-diffusion}
\end{equation}
Velocity space instabilities set an upper limit on the electron thermal conductivity in a turbulent plasma.   Physically, as the magnetic field in the accretion disc fluctuates in time, this generates pressure anisotropy, which is resisted by velocity space instabilities that isotropise the distribution function and thus limit the magnitude of the thermal diffusivity.   \citet{ManiModel} show that the theory of relativistic anisotropic viscosity implies that the pressure anisotropy and scattering rate are related by
\begin{equation}
\nu_e \, \frac{\Delta P_e}{P_e} \simeq \ u^\mu \p_\mu \ln\left[\frac{B^3}{\rho^2}\right].
\label{eq:dPad-rel}
\end{equation}
where $\Delta P_e = P_\perp - P_\parallel$ and we have neglected some general relativistic terms for simplicity.

Electrons satisfy limits on pressure anisotropy of
\begin{equation}
\frac{\Delta P_e}{P_e} \gtrsim -\frac{1.3}{\beta_e} \hspace{0.45in}  \frac{\Delta P_e}{P_e} \lesssim \frac{0.25}{\beta_e^{0.8}}
\label{eq:dPe-limits}
\end{equation}
The second term on the right hand side of equation \eqref{eq:dPe-limits} is a fit to the whistler instability threshold for relativistically hot electrons (based on numerical solutions of the dispersion relation derived in \citealt{Gladd1983}).   The coefficient in the numerator technically depends weakly on $\Theta_e$, varying from $\simeq 0.125$ for non-relativistic electrons to $\simeq 0.25$ for $\Theta_e \simeq 10$ \citep{Lynn2014}.   Note that the slope of the $\beta_e$ term for the whistler instability in equation \eqref{eq:dPe-limits} is a fit for $\beta_e \simeq 0.1-30$.  \citet{Gary1996} and \citet{Sharma2007} found a somewhat shallower slope $\propto \beta_e^{-1/2}$ in non-relativistic calculations but this is not a good fit over the large dynamic range of $\beta_e$ considered here.
The first limit in equation \eqref{eq:dPe-limits} is the electron firehose instability which is an electron-scale resonant analogue of the fluid firehose instability \citep{Gary2003}.   This limit is based on non-relativistic calculations and should to be extended to the relativistic limit in future work.  However, based on our whistler calculations this is unlikely to be a significant effect.

\citet{Sharma2007} found that the typical pressure anisotropy satisfied $\Delta P/P \ge 0$ in simulations that explicitly evolved a pressure tensor. Physically, this sign of the pressure anisotropy corresponds to outward angular momentum transport.   Assuming that the RHS of equation \eqref{eq:dPad-rel} is $\sim \Omega$, the whistler instability limit in equation \eqref{eq:dPe-limits} thus implies $\chi_e \sim c r_g (r/r_g)^{3/2} (4 \beta_e)^{-0.8}$.   This is not a significant constraint on the conductivity relative to the saturated value ($\chi_e\sim c r_g$), for $\beta_e \lesssim 1$, which can occur either in the corona/outflow or because $T_e \ll T_p$.  However,  this estimate does suggest that the electron conductivity may be modest in the bulk of the disc at $\sim 10 r_g$ if $\beta_e \gg 1$.

Equation \eqref{eq:dPe-limits} can be implemented by calculating $\Delta P_e/P_e$ using equation \eqref{eq:dPad-rel} given an assumed $\chi_e$ (and using equation \eqref{eq:e-diffusion} to relate $\nu_e$ and $\chi_e$).   If equation \eqref{eq:dPe-limits} is violated, $\nu_e$ should be increased and $\chi_e$ decreased such that  equation \eqref{eq:dPe-limits} is satisfied.   Alternatively, an even simpler first approximation would be to simply limit 
\begin{equation}
\chi_e \lesssim c r_g (r/r_g)^{3/2} (4 \beta_e)^{-0.8}\equiv \chi_{\rm max}
\label{eq:chimax}
\end{equation} motivated by the estimate in the preceding paragraph for the whistler instability. This is the limit we have used in the accretion disc simulations in \S\ref{sec:app} of the main text.

\label{app:whist}

\subsection{Electron Conduction Numerical Stability}
Non-relativistically, an explicit implementation of thermal conduction is stable only if the time step, $\Delta t$, satisfies the condition $\Delta t \lesssim \Delta x^2 /\chi$, where $
\Delta x$ is the grid spacing in 1-dimension and $\chi$ is the thermal diffusivity.  The relativistic theory outlined in $\S\ref{sec:cond}$, however, where the heat flux $\phi$ is evolved according to equation~\eqref{eq:phiev}, differs from the non-relativistic case in that it is no longer diffusive.  This alters the criterion for stability to be a condition on the relaxation time, $\tau$, given by equation~\eqref{eq:taulim}, which we derive here.

To check the numerical stability of our conduction theory we assume that we are in Minkowski space in the rest frame of the fluid, and further simplify our analysis to one dimension in which $\hat b = \hat i$.

Under these assumptions, a Von Neumann stability analysis on equations~\eqref{eq:phiev} and~\eqref{eq:electron} leads to a quadratic equation for the amplification factor, $G$, with the following solutions:

\begin{equation}
\begin{aligned}
G = 1&-\mathcal{C}\left[1-\cos(k)\right] \\ &-\frac{1}{2}\frac{\Delta t}{\tau}\left( 1\pm  \sqrt{1-4(\gamma_e-1)  \frac{\chi_e \tau}{ \Delta x^2} \sin\left(k\right)^{2}}\right),
\end{aligned}
\label{eqn:G}
\end{equation}
with the condition for stability being that $|G|\le1$.  Here, as before, $\mathcal{C}$ denotes the Courant factor. To analyse equation~\eqref{eqn:G}, we consider two cases: 1) when the square root term is real, and 2) when the square root term is imaginary.  

When the square root term is real, the condition for stability becomes:
\begin{equation}
\tau \ge \Delta t \left[ \frac{2-\mathcal{C}\left[1-\cos(k)\right]-(\gamma_e-1)  \frac{\chi_e \Delta t}{ \Delta x^2} \sin\left(k\right)^{2}}{\left(2-\mathcal{C}[1-\cos(k)]\right)^2}\right]
\end{equation}
The right hand side is a maximum for $k=\pi$ modes, which gives, simply:
\begin{equation}
\tau \ge \frac{\Delta t}{2(1-\mathcal{C})}.
\end{equation}

The more interesting case is when the square root term in Equation~\eqref{eqn:G} is imaginary, where the criterion for stability becomes:
\begin{equation}
\tau \ge \Delta t \left[ \frac{(\gamma_e-1)  \frac{\chi_e \Delta t}{ \Delta x^2} \sin\left(k\right)^{2}+\mathcal{C}\left[1-\cos(k)\right]-1}{\mathcal{C}\left(2-\mathcal{C}[1-\cos(k)]\right)\left[1-\cos(k)\right]}\right],
\end{equation}
which, defining $K\equiv (\gamma_e-1)\Delta t \chi_e/\Delta x^2$, has a maximum at 
\begin{equation}
\cos(k) = 1 -\frac{\mathcal{C}- \sqrt{4K(1-\mathcal{C})-\mathcal{C}^2}}{a^2-2K(1-\mathcal{C})}
\end{equation}
if 
\begin{equation}
\begin{aligned}
\Delta t &> \frac{\Delta x^2}{(\gamma_e-1)\chi_e} \frac{1-4\mathcal{C}(1-\mathcal{C}) + \sqrt{1-4\mathcal{C}(2\mathcal{C}-1)}}{8(1-\mathcal{C})} \\ &\equiv \Delta t_{crit},
\end{aligned}
\label{eqn:dtcon}
\end{equation} and a maximum at $k=\pi$ otherwise. So if $\Delta t < \Delta t_{crit}$, our criterion becomes:
\begin{equation}
\tau \ge \Delta t \left[ \frac{2\mathcal{C}-1}{4\mathcal{C}\left(1-\mathcal{C}\right)}\right] \equiv \tau_{max,1}.
\label{eq:taumax1}
\end{equation}
Finally, if $\Delta t> \Delta t_{crit}$, then we have
\begin{equation}
\begin{aligned}
\tau &\ge  \\ & \Delta t \times\left[ \frac{2K\left(\mathcal{C}^2-4C(1-\mathcal{C})\right)}{{4K\mathcal{C}(1-\mathcal{C})\left(\sqrt{4K(1-\mathcal{C})-\mathcal{C}^2}-2\mathcal{C}\right)+2\mathcal{C}^4}} \right.\\ &\left.+\frac{\left(4K^2(1-\mathcal{C})+4K\mathcal{C}(1-\mathcal{C})-\mathcal{C}^3\right)\sqrt{4K(1-\mathcal{C})-\mathcal{C}^2}}{4K\mathcal{C}(1-\mathcal{C})\left(\sqrt{4K(1-\mathcal{C})-\mathcal{C}^2}-2\mathcal{C}\right)+2\mathcal{C}^4}\right] \\ &\equiv \tau_{max,2}.
\end{aligned}
\label{eqn:taustabi}
\end{equation}
The general behaviour of Equation~\eqref{eqn:taustabi} is complicated, but the result is roughly consistent with 
\begin{equation}
\tau \gtrsim(\gamma_e-1)\left(\frac{\Delta t}{\Delta x}\right)^2\chi_e.
\end{equation}
for most reasonable choices of the Courant factor.  This is the result quoted in equation~\eqref{eq:taulim} of the main text.

To summarise, our scheme is stable when:
\begin{displaymath}
\tau \ge \left\{\begin{array}{ll}
    \max\biggl[\displaystyle\frac{\Delta t}{2(1-\mathcal{C})},  \tau_{max,1} \biggr]&: \Delta t<\Delta t_{crit}\\
\max\biggl[\displaystyle\frac{\Delta t}{2(1-\mathcal{C})}, \tau_{max,2}\biggr] &: \Delta t\ge\Delta t_{crit},
 \end{array}
\right.
\end{displaymath}
for $\Delta t_{crit}$, $\tau_{max,1}$, and $\tau_{max,2}$ as defined in equations~\eqref{eqn:dtcon}, \eqref{eq:taumax1}, and \eqref{eqn:taustabi}, respectively. 

\label{app:stab}

\subsection{Electron Heating in a 1D Shock }

Formally, for an ideal shock in a zero-viscosity fluid there is no unique path in $(P,\rho)$ space that connects the pre and post-shock values given by the Rankine-Hugoniot conditions, meaning that the dissipation per unit volume, $\int \rho T ds$, is not a well-defined quantity.  However, by introducing \emph{any} non-zero viscosity, the degeneracy is broken and there exists a unique path in $(P,\rho)$ space and hence a well-defined dissipation.  To see this, we take the 1D Rankine-Hugoniot relations for a static shock, given some prescription for the viscous stress, $\tau \equiv 4/3 \mu \vec{\nabla} \cdot v$ ($\mu$ is the dynamic viscosity coefficient, and can be an arbitrary function of plasma parameters), 
\begin{equation}
\begin{aligned}
\dot m  &= \rho v \\
\dot p &= \rho v^2 + P  + \tau   \\
\dot E &= \frac{1}{2}\rho v^3 + \frac{\gamma}{\gamma-1}P v + \tau v  ,
\end{aligned}
\end{equation}
where $\dot m$, $\dot p$, and $\dot E$ are constants representing the mass, momentum, and energy flux across the shock. Absent $\tau$, we could combine these three equations in several different ways to get a relationship of the form $P = P(\rho)$.  With non-zero viscosity, however, there is only one unique way to do this, namely, by taking $\dot p v - \dot E$ and solving for $\dot m$, which gives:
\begin{equation}
P(\rho) = \left(\gamma-1\right)\left(\frac{1}{2}\frac{\dot m^2}{\rho} -\dot p + \frac{\dot E}{\dot m}\rho\right),
\label{eq:Psubrho}
\end{equation}
or, in terms of $\kappa \equiv P \rho^{-\gamma}$,
\begin{equation}
\kappa_{g}(\rho) = \left(\gamma-1\right)\left(\frac{1}{2}\frac{\dot m^2}{\rho^{\gamma+1}} -\frac{\dot p}{\rho^\gamma}+ \frac{\dot E}{\dot m \rho^{\gamma-1}} \right).
\end{equation}
We assume that the electrons receive a constant fraction of the total heat:
\begin{equation}
\begin{aligned}
&\rho T_e u^\mu \partial_\mu s_e = f_e \rho T_{g} u^\mu \p_\mu s_{g} \\
&\Rightarrow \frac{\rho^{\gamma_e}}{\gamma_e-1} u^\mu \partial_\mu \kappa_e = f_e \frac{\rho^{\gamma}}{\gamma-1} u^\mu \p_\mu \kappa_{g} ,
\end{aligned}
\end{equation}
or in quasi-conservative form (using the mass continuity equation and assuming a flat space metric):
\begin{equation}
\frac{\p}{\p x^\mu} \left(\rho u^\mu \kappa_e \right)  = f_e \frac{\gamma_e-1}{\gamma-1}\rho^{\gamma-\gamma_e}\frac{\p}{\p x^\mu} \left(\rho u^\mu \kappa_{g} \right).
\end{equation} The final electron entropy is given by integrating this equation from the initial to the final density, which, for a 1D shock reduces to
\begin{equation}
\int\limits^{\infty}_{-\infty} \frac{\p}{\p x} \left(\dot m \kappa_e \right) = f_e \frac{\gamma_e-1}{\gamma-1}\int \limits ^{\rho_f}_{\rho_i}\rho^{\gamma-\gamma_e} \dot m \frac{\p \kappa}{\p \rho} d \rho,
\end{equation}
giving:
\begin{equation}
\begin{aligned}
u_{e}^f &=  u_{e}^i\left(\frac{\rho_f}{\rho_i}\right)^{\gamma_e} \\
&+ \frac{f_e}{\gamma-1}\left( \frac{\dot m }{2\rho_f}\frac{\gamma+1}{\gamma_e+1} - \dot p\frac{\gamma}{\gamma_e} + \frac{\dot E \rho_f}{\dot m } \frac{\gamma-1}{\gamma_e-1}\right)\\
&- \frac{f_e}{\gamma-1} \left(\frac{\rho_f}{\rho_i}\right)^{\gamma_e}\left(\frac{\dot m }{2\rho_i}\frac{\gamma+1}{\gamma_e+1}- \dot p\frac{\gamma}{\gamma_e} + \frac{\dot E \rho_i}{\dot m } \frac{\gamma-1}{\gamma_e-1}\right),
\end{aligned}
\label{eq:ugesubrho}
\end{equation}
where $\rho_f$ is determined from the Rankine-Hugoniot conditions. For a strong shock with Mach number $\gg 1$, this simplifies to 
\begin{equation}
u_{e}^f = \dot m v_i \frac{f_e}{\gamma_e^2-1} \left[ \left(\frac{\gamma + 1}{\gamma - 1}\right)^{\gamma_{e}}\left(1-\frac{\gamma}{\gamma_e}\right)+ 1 + \frac{\gamma}{\gamma_e}  \right],
\label{eq:Mache}
\end{equation}
where $v_i$ is the pre-shock fluid velocity in the shock's rest frame. Dividing by $u_{g}^f = 2 \dot m v_i \left(\gamma^2-1\right)^{-1}$ yields equation~\ref{eq:shockrat}.

\label{App:shock}
\section{Electron Heating in a Viscous Shock}

In this appendix we show that by introducing an explicit bulk viscosity to the non-relativistic hydrodynamic equations, our electron heating calculation outlined in \S\ref{sec:eheatnum} give an electron internal energy that converges to the analytic result derived in Appendix~\ref{App:shock} for electron heating at a shock.

We treat viscosity by explicitly adding the 1D viscous energy and momentum fluxes to the ideal MHD fluxes for a constant kinematic viscosity, $\nu$:
\begin{equation}
  \begin{aligned}
    F_{E,visc} = -\frac{4\nu}{3} \rho v \frac{{\rm d}v}{{\rm d}x} \\
    F_{p,visc} = -\frac{4\nu}{3} \rho \frac{{\rm d}v}{{\rm d}x}. \\
  \end{aligned}
  \label{eq:fluxnu}
\end{equation}
Note that these are non-relativistic fluxes which are formally inconsistent with the relativistic code in which they are used.  However, our goal here is simply to show that with a resolved shock structure the electron heating calculation converges to the correct answer.  The non-relativistic limit is fine for this purpose. The fluxes in equation~\eqref{eq:fluxnu} smooth out discontinuities to a continuous profile of finite width, determined by $\nu$ and the velocity scale.  The solution for the profile of a viscous shock, now defined as a smooth transition from an initial to final state as opposed to a discontinuity, can be computed analytically for a constant kinematic viscosity, $\nu$.  In the shock frame, taking $x \rightarrow -\infty$ as the initial state, this solution takes the form:
\begin{equation}
  v(x) = \frac{{\left(\gamma +\displaystyle\frac{2}{M}  -1\right)} +  \exp\left[-\displaystyle\frac{3(x-x_0) v_i}{4\nu}\left(1  - \frac{1}{ M}\right)\right]}{ \exp\left[-\displaystyle\frac{3(x-x_0) v_i}{4\nu}\left(1  - \frac{1}{ M}\right)\right]+ (\gamma +1)},
  \label{eq:vsubx}
\end{equation}
where $M$ is the pre-shock Mach number, $v_i$ is the pre-shock speed at $x\rightarrow -\infty$, and $x_0$ is a constant determining the location of the shock. For a pre-shock density $\rho_i$, the density profile is obtained from the mass conservation equation: $\rho_i v_i / v(x)$, which determines the pressure profile from equation~\eqref{eq:Psubrho}. Similarly, the profile for the internal energy of the electrons in terms of $\rho(x)$ is given by equation~\eqref{eq:ugesubrho} with the substitution $\rho_f \rightarrow \rho(x)$.

\begin{figure}
\includegraphics[scale = .4]{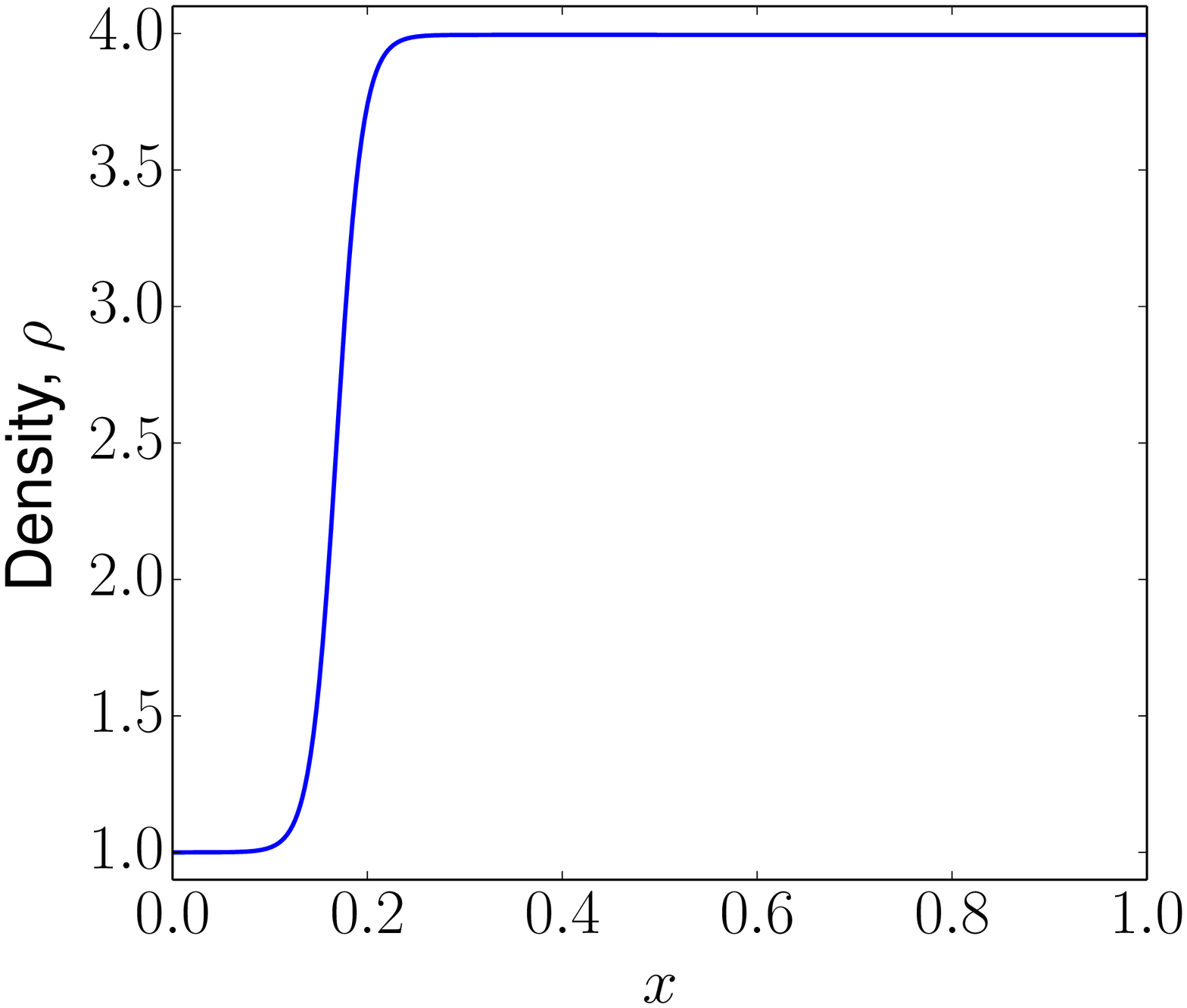}\\
\includegraphics[scale = .4]{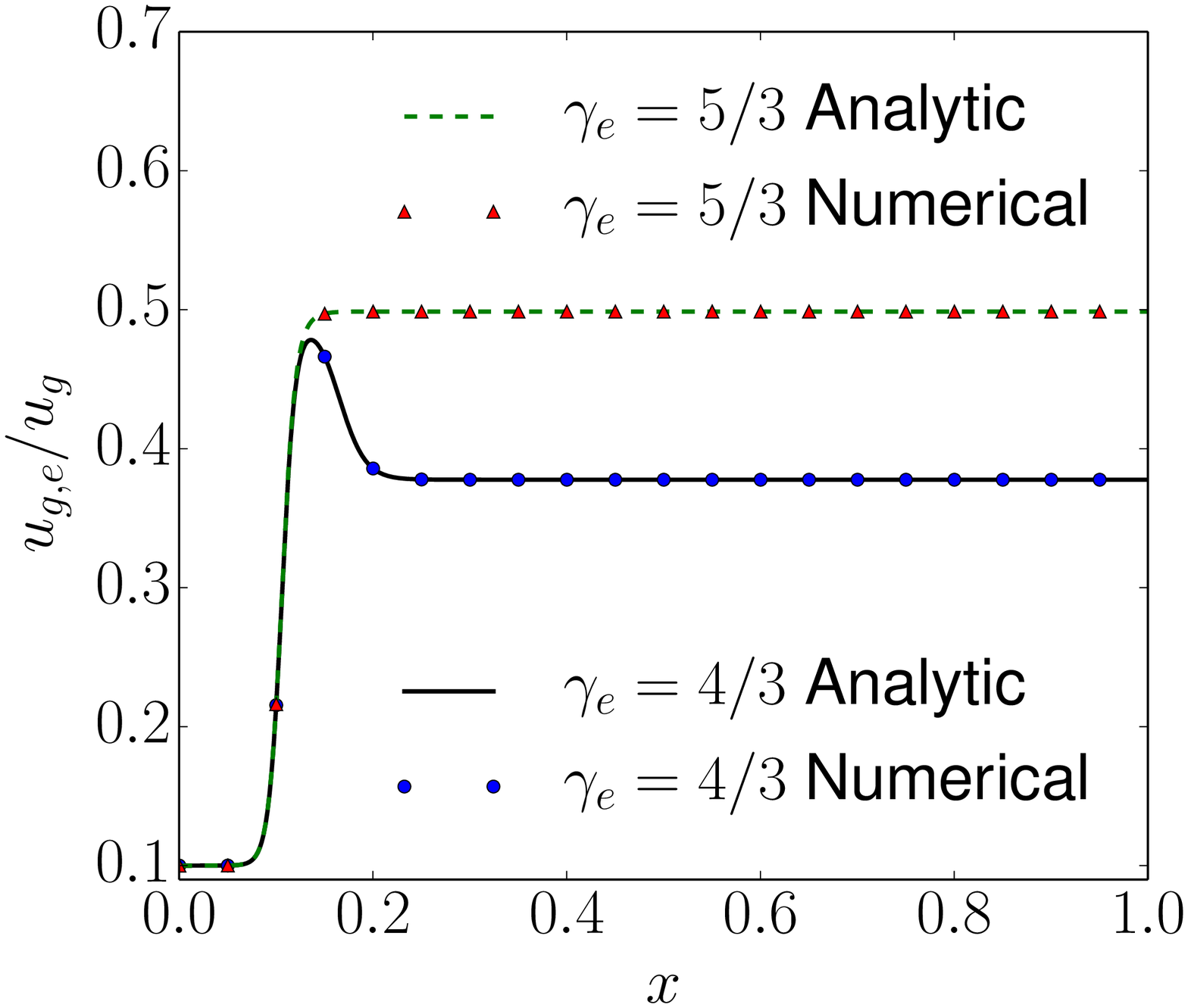}
\caption{High Mach number ($\sim 49$), stationary, viscous shock results for an electron heating fraction $f_e = 0.5$ at a resolution of $2000$ cells. Top: fluid density. Bottom: electron internal energy relative to total fluid internal energy.  Both the $\gamma_e = 4/3$ and $\gamma_e = 5/3$ electrons display good agreement with the analytic solution, converging at 2nd order (see Figure~\ref{fig:viscshockcon}). This is in contrast to the formulation without explicit viscosity used in \S\ref{sec:testshock}, in which the shock structure is always just a few grid points.   An accurate calculation of the shock heating requires a well-resolved shock structure (i.e., a shock with a finite width), which is provided by adding explicit bulk viscosity to the fluid equations. Given that the error incurred by our numerical scheme without explicit viscosity ($\sim3\%$) is acceptable for our purposes, we do not use explicit viscosity in our calculations. }
\label{fig:viscshock}
\end{figure}

\begin{figure}
\includegraphics[scale = .4]{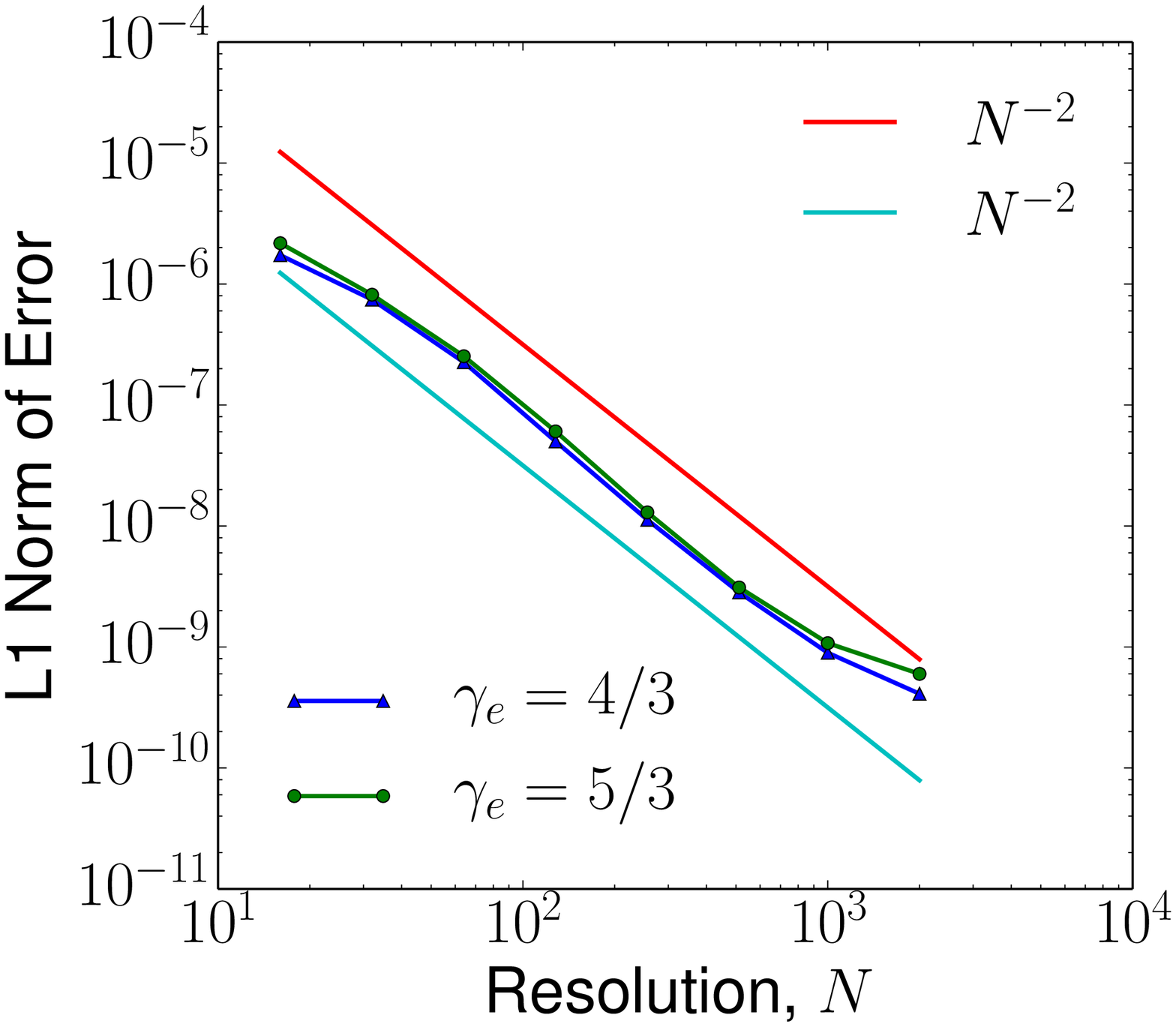}
\caption{Convergence results for the electron internal energy in a steady-state, 1D, high Mach number, viscous shock as compared to the analytic solution (see Appendix~\ref{App:viscshock}).  Both the $\gamma_e = 5/3$ and $\gamma_e = 4/3$ electrons converge at 2nd order, as opposed to the non-viscous shock of \S\ref{sec:testshock} where only the $\gamma_e = 5/3$ electrons converged to the analytic solution.  Second order convergence is achieved in this test problem because the shock profile is well-resolved and continuous.  This shows that our method correctly captures the dissipation in strong shocks when the shock profile can be resolved. At the highest resolution, relativistic corrections to the (non-relativistic) analytic solution become important so the error no longer converges at second order.}
\label{fig:viscshockcon}
\end{figure}

For our numerical test, we do not use the standard Noh test as outlined in \S\ref{sec:testshock} due to the problems noted by the original paper \citep{Noh1987}.  For any numerical scheme that gives the shock a finite width, the formation of the shock from the converging flow undershoots the density at the center of the grid by a finite amount that does not disappear at higher resolution.  Given this difficulty, our numerical test is instead to set the initial and boundary conditions of both the fluid and electron variables equal to the analytic solution for a stationary shock (e.g., equation~\ref{eq:vsubx}) and evolve for a dynamical time of $L/v_i$, where $L$ is the grid size. We choose $\gamma=5/3$, $v_i = 10^{-2} c$, $M \sim 49$, and $\nu = 0.01 v_i L$.  Figure~\ref{fig:viscshock} shows both the density profile and the ratio of the electron internal energy to the total internal energy for both $\gamma_e = 4/3$ and $\gamma_e = 5/3$ electrons at the end of the run as compared to the analytic solution (equation~\ref{eq:ugesubrho}).  We find good agreement with the analytic solution and second order convergence (Figure~\ref{fig:viscshockcon}) up to the resolution at which relativistic errors in the analytic solution become important ($\delta u_g /u_g \sim (v/c)^2 \sim 10^{-4}$). \label{App:viscshock}

\onecolumn

\section{Torus Initial Conditions}
\label{app:fishbone}
In this appendix we describe in more detail the initial configuration of the torus in our simulations of an accreting black hole. In all expressions that follow we measure radii in units of the gravitational radius $r_g \equiv GM/c^2$ (or equivalently set $G=M=c=1$).  

\citet{Fishbone1976} derived an equilibrium solution (their equation 3.6) of the general relativistic hydrodynamic equations in the Kerr metric in terms of the relativistic enthalpy, $h \equiv (\rho + P_g + u_g)/\rho$, and the constant angular momentum per unit mass, $l \equiv u_{\varphi}u^t$. We use their equation 3.6 exactly as presented when $r>r_{in}$ and when the right-hand side is positive, otherwise we set $\rho = P = 0$. Additionally, we assume an adiabatic equation of state, $P = \kappa_0 \rho^\gamma$, for some choice of $\kappa_0$, and fix $l$ such that the density maximum occurs at $r_{\rm max}$:
\begin{equation}
  \begin{aligned}
  l =& \left\{\frac{\left[a^2-2a\sqrt{r_{\rm max}} + r_{\rm max}^2\right]\left[-2ar_{\rm max}\left(a^2-2a\sqrt{r_{\rm max}} + r_{\rm max}^2\right)\right]}{\sqrt{2a\sqrt{r_{\rm max}}+r_{\rm max}^2-3r_{\rm max}}} +\frac{\left(a+\sqrt{r_{\rm max}}(r_{\rm max}-2)\right)\left(r_{\rm max}^3+a^2(r_{\rm max}+2)\right)}{\left(a^2 + r_{\rm max}^2 -2r_{\rm max}\right)\sqrt{1+2ar_{\rm max}^{-3/2}-3/r_{\rm max}}}\right\} \\&\times \frac{1}{r_{\rm max}^3\sqrt{2a\sqrt{r_{\rm max}}+r_{\rm max}^2 - 3r_{\rm max}}},
  \end{aligned}
\end{equation}
where $a$ is the dimensionless spin parameter of the black hole. This expression for $l$ is equivalent to the Keplerian value at $r =r_{\rm max}$.
 
\label{lastpage}
\end{document}